\documentclass[journal]{IEEEtran}

\usepackage[english]{babel}
\usepackage[utf8x]{inputenc}
\usepackage{array, xcolor}         
\usepackage{amsmath,amsfonts,graphicx, amssymb, placeins}
\usepackage{enumerate, url} 
\usepackage{xargs}
\usepackage{cite}
\usepackage{subfig} 
\usepackage{dsfont}
\usepackage{enumitem}
\usepackage[colorinlistoftodos]{todonotes}
\usepackage{filecontents}
\usepackage{algorithm}
\usepackage[noend]{algpseudocode}
\usepackage{overpic}
\usepackage{float}
\usepackage[
  pass,
]{geometry}
\usepackage{soul}
\soulregister\cite7
\soulregister\ref7
\soulregister\pageref7
\newenvironment{myalign}{\par\nobreak\small\noindent\align}{\endalign}

\newtheorem{theorem}{Theorem}[section]
\newtheorem{lemma}[theorem]{Lemma}
\newtheorem{proposition}[theorem]{Proposition}

\newtheorem{corollary}[theorem]{Corollary}
\newtheorem{observation}[theorem]{Observation}
\newtheorem{definition}[theorem]{Definition}
\newenvironment{IEEEproof-sketch}{{\bf Proof Sketch:}}{\hfill\rule{2mm}{2mm}}
\makeatletter
\newsavebox\myboxA
\newsavebox\myboxB
\newlength\mylenA

\newcommand*\xoverline[2][0.75]{%
    \sbox{\myboxA}{$\m@th#2$}%
    \setbox\myboxB\null
    \ht\myboxB=\ht\myboxA%
    \dp\myboxB=\dp\myboxA%
    \wd\myboxB=#1\wd\myboxA
    \sbox\myboxB{$\m@th\overline{\copy\myboxB}$}
    \setlength\mylenA{\the\wd\myboxA}
    \addtolength\mylenA{-\the\wd\myboxB}%
    \ifdim\wd\myboxB<\wd\myboxA%
       \rlap{\hskip 0.5\mylenA\usebox\myboxB}{\usebox\myboxA}%
    \else
        \hskip -0.5\mylenA\rlap{\usebox\myboxA}{\hskip 0.5\mylenA\usebox\myboxB}%
    \fi}
\makeatother

\makeatletter
\g@addto@macro\normalsize{%
  \setlength\abovedisplayskip{3pt}
  \setlength\belowdisplayskip{3pt}
  \setlength\abovedisplayshortskip{3pt}
  \setlength\belowdisplayshortskip{3pt}
}
\makeatother

\graphicspath{{imgs/}}

\usepackage{soul}
\newtheorem{condition}[theorem]{Condition}

\title{Resource Allocation and Rate Gains\\ in Practical Full-Duplex Systems}
\author{Jelena Mara\v{s}evi\'{c}, Jin Zhou, Harish Krishnaswamy, Yuan Zhong, Gil Zussman
\thanks{A partial and preliminary version of this paper appeared in the Proceedings of ACM SIGMETRICS'15 \cite{full-duplex-sigmetrics}.}
  \thanks{J. Mara\v{s}evi\'{c}, J. Zhou, H. Krishnaswamy, and G. Zussman are with the Electrical Engineering dept. and Y. Zhong is with the Industrial Engineering and Operations Research dept., Columbia University, New York, NY, 10027, USA, email: {\{jelena@ee., jz2495@, harish@ee., yz2561@, gil@ee.\}columbia.edu.}
}
} 

\begin{document}

\belowdisplayskip=8pt plus 1pt minus 9pt
\belowdisplayshortskip=5pt plus 3pt minus 4pt
\maketitle
\begin{abstract}
Full-duplex communication has the potential to substantially increase the throughput in wireless networks. 
However, the benefits of full-duplex are still not well understood. In this paper, we characterize the full-duplex rate gains in both single-channel and multi-channel use cases. For the single-channel case, we \emph{quantify the rate gain as a function of the remaining self-interference and SNR values}. We also provide a sufficient condition under which the sum of uplink and downlink rates on a full-duplex channel is concave in the transmission power levels. Building on these results, we consider the multi-channel case. For that case, we \emph{introduce a new realistic model of a compact (e.g., smartphone) full-duplex receiver} and demonstrate its accuracy via measurements. 
We study the problem of jointly allocating power levels to different channels and selecting the frequency of maximum self-interference suppression, where the objective is maximizing the sum of the rates over uplink and downlink OFDM channels. We develop a \emph{polynomial time algorithm which is nearly optimal {in practice} under very mild restrictions}. To reduce the running time, we develop an efficient  nearly-optimal algorithm under the high SINR approximation. Finally, we demonstrate via numerical evaluations the capacity gains in the different use cases and obtain insights into the impact of the remaining self-interference and wireless channel states on the performance. 

\end{abstract}
\section{Introduction}\label{section:introduction}
Full-duplex (FD) communication -- simultaneous transmission and reception on the same frequency channel -- 
holds great promise of substantially improving the throughput 
in wireless networks. 
The main challenge hindering the implementation of practical FD devices is high self-interference (SI) caused by signal leakage from the transmitter into the receiver circuit. The SI signal is usually many orders of magnitude higher than the desired signal at the receiver's input, requiring over 100dB (i.e., by $10^{10}$ times) of self-interference cancellation (SIC).

Cancelling SI is a very challenging problem. Even though different techniques of SIC were proposed over a decade ago, only recently receiver designs that provide sufficient SIC to be employed in Wi-Fi and cellular networks emerged (see \cite{FullDuplex_RiceU_JSCAinvited14} and references therein for an overview). Exciting progress was made in the last few years by various research groups demonstrating that a combination of SIC techniques employed in both analog and digital domains can provide sufficient SIC to support practical applications \cite{choi2010achieving, choi2012beyond, jain2011practical, duarte2012characterization, everett2013passive, khojastepour2011case, aryafar2012midu, nec2012wideband, knox2012single, bharadia2013full, FD_MultiANT_TVT14, FD_MultiANT_Khandani10, FD_AnalogSIC_McMichael12, FD_PhaNoise_Rice2012}.

\begin{figure}[t!]
\center
\subfloat[]{\label{fig:dual}\includegraphics[height=0.83in, ]{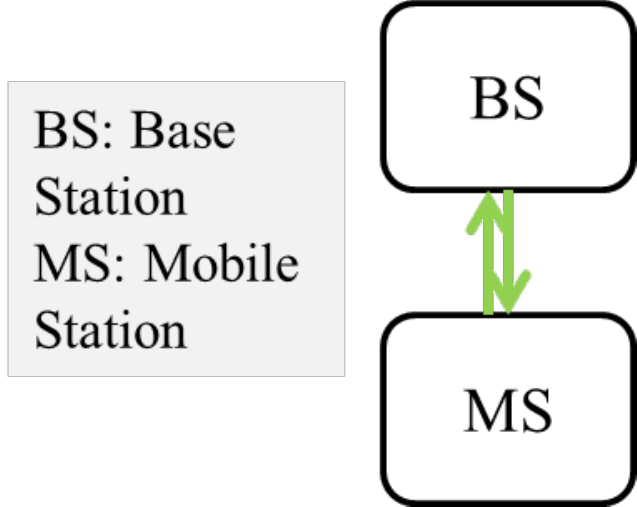}}\hspace{\fill}
\subfloat[]{\label{fig:cross}\includegraphics[height = 0.85in]{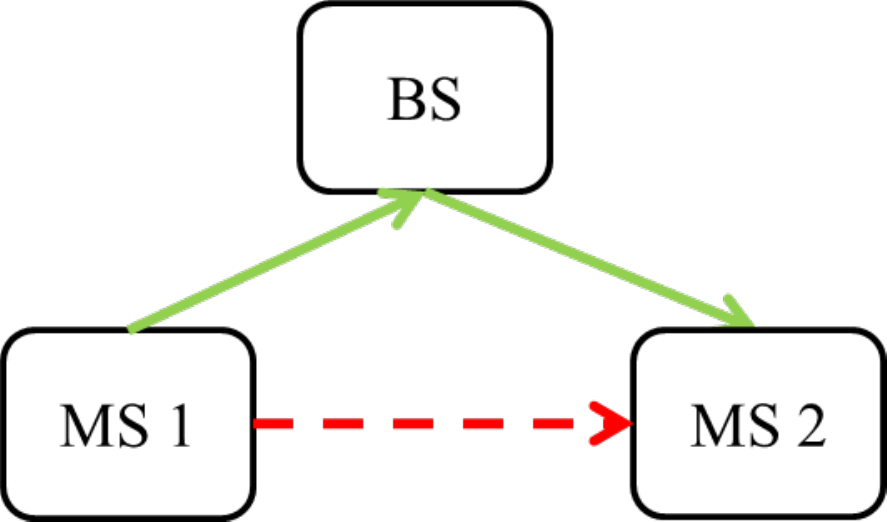}}\hspace{\fill}
\subfloat[]{\label{fig:multi}\includegraphics[height = 0.85in]{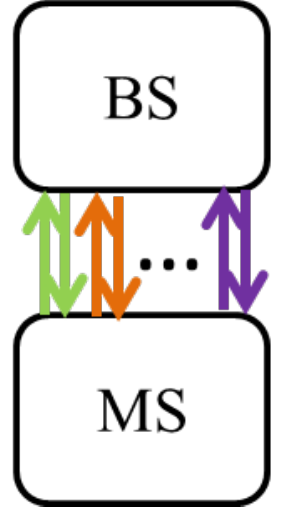}}\hspace{\fill}
\caption{Some possible uses of full-duplex: \protect\subref{fig:dual} simultaneous UL and DL for one MS; \protect\subref{fig:cross} UL and DL used by two different MSs and caused inter-node interference (red dashed line), \protect\subref{fig:multi} simultaneous UL and DL over {OFDM} channels.}\vspace{-10pt}
\label{fig:full-duplex-links}
\end{figure}
 
While there has been significant interest in FD from both industry and academia \cite{singh2011efficient, choi2010achieving, choi2012beyond, jain2011practical, duarte2012characterization, everett2013passive, khojastepour2011case, aryafar2012midu, nec2012wideband, knox2012single, bharadia2013full, FD_MultiANT_TVT14, FD_MultiANT_Khandani10, FD_AnalogSIC_McMichael12, FD_PhaNoise_Rice2012, sahai2013uplink, bai2013distributed, goyal2014improving, xie2014does, ahmed2013rate, li2014rate, cheng2013optimal}, \emph{the exact rate gains resulting from the use of FD are still not well understood}. 
The first implementations of FD receivers optimistically envisioned 100\% rate improvement (e.g., \cite{jain2011practical, bharadia2013full}). To achieve such an increase in data rates, the FD receiver would need perfect SIC, namely, to cancel SI to at least one order of magnitude below the noise floor to render it negligible. The highest reported SIC \cite{bharadia2013full}, however,  suppresses the SI to the level of noise.

Despite this insufficient cancelling capabilities, much of the work on FD rate improvement assumes perfect SIC in the FD receiver \cite{sahai2013uplink, bai2013distributed, goyal2014improving, xie2014does}. While non-negligible SI has also been considered \cite{ahmed2013rate, li2014rate, cheng2013optimal}, \emph{there are still no explicit bounds on the rate gains for given FD circuit parameters and parameters of the wireless signal}. 
Moreover, from a modeling perspective, the frequency selectivity of SIC has not been considered in any 
analytical work. This is \emph{an important feature that is inherent in conventional 
compact implementations of an FD receiver, 
such as that found in small-form factor mobile devices (e.g., smartphones and tablets)}, where frequency selectivity is mainly a consequence of the cancellation in the RF domain.\footnote{See our recent work \cite{ZhouActiveTXCancISSCC14,Zhou_NCSIC_JSSC14} and Section \ref{section:circuit-model} for more details.}      
\vspace{-10pt}
{\subsection{Summary of Contributions}}
\emph{{T}he main contribution of this paper is a thorough analytical study of rate gains from FD under non-negligible SI.} 
We consider both \emph{single-channel and multi-channel orthogonal frequency division multiplexing (OFDM) scenarios}. 
For the multi-channel case, \emph{we develop a new model for frequency-selective SIC 
in small-form factor receivers}. 
Our results provide explicit guarantees on the rate gains of FD, as a function of 
receivers' signal-to-noise ratios (SNRs) and SIC profile. Our analysis provides several insights into the structure of the sum of uplink (UL) and downlink (DL) rates under FD, which will be useful for future work on FD MAC layer algorithm design.

Specifically, we consider three different use cases of FD, as illustrated in Fig.~\ref{fig:full-duplex-links}:~(i) a single channel bidirectional link, where one mobile station (MS) communicates with the base station (BS) both on the UL and on the DL (Fig.~\ref{fig:full-duplex-links}\subref{fig:dual}); (ii) two single channel unidirectional links, where one MS communicates with the BS on the UL, while another MS communicates with the BS on the DL (Fig.~\ref{fig:full-duplex-links}\subref{fig:cross}); and (iii) 
a multi-channel bidirectional link, where one MS communicates with the BS over multiple {OFDM} channels, both on the UL and on the DL (Fig.~\ref{fig:full-duplex-links}\subref{fig:multi}).

\subsubsection{{Models of Residual SI}} 
For SI, we consider two different models. 
For the BS in all use cases and the MS in use case (i), we model the remaining SI after cancellation as a constant fraction of the transmitted signal.  
Such design is possible for devices that do not require a very small form factor (e.g., base stations), and was demonstrated in \cite{bharadia2013full}. 

In the multi-channel case, we rely on the characteristics of RFIC receivers that we recently designed \cite{ZhouActiveTXCancISSCC14,Zhou_NCSIC_JSSC14} and \emph{develop a frequency selective model for the remaining SI in a small form-factor device} {(Section \ref{section:circuit-model})}. We demonstrate the accuracy of the developed model via measurements with our receivers \cite{ZhouActiveTXCancISSCC14,Zhou_NCSIC_JSSC14}. 
We note that a frequency-selective profile of SIC that we model is inherent to RF cancellers with flat amplitude and phase response (see Section \ref{section:circuit-model}). {A} mixed-signal SIC architecture \cite{FD_PhaNoise_Rice2012} {where the digital TX signal is processed and upconverted to RF for cancellation} does not necessarily have flat amplitude and phase response{. However,} we do not consider this architecture because it requires a{n} {additional} up-conversion path {compared to the architecture of this work,} {and this additional path} introduces its own noise and distortion, limiting the resultant RF SIC. 

\subsubsection{{Sum Rate Maximization}}
We focus on the problem of maximizing the sum of UL and DL rates under FD (referred to as the sum rate in the rest of the paper).  
This problem, in general, is neither concave nor convex in the transmission power levels, 
since the remaining SI after cancellation depends on the transmission power level. 
Due to the lack of  {a} good problem structure, existing analytical results (see e.g., \cite{ahmed2013rate, cheng2013optimal, li2014rate}) are often restricted to specialized settings. 
Yet, we obtain several analytical results on the FD rate gains, often under mild restrictions, 
by examining closely the structural properties of the sum rate function. 

{\noindent\textbf{Single-Channel Results. }}In the single-channel cases, we prove that if any rate gain can be achieved from FD, then the gain is maximized by setting the transmission power levels to their respective maximum values. This result is somewhat surprising because of the lack of good structural properties of the sum rate. 
We then derive a sufficient condition under which the sum rate is {bi}concave{\footnote{{A function is biconcave, if there exists a partition of variables into two sets, such that the function is concave when variables from either set are fixed.}}} in both transmission power levels, and show that when this condition is not satisfied, one cannot gain more than 1b/s/Hz (additively) from FD as compared to time-division duplex (TDD). We note that although the model for the remaining SI in the single channel case is relatively simple, it nonetheless captures the main characteristics of the FD receivers. Moreover, the results for the single channel case under this model are fundamental for analyzing the multi-channel setting, and often extend to this more general setting.

{\noindent\textbf{Multi-Channel Results. }}In the multi-channel case, we use the frequency-selective SI model for the MS receiver { that is introduced in Section \ref{section:remaining-si-model} and motivated by FD implementation challenges discussed in Section \ref{section:circuit-model}}{. Based on this model, we} study the problem of transmission power allocation over {OFDM} channels and frequency selection, where the objective is to  maximize the sum of the rates over UL and DL {OFDM} channels (in this case, frequency refers to the frequency of maximum SIC of the SI canceller). Although in general it is hard to find an optimal solution to this problem, we develop {an algorithm that converges to a stationary point (in practice, a global maximum)} under two mild technical conditions. One condition ensures that the sum rate is {bi}concave in transmission power levels. This restriction is mild, since we prove that when it does not hold, the possible gains from FD are small. The other condition imposes bounds on the magnitude of the first derivative of the sum rate in terms of maximum SIC frequency, and has a negligible impact on the sum rate in {OFDM} systems with a large number of channels, because it can only affect up to 2 {OFDM} channels (see Section \ref{section:OFDM-sum-rate} for more details). 

Although the algorithm {in practice converges to a near-optimal solution} and runs in polynomial time, its running time is relatively high. 
Therefore, 
we consider a high SINR approximation of the sum rate, and derive fixed optimal power allocation and maximum SIC frequency setting that maximizes the sum rate up to an additive $\epsilon$ in time $O(K\log(1/\epsilon))$, for any given $\epsilon$, where $K$ is the number of channels. 

{\noindent\textbf{Numerical Results. }}Finally, we note that throughout the paper, we provide numerical results that quantify the rate gains in various use cases and illustrate the impact of different parameters on these gains. For example, for the multi-channel case, we evaluate the rate gains using measured SI of our RFIC receiver \cite{ZhouActiveTXCancISSCC14, Zhou_NCSIC_JSSC14}. We use algorithms for the general SINR regime and for the high SINR regime {and compare their results to those obtained by allocating power levels equally among the OFDM channels. Our results suggest that whenever the rate gains from FD are non-negligible, all considered power allocation policies yield similar rate gains. Therefore, one of the main messages of our work is that  \emph{whenever it is beneficial to use FD, simple power allocation policies are near-optimal}.} 
{\subsection{Organization of the Paper}}
The rest of the paper is organized as follows. Section \ref{section:related-work} reviews related work and Section \ref{section:circuit-model} outlines the challenges in implementing FD receivers. Section \ref{section:preliminaries} introduces the new model of a small form factor FD receiver, and the model for the various use cases. Section \ref{section:single-channel} provides analysis and numerical evaluation for the sum rate maximization on a single channel for use case{s} (i) {and (ii)}.  Sections \ref{section:OFDM} and \ref{section:OFDM-num-results} provide analysis, algorithms, and numerical evaluation for use case (iii). We conclude in Section \ref{section:conclusion}.  
\section{Related Work}\label{section:related-work}

Possible rate gains from FD have been studied in \cite{sahai2013uplink, bai2013distributed, goyal2014improving, ahmed2013rate, li2014rate, xie2014does, cheng2013optimal}, with much of the work \cite{sahai2013uplink, bai2013distributed, goyal2014improving, xie2014does} focusing on perfect SIC. 
Unlike this body of work, we focus on rate gains from FD communication under imperfect SIC.

Non-negligible SI has been considered in \cite{ahmed2013rate, cheng2013optimal, li2014rate}. A sufficient condition for achieving positive rate gains from FD on a bidirectional link has been provided in \cite{ahmed2013rate}, 
for the special case of equal SINRs on the UL and DL. This condition does not quantify the rate gains. 

Power allocation over orthogonal bidirectional links was considered in \cite{cheng2013optimal} and \cite{li2014rate} for MIMO and OFDM systems, respectively. 
The model used in \cite{cheng2013optimal} assumes the same amount of SIC and equal power allocation on all channels, which is a {\emph{less general model than the one that we consider}}.

A more detailed model with different SIC over OFDM channels was considered in \cite{li2014rate}. The model from \cite{li2014rate} does not consider dependence of SIC in terms of canceller frequency (although, unlike our work, it takes into account the transmitter's phase noise). Optimal power allocation that maximizes one of the rates when the other is fixed is derived for \emph{equal power levels} across channels, while for the general case of unequal power levels, \cite{li2014rate} only provides a heuristic solution. 

Our work relies on structural properties of the sum rate to derive near-optimal power allocation and maximum SIC frequency setting that maximizes the sum rate.~While the model we consider is different than \cite{ahmed2013rate, li2014rate}, we provide a more specific characterization of achievable rate gains, and derive results that provide insights into the rate dependence on the power allocation. These results allow us to solve a very general problem of rate maximization.

\section{FD Implementation Challenges}\label{section:circuit-model}
\begin{figure}
	\centering
	\includegraphics[scale=0.65]{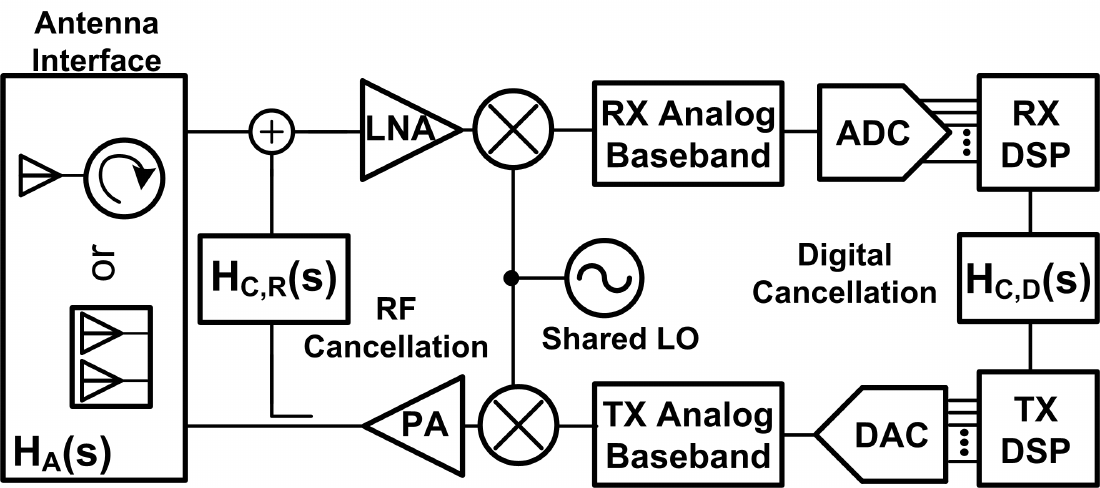}
	\caption{Block diagram of a full-duplex transceiver employing RF and digital cancellation.}
	\label{fig:Diagram_FDradio}\vspace{-10pt}
\end{figure}

{In this section, we overview the challenges associated with the implementation of compact FD radios. These challenges motivate the model of remaining SI that is introduced in Section \ref{section:remaining-si-model} and used in the design of sum-rate maximization algorithms (Section \ref{section:OFDM}).}

Fig.~\ref{fig:Diagram_FDradio} shows the block diagram of a full-duplex transceiver. There are two antenna interfaces that are typically considered for full-duplex operation: (i) an antenna pair and (ii) a circulator. The advantage of using a circulator is that it allows a single antenna to be shared between the transmitter (TX) and the receiver (RX). SIC must be performed in both the RF and digital domains to achieve in excess of 100dB SI suppression. The RF canceller taps a reference signal at the output of the power amplifier (PA) and performs SIC at the input of the low-noise amplifier (LNA) at the RX side \cite{FullDuplexJSAC14}. 

Typically, 20-30dB of SIC is required from the RF, given that the antenna interface typically has a TX/RX isolation of 20-30dB\cite{Circulator}. Thus, an overall 50-60dB RF TX/RX isolation is achieved before digital SIC is engaged. This amount of RF TX/RX isolation is critical to alleviate the RX linearity and the analog-to-digital conversion (ADC) dynamic range requirements \cite{FullDuplex_RiceU_JSCAinvited14,FullDuplexJSAC14}. Digital cancellation further cancels the linear SI as well as the non-linear distortion products generated by the RX or the RF canceller. 

A mixed-signal SIC architecture has been proposed in \cite{FD_PhaNoise_Rice2012}, where the digital TX signal is processed and upconverted to RF for cancellation. However, this requires a separate up-conversion path which introduces its own noise and distortion. Moreover, the noise and distortion of the TX analog and RF circuits (such as the power amplifier) are not readily captured in the cancellation signal, limiting the resultant RF SIC. In addition, the dedicated up-conversion path results in area and power overhead. Because of these reasons, we are not considering this SIC architecture in this paper.

\begin{figure}
	\centering
	\subfloat[]{
		\includegraphics[keepaspectratio,height=0.2\linewidth]{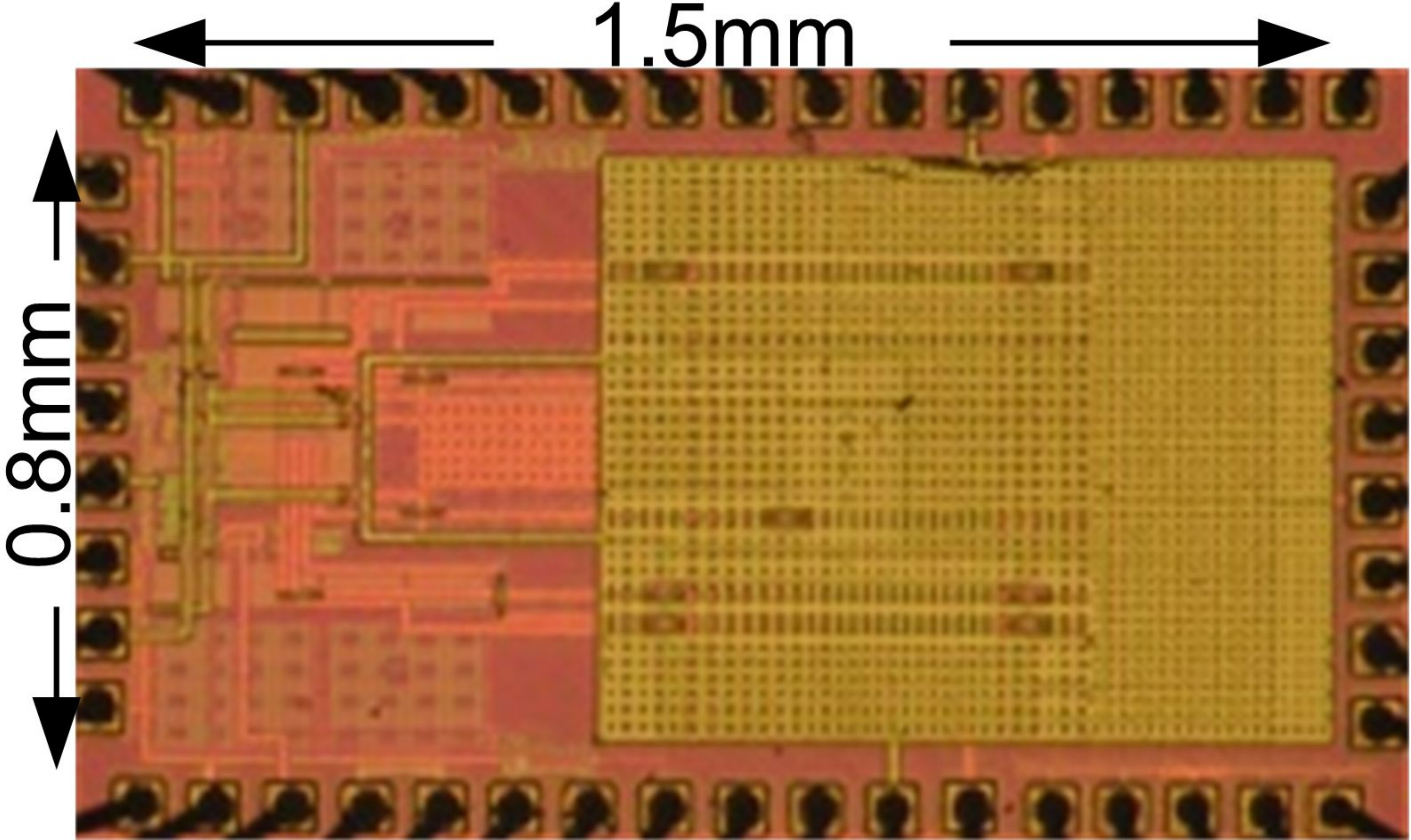}
		\label{fig:ChipPhoto}\
	}	\hspace{\fill}\vspace{-2pt}
	\subfloat[]{
		\includegraphics[keepaspectratio,height=0.2\linewidth]{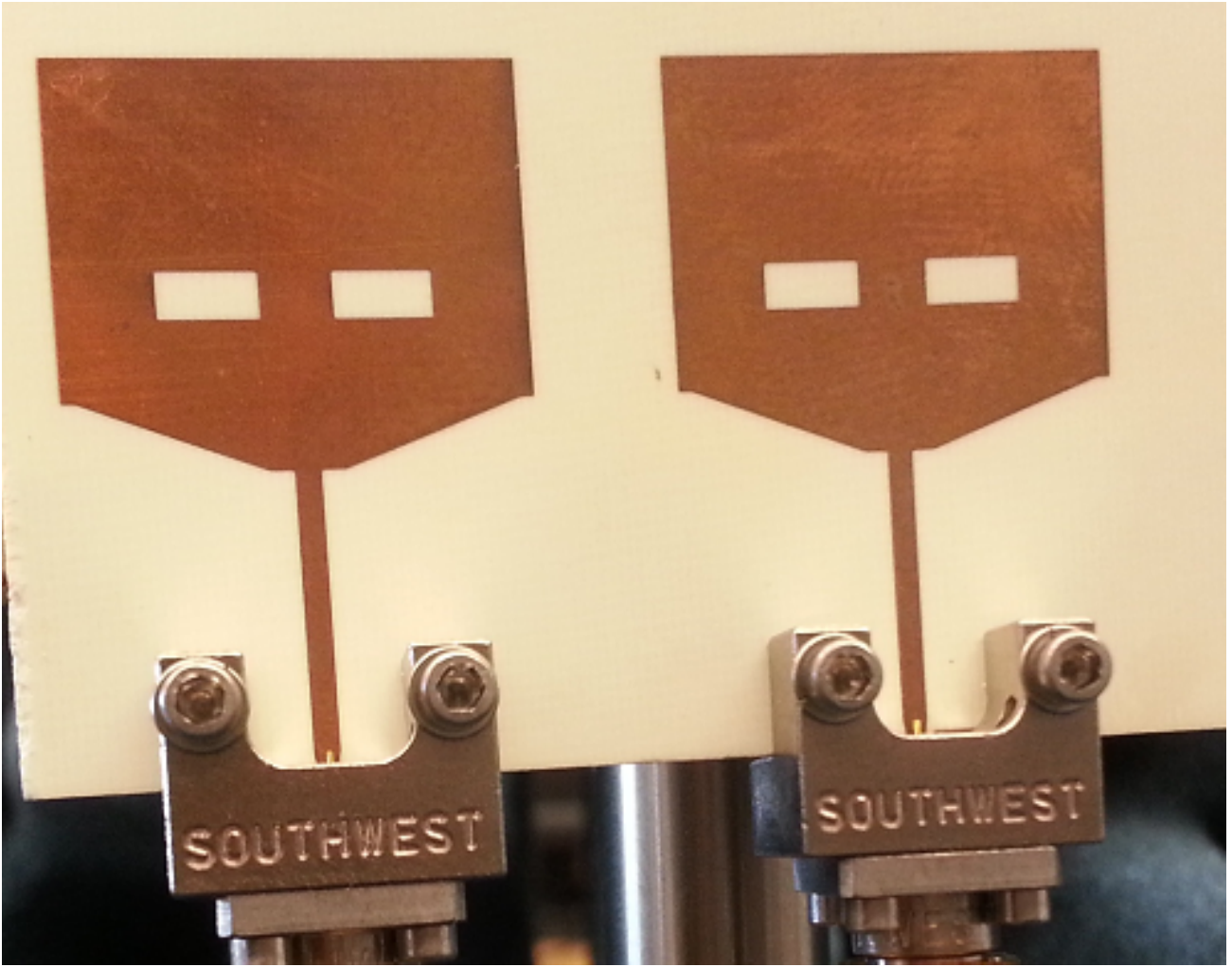}
		\label{fig:AntennaPair}\
	} \hspace{\fill}\vspace{-2pt}
	\subfloat[]{
		\includegraphics[keepaspectratio,height=0.2\linewidth]{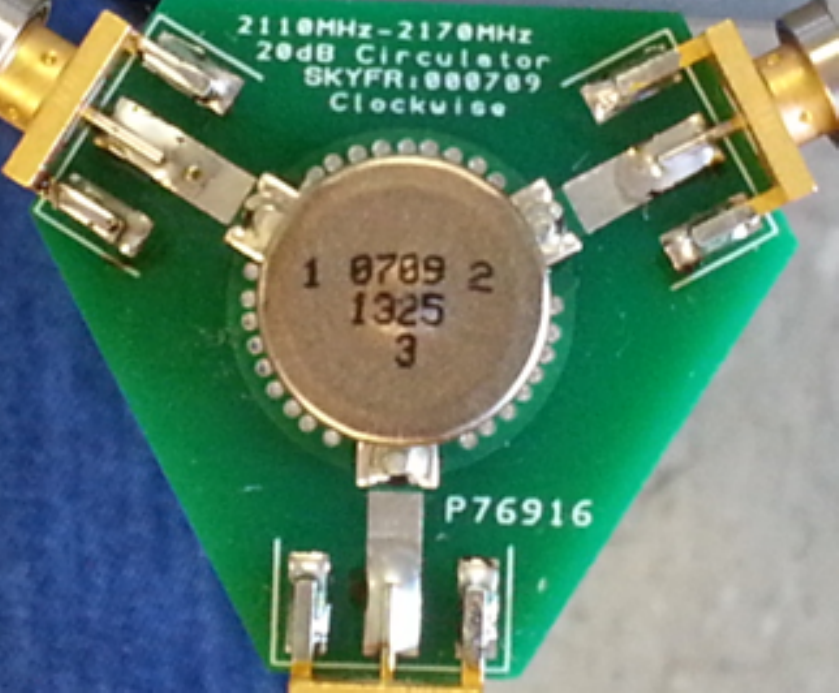}
		\label{fig:Circulator}
	}\vspace{-2pt}
	\caption{\protect\subref{fig:ChipPhoto} RFIC receiver with RF SI cancellation \cite{ZhouActiveTXCancISSCC14,Zhou_NCSIC_JSSC14} and the two antenna interfaces used in our measurements: \protect\subref{fig:AntennaPair} an antenna pair and \protect\subref{fig:Circulator} a circulator.}
	\label{fig:MeasurementSetup}\vspace{-20pt}
\end{figure}

For wideband SIC, the transfer function of the canceller must closely track that of the antenna interface across frequency. However, the frequency dependence of the inherent antenna interface isolation together with selective multi-path-ridden SI channels render this challenging for the RF canceller in particular. The net antenna interface isolation amplitude and phase response can vary significantly with frequency. A rapidly-varying phase response is representative of a large group delay, requiring bulky delay lines to replicate the selectivity in the RF canceller \cite{bharadia2013full,FullDuplexJSAC14}.

\emph{The fundamental challenge associated with wideband SIC at RF in a small form-factor and/or using integrated circuits is the generation of large time delays}. The value of true time delay is linearly proportional to the dimension of the delay structure and inversely proportional to the wave velocity in the medium. To generate 1ns delay in a silicon integrated circuit, a transmission line of 15cm length is required as the relative dielectric constant of silicon oxide is 4. A conventional integrated RF SI canceller with dimensions less than 1$\mathrm{mm}^2$ will therefore exhibit negligible delay{. Note that the
canceller phase response can be calculated by integrating the
delay with respect to frequency, and conventional integrated
RF SI cancellers typically have a flat amplitude response
\cite{ZhouActiveTXCancISSCC14,Zhou_NCSIC_JSSC14}. Therefore,} \emph{the amplitude and phase response of the canceller can be assumed to be flat with respect to frequency} when compared with antenna interface isolation, limiting the cancellation bandwidth\cite{FullDuplex_RiceU_JSCAinvited14,Zhou_NCSIC_JSSC14}.

While achieving wideband RF SI cancellation using innovative RFIC techniques is an active research topic (e.g., frequency domain equalization based RF SI cancellation in \cite{Zhou_WBSIC_ISSCC15}), in this paper we focus on compact flat amplitude- and phase-based RF cancellers, such as the one we implemented in the RFIC depicted in Fig.~\ref{fig:MeasurementSetup}\subref{fig:ChipPhoto} \cite{ZhouActiveTXCancISSCC14,Zhou_NCSIC_JSSC14}.

In \cite{ZhouActiveTXCancISSCC14} and \cite{Zhou_NCSIC_JSSC14}, the RF canceller is embedded in the RX’s LNA, and consists of a variable amplifier and a phase shifter. The RF canceller adjusts the amplitude and the phase of a TX reference signal tapped from the PA's output performing SIC at the RX input. Thanks to the co-design of RF canceller and RX in a noise-cancelling architecture, the work in \cite{ZhouActiveTXCancISSCC14} and \cite{Zhou_NCSIC_JSSC14} is able to support antenna interface with about 20dB TX/RX isolation with minimum RX sensitivity degradation.

\begin{figure}[t!]
	\centering	
	\subfloat[]{
		\includegraphics[keepaspectratio,width=0.46\linewidth]{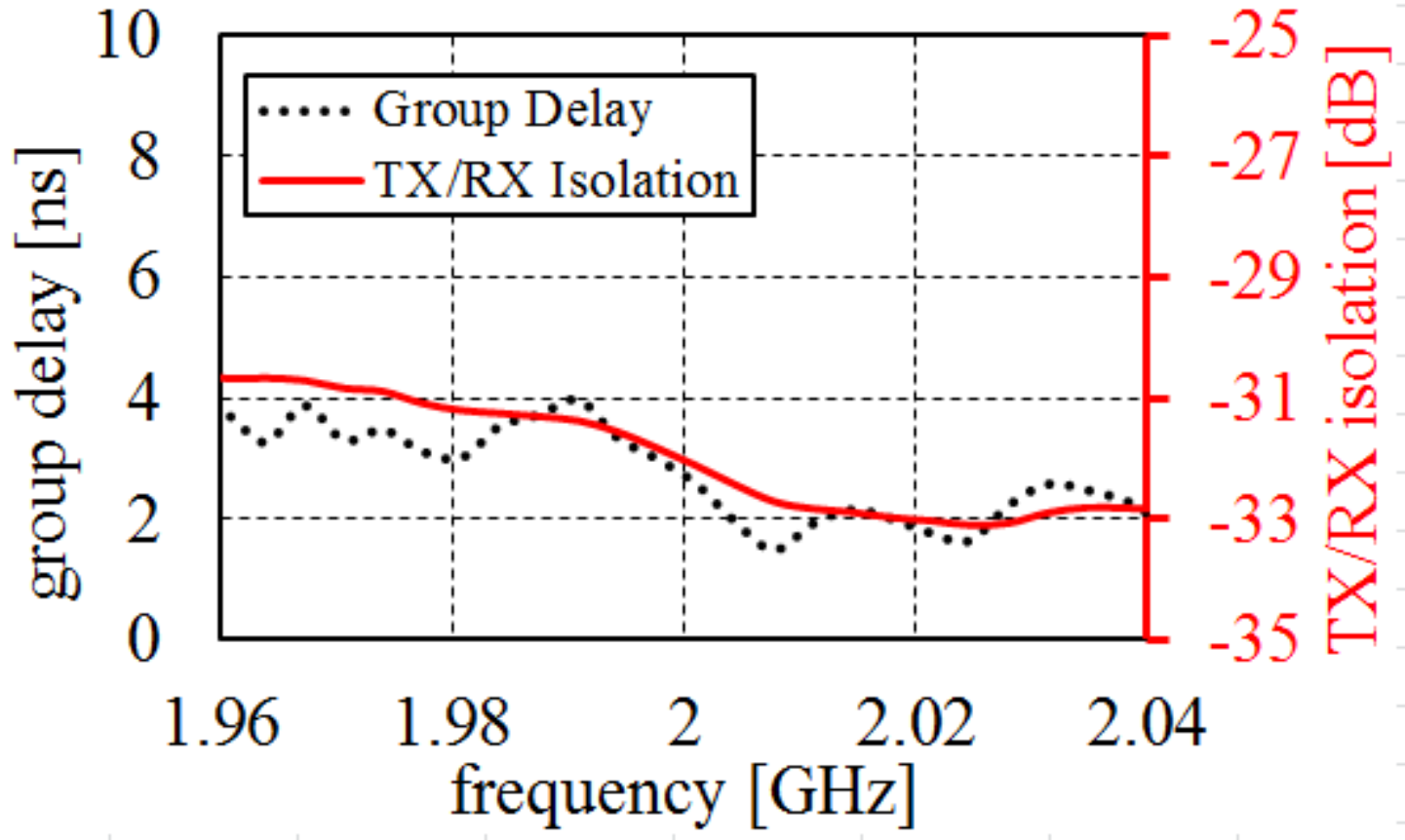}
		\label{fig:GD_ISO_ANT}
	}\hspace{\fill}
	\subfloat[]{
		\includegraphics[keepaspectratio,width=0.46\linewidth]{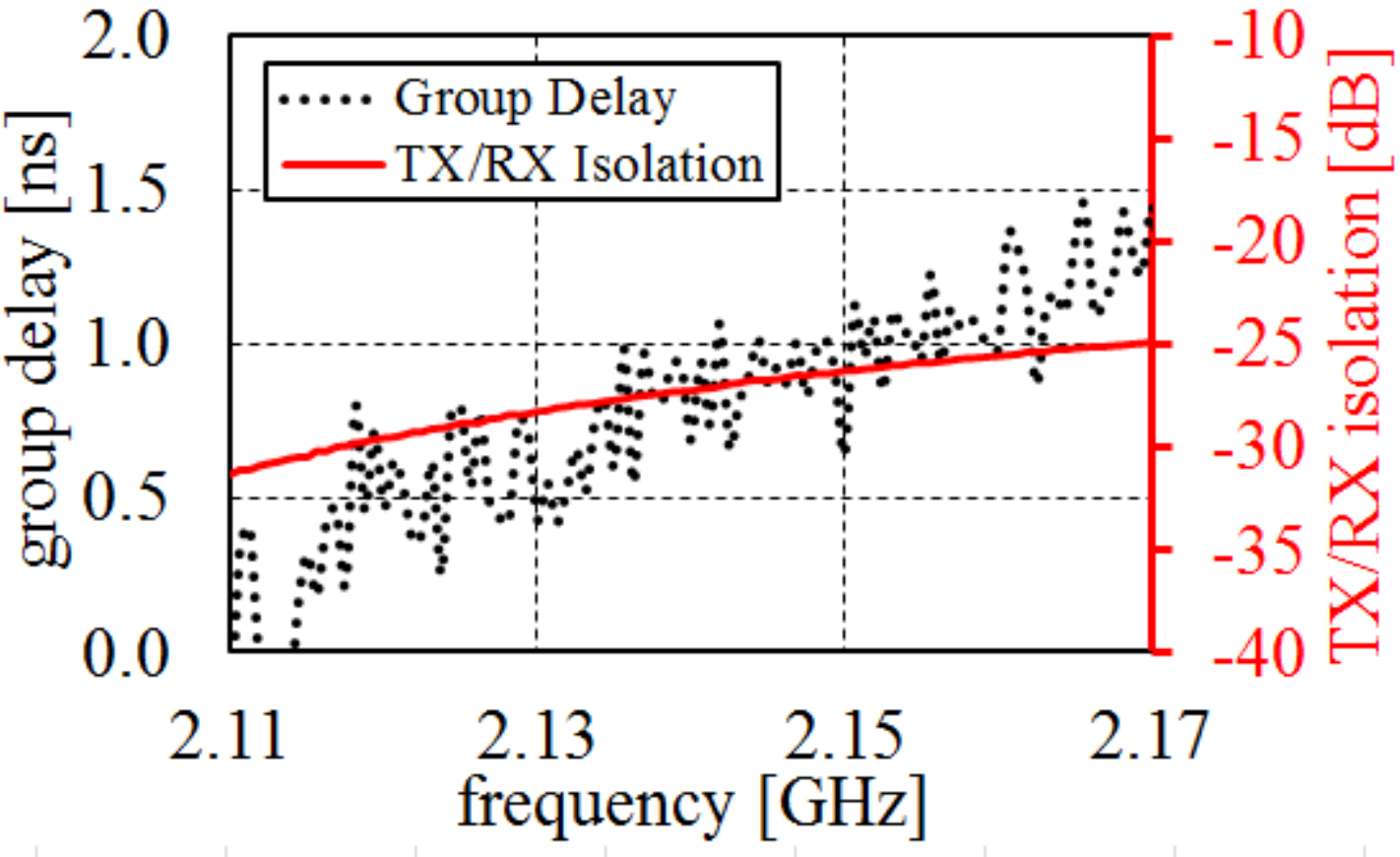}
		\label{fig:GD_ISO_Circ}
	}\\
    \vspace{-10pt}
    \subfloat[]{
		\includegraphics[keepaspectratio,width=0.46\linewidth]{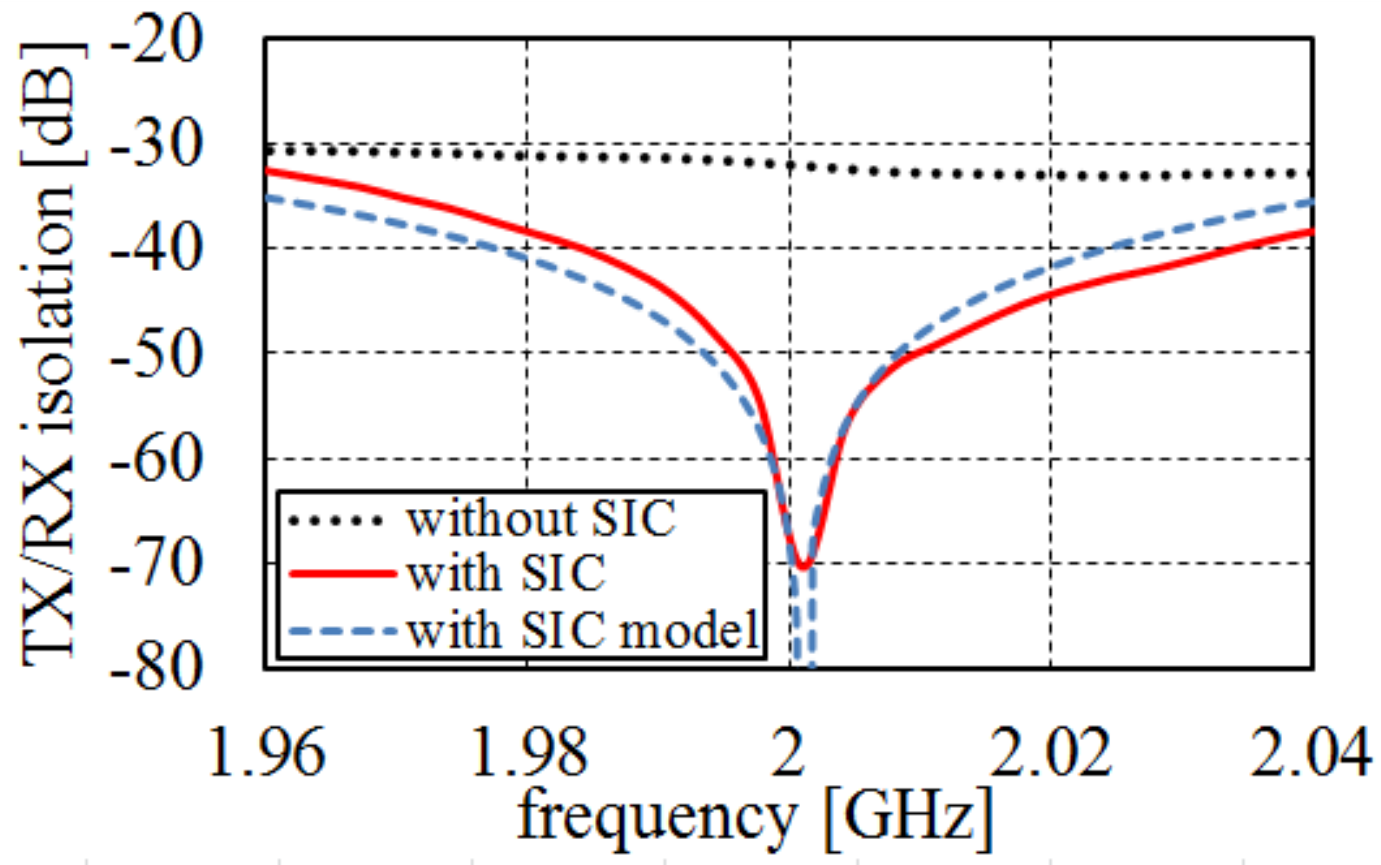}
		\label{fig:SIC_BW_ANT}
	}\hspace{\fill}
	\subfloat[]{
		\includegraphics[keepaspectratio,width=0.46\linewidth]{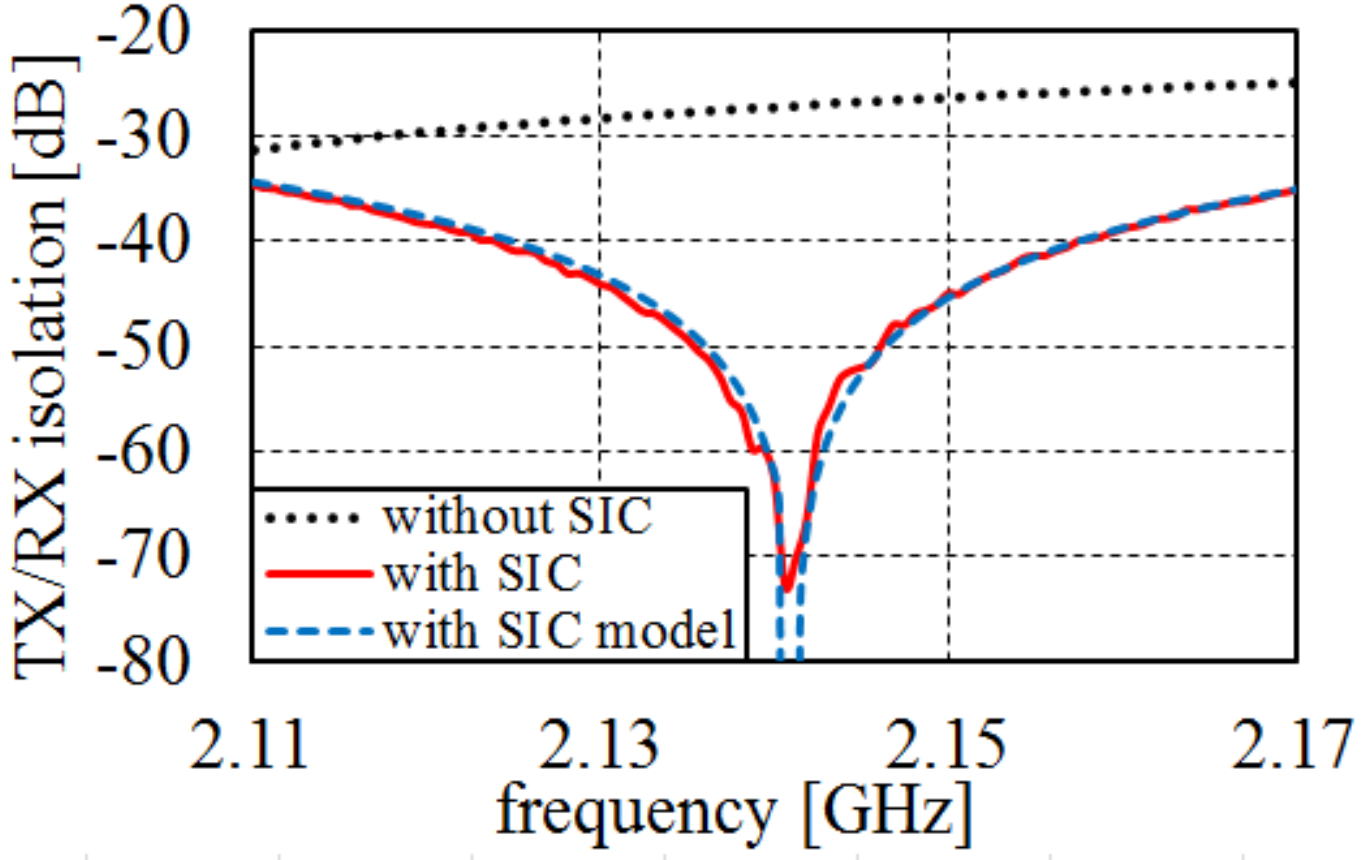}
		\label{fig:SIC_BW_Circ}
	}
	\caption{Measured isolation amplitude and group delay of \protect\subref{fig:GD_ISO_ANT} a PCB antenna pair and \protect\subref{fig:GD_ISO_Circ} a commercial 2110-2170~MHz miniature circulator from Skyworks \cite{Circulator}, and the resultant TX/RX isolation using the integrated RF canceller with flat amplitude and phase response from \cite{ZhouActiveTXCancISSCC14,Zhou_NCSIC_JSSC14} with \protect\subref{fig:SIC_BW_ANT} the antenna pair and \protect\subref{fig:SIC_BW_Circ} the circulator compared to the SIC model.} 
	\label{fig:SIC_Results}\vspace{-10pt}
\end{figure}

We measured isolation amplitude and group delay response of (i) a PCB antenna pair (see Fig.~\ref{fig:MeasurementSetup}\subref{fig:AntennaPair}) and (ii) a commercial 2110-2170MHz miniature circulator from Skyworks \cite{Circulator} (see Fig.~\ref{fig:MeasurementSetup}\subref{fig:Circulator}). The results are shown in Fig.~\ref{fig:SIC_Results}\subref{fig:GD_ISO_ANT} and Fig.~\ref{fig:SIC_Results}\subref{fig:GD_ISO_Circ}, respectively. The resultant TX/RX isolations using an RF canceller with flat amplitude and phase response after the antenna interfaces (i) and (ii) are shown in Fig.~\ref{fig:SIC_Results}\subref{fig:SIC_BW_ANT} and Fig.~\ref{fig:SIC_Results}\subref{fig:SIC_BW_Circ}, respectively. As Fig.~\ref{fig:SIC_Results}\subref{fig:SIC_BW_ANT} and Fig.~\ref{fig:SIC_Results}\subref{fig:SIC_BW_Circ} suggest, for -60dB TX/RX isolation after RF cancellation, the bandwidths are about 4MHz and 2.5MHz, respectively.

\section{Model}\label{section:preliminaries}

We consider three use cases of FD: (i) a bidirectional link, where one mobile station (MS) communicates with the base station (BS) both on the UL and on the DL (Fig.~\ref{fig:full-duplex-links}\subref{fig:dual}), (ii) two unidirectional links, where one MS is communicating with the BS on the UL, while another MS is communicating with the BS on the DL (Fig.~\ref{fig:full-duplex-links}\subref{fig:cross}), and (iii) multiple orthogonal bidirectional links (Fig.~\ref{fig:full-duplex-links}\subref{fig:multi}). Note that in (ii) only the BS is operating in FD. 

For the multi-channel FD (use case (iii)), we assume that the network bandwidth of size $B$ is subdivided into $K$ orthogonal frequency channels of width $B/K$ each, and index the frequency channels with $k\in\{1,...,K\}$. An example of such sub-channelization is OFDM with each frequency channel consisting of an integral number of subcarriers.

For all notation that relates to the BS, we use $b$ in the subscript. For the notation that relates to the MS in use cases (i) and (iii), we use $m$ in the subscript, while in the use case (ii) we use $m_1$ and $m_2$ to refer to MS~1 and MS~2, respectively. Summary of the main notation is provided in Table~\ref{table:notation}.

The transmission power of a station $u\in\{b, m, m_1, m_2\}$ on channel $k$ is denoted by $P_{u, k}$, where $k\in\{1,...,K\}$. In use cases (i) and (ii), $k$ is omitted from the subscript, since we consider a single channel.

\subsection{Remaining SI}\label{section:remaining-si-model}

\noindent\textbf{Single-channel FD.} For single-channel FD, we assume that the remaining SI both at the BS and at an MS can be expressed as a constant fraction of the transmitted power. In particular, if the BS transmits at the power level $P_b$, the remaining SI is $RSI_b = g_bP_b$, where $g_b$ is a constant determined by the hardware. Similarly, if an MS transmits at the power level $P_m$, its remaining SI is $RSI_m = g_mP_m$.

\noindent\textbf{Multi-channel FD.} 
We assume that the FD receiver at the BS has frequency-flat SIC profile, meaning that the remaining SI at the BS on channel $k$ is $RSI_{b, k}=g_bP_{b, k}$, where $g_b$ is a constant. We note that such FD receiver design is possible to implement in devices that do not require small form factor of the circuit (e.g., a BS or an access point (AP)), and has been reported in \cite{bharadia2013full}.

{In the rest of this section, we describe the mathematical model of the remaining SI for a small form factor device (MS).}   
We consider a compact/RFIC FD receiver with a \emph{circulator} at the antenna interface, described in Section \ref{section:circuit-model}, and assume a frequency-flat amplitude and phase response of the canceller, denoted by $|H_{C, R}|$ and $\angle H_{C, R}$, respectively.  The amplitude and phase response of the canceller are assumed to be programmable but constant with frequency. 

For the antenna interface's TX/RX isolation, we assume a flat amplitude response $|H_{A}(f)|=const =|H_{A}|$ and a constant group delay equal to $\tau$, so that $H_{A}(f)=|H_{A}|e^{-j2\pi f \tau}$ {(recall that the measured amplitude and group delay response are shown in Fig.~\ref{fig:SIC_Results}\subref{fig:GD_ISO_Circ})}. For the digital SIC, denoted by $SIC_D$, we assume that the amount of cancellation is constant across frequency, as delay can be easily generated in the digital domain. 
Let $f_k$ denote the central frequency of the $k^{\text{th}}$ channel, so that $f_k = f_1 + (k-1)B/K$. Then, the remaining SI after cancellation can be written as:
\begin{align}
RSI_{m, k} =& |P_{m, k}(H_A-H_{C,R}) SIC_D^{-1}|\notag\\
            =& P_{m, k} |(|H_A|e^{-j\angle{H_{A}(f_k)}}-|H_{C,R}|e^{-j\angle{H_{C,R}}})| SIC_D^{-1}\notag\\
            =& P_{m, k} |(|H_A|^2 + |H_{C,R}|^2-2|H_A| |H_{C,R}|\notag\\
            &\cdot\cos(\angle H_A(f_k)+\angle H_{C,R}))| SIC_D^{-1}.\label{eq:remaining-SI}
\end{align}
Note that in (\ref{eq:remaining-SI}), $P_{m, k}$ is the MS transmission power on channel $k$, $P_{m, k}(H_A-H_{C,R})$ is the remaining SI after the RF SIC, and $P_{m, k}(H_A-H_{C,R}) SIC_D^{-1}$ is the remaining SI after both the RF and digital SIC.

We assume a common oscillator for the TX and RX, with the phase noise of the oscillator being good enough so that it does not affect the remaining SI.

The RF canceller's settings can be programmed in the field to adjust the frequency at which peak SIC is achieved \cite{ZhouActiveTXCancISSCC14,Zhou_NCSIC_JSSC14}. With the amplitude ($|H_{C,R}|$) and the phase ($\angle H_{C,R}$) of the RF canceller set to $|H_A|$ and $-\angle H_A(f_{c})$, respectively, peak SIC is achieved at frequency $f_{c}$.  
Therefore, the total remaining SI at the MS on channel $k$ can be written as:
\begin{equation*}
RSI_{m, k} = 2|H_A|^2P_{m, k}(1-\cos(2\pi\tau(f_k - f_{c}))) SIC_D^{-1},
\end{equation*} 
where $\tau$ is the group delay from the antenna interface with a typical value at the order of 1ns (which agrees with the measured group delay in Fig.~\ref{fig:SIC_Results}\subref{fig:GD_ISO_Circ}). Frequency bands used by commercial wireless systems are at most 10s of MHz wide. It follows that  $2\pi\tau(f_k - f_{c})<<1$, and using the standard approximation $\cos(x)\approx 1-x^2/2$ for $x<<1$, we further get: 
\begin{equation*}
RSI_{m, k} \approx |H_A|^2P_{m, k}(2\pi\tau)^2(f_k - f_{c})^2 SIC_D^{-1}.
\end{equation*}

Recalling that $f_k = f_1 + (k-1)B/K = f_0 + kB/K$ for $f_0 = f_1 - B/K$, and writing $f_{c}$ as $f_{c} = f_0 + cB/K$, for $c\in\mathbb{R}$, we can combine all the constant terms and represent the remaining SI as:
\begin{equation}
RSI_{m, k} = g_m P_{m, k} (k-c)^2,\label{eq:RSI-MS}
\end{equation}
where $g_m = |H_A|^2(2\pi\tau)^2(B/K)^2SIC_D^{-1}$. Note that even though in this notation we allow $c$ to take negative values, we will later show that in any solution that maximizes the sum rate it must be $c\in(1, K)$ (Lemma \ref{lemma:ci-maxima-localization}).

Fig.~\ref{fig:SIC_Results}\subref{fig:SIC_BW_Circ} shows the TX/RX isolation based on Eq.~(\ref{eq:RSI-MS}) and based on measurement results. The parameter $g_m$ in Eq.~(\ref{eq:RSI-MS}) was determined via a least square estimation. The modeled TX/RX isolation based on Eq.~(\ref{eq:RSI-MS}) is also compared to the measured TX/RX isolation of the canceller with the antenna pair interface in Fig.~\ref{fig:SIC_Results}\subref{fig:SIC_BW_ANT}. \emph{As Fig.~\ref{fig:SIC_Results} shows, our model of the remaining SI closely matches the remaining SI that we measured with the RFIC FD receiver presented in} \cite{ZhouActiveTXCancISSCC14,Zhou_NCSIC_JSSC14}. 

\subsection{Sum Rate}\label{section:system-model}

\renewcommand{\arraystretch}{1}
\begin{table}[t!]
\small
\caption{Nomenclature.}
\centering
\renewcommand{\arraystretch}{0.9}
\begin{tabular}{|m{0.09\linewidth}| m{0.81\linewidth}|}
\hline
& \\[-3pt]
$m$ & Subscript notation for an MS\\
$b$ & Subscript notation for the BS\\
$K$ & Total number of OFDM channels\\
$k$ & Channel index, $k\in\{1,..., K\}$\\
$u, v$ & Station indexes, $u, v \in\{b, m, m_1, m_2\}$\\
$P_{u, k}$ & Transmission power of station $u$ on channel $k$\\
$\xoverline{P_u}$ & Maximum total power: $\sum_{k=1}^K P_{u, k}\leq \xoverline{P_u}$\\
$g_b$ & Remaining SI at the BS per unit transmitted power\\
$c$ & Position of the maximum SIC frequency, \hbox{$c\in\mathbb{R}$}\\
$g_m$ & Remaining SI at an MS: (i) per unit transmitted power for $K=1$; (ii) per unit transmitted power at unit distance from the maximum SIC frequency for $K>1$\\
$h_{uv, k}$ & Wireless channel gain for signal from $u$ to $v$ on channel $k$, for $u \neq v$\\
$N_u$ & Thermal noise at station $u$\\
$\gamma_{uv, k}$ & SNR of signal from $u$ to $v$ on channel $k$, where $u \neq v$\\
$\gamma_{uu, k}$ & XINR at station $u$, channel $k$\\
$\gamma_{uv, k}^{\max}$ & $\gamma_{uv, k}$ for $P_{u} = \xoverline{P_{u}}$\\
$r_k$ & Total rate on channel $k$\\
$r$ & Sum rate: $r = \sum_{k=1}^K r_k$\\[-4pt]
& \\
\hline
\end{tabular}\vspace{-10pt}
\label{table:notation}
\end{table}

The total transmitted power of each station is assumed to be bounded as follows. In use cases (i) and (ii): $P_b\leq \xoverline{P_b}$, and each $P_m, P_{m_1}, P_{m_2}\leq \xoverline{P_m}$. In use case (iii): $\sum_{k=1}^K P_{u, k}\leq \xoverline{P_u}$, where $u\in\{b, m\}$, $\xoverline{P_u}>0$. The channel gain from station $u$ to station $v$ on channel $k$ is denoted by $h_{uv, k}$ in use case (iii) and by $h_{uv}$ in use cases (i) and (ii). The noise level at station $u$ is assumed to be equal over channels and is denoted by $N_u$. {We assume that the channel states and noise levels are known.}

For the signal transmitted from $u$ to $v$, where $u, v\in\{b, m, m_1, m_2\}$, $u\neq v$, and either $u=b$ or $v=b$, we let $\gamma_{uv, k}=\frac{h_{uv, k}P_{u, k}}{N_v}$ denote signal to noise ratio (SNR) at $v$ on channel $k$. Similarly as before, in use cases (i) and (ii), index $k$ is omitted from the notation. In the use case (ii), $\gamma_{m_1m_2}$ denotes the (inter-node-)interference to noise ratio (INR). 
Self-interference to noise ratio (XINR) at the BS is denoted by $\gamma_{bb} = \frac{g_bP_b}{N_b}$ in use cases (i) and (ii), and by $\gamma_{bb, k} = \frac{g_bP_{b, k}}{N_b}$ in use case (iii). XINR at the MS is denoted by $\gamma_{mm} = \frac{g_mP_m}{N_m}$ and $\gamma_{mm, k} = \frac{g_m(k-c)^2P_{m, k}}{N_m}$ in use cases (i) and (iii), respectively.

We use Shannon's capacity formula for spectral efficiency, and let $\log(.)$ denote the base 2 logarithm, $\ln(.)$ denote the natural logarithm. We use the terms ``spectral efficiency'' and ``rate'' interchangeably, as the spectral efficiency on a channel is the rate on that channel normalized by $B/K$.

In use case (i), the sum rate on the channel is given as:
\begin{align}
r =& \log\Big(1 + \frac{\gamma_{mb}}{1 + \gamma_{bb}}\Big) + \log\Big(1 + \frac{\gamma_{bm}}{1 + \gamma_{mm}}\Big).\label{eq:bidirectional-single-SE}
\end{align}
Observe that $\frac{\gamma_{mb}}{1 + \gamma_{bb}}$ and $\frac{\gamma_{bm}}{1 + \gamma_{mm}}$ are signal to interference-plus-noise ratios (SINRs) on the UL and DL, respectively. We will refer to $r_m = \log\Big(1 + \frac{\gamma_{mb}}{1 + \gamma_{bb}}\Big)$ as the UL rate and 
$r_b = \log\Big(1 + \frac{\gamma_{bm}}{1 + \gamma_{mm}}\Big)$ as the DL rate.

Similarly as for (i), the sum rate for use case (ii) is:
\begin{align}
r
=& \log\Big(1 + \frac{\gamma_{m_1b}}{1 + \gamma_{bb}}\Big) + \log\Big(1 + \frac{\gamma_{bm_2}}{1 + \gamma_{m_1m_2}}\Big).\label{eq:two-uni-SE}
\end{align}

Finally, in use case (iii), the rate on channel $k$ is given as: 
\begin{align}
r_{k}
= &\log\Big(1 + \frac{\gamma_{mb, k}}{1 + \gamma_{bb, k}}\Big) + \log\Big(1 + \frac{\gamma_{bm, k}}{1 + \gamma_{mm, k}}\Big),\label{eq:bidirectional-SE}
\end{align}
while the sum rate (on all channels) is $r = \sum_{k=1}^Kr_k$.

\textbf{The objective in all problems considered 
is to maximize $r$ subject to the upper bound on total transmitted power and non-negativity constraints.} In use cases (i) and (ii), the variables are $P_b$ and $P_m$, while in the use case (iii), the variables are $c$, $P_{b, k}$, and $P_{m, k}$, for $k\in\{1,...,K\}$.

For the purpose of comparison to TDD systems, we will sometimes also consider TDD rates. We denote by $r_{\mathrm{TDD}, m}^{\max}\equiv \log(1+\gamma_{mb}^{\max})$ and $r_{\mathrm{TDD}, b}^{\max}\equiv \log(1+\gamma_{bm}^{\max})$ the maximum UL and DL TDD rates, respectively, where $\gamma_{mb}^{\max} = \gamma_{mb}(\xoverline{P_m})$, $\gamma_{bm}^{\max} = \gamma_{bm}(\xoverline{P_b})$. The maximum achievable TDD rate can then be written as $r_{\mathrm{TDD}}^{\max} = \max\{r_{\mathrm{TDD}, m}^{\max}, r_{\mathrm{TDD}, b}^{\max}\}$.%

\section{{Single Channel FD}}\label{section:single-channel}
\subsection{{A Bidirectional FD Link}\label{section:bidirectional}}

In this section, we derive general properties of the sum rate function for use case (i) (Fig.~\ref{fig:full-duplex-links}\subref{fig:dual}). 

First, we show that if it is possible for the FD sum rate to exceed the maximum TDD rate, it is always optimal for the MS and the BS to transmit at their maximum respective power levels (Lemma~\ref{lemma:dependence-on-subch-power}). 
This result is somewhat surprising, because in general, the FD sum rate function 
does not have good structural properties, i.e., it need not 
be convex or concave in the transmission power variables. 

Building upon this insight, we quantify the FD rate gains  
by comparing the FD sum rate to corresponding TDD rates (Section~\ref{section:bidirectional-SINR-region}). 
More specifically, we define a metric that characterizes by how much the FD capacity region 
extends the corresponding TDD capacity region, and provide a sufficient condition
on the system parameters for rate gains to hold.

Finally, we establish a sufficient condition for the FD sum rate function to be {bi}concave in transmission power levels (Section \ref{section:bidirectional-concavity}). This condition imposes very mild restrictions on the XINRs at the BS and the MS. Moreover, the established condition extends to the multi-channel scenario (use case (iii)), where it plays a crucial role in deriving a{n} algorithm for the sum rate maximization {that converges to a stationary point that is a global maximum in practice} (Section \ref{section:OFDM-OPS-general}). {Without such a condition, the problem would not have enough structure to be amenable to efficient optimization methods.

\subsubsection{{Power Allocation}}\label{section:bidirectional-PA}

\begin{lemma}\label{lemma:dependence-on-subch-power}
If there exists an FD sum rate $r$ that is higher than the maximum TDD rate, then $r$ is maximized for $P_m = \xoverline{P_m}$, $P_b = \xoverline{P_b}$.
\end{lemma}

\begin{IEEEproof}
From (\ref{eq:bidirectional-single-SE}), the sum rate can be written as:
\begin{align*}
r &= \log\Big(1+\frac{h_{mb}P_{m}}{N_{b} + g_b P_{b}}\Big) + \log\Big(1+\frac{h_{bm}P_{b}}{N_{m} + g_mP_m}\Big)
\end{align*}
Taking partial derivatives of $r$ directly does not provide conclusive information about the optimal power levels. Instead, we write $r$ as an increasing function of another function that is easier to analyze. Specifically: 
\begin{align*}
r =& \log\Big(\Big(1+\frac{h_{mb}P_{m}}{N_{b} + g_b P_{b}}\Big)\cdot \Big(1+\frac{h_{bm}P_{b}}{N_{m} + g_mP_m}\Big)\Big)\\
=& \log(1 + \gamma), \;\;\text{ where}\\
\gamma =& \frac{h_{mb}P_{m}}{N_{b} + g_b P_{b}}
+ \frac{h_{bm}P_{b}}{N_{m} + g_mP_m} + \frac{h_{mb}P_{m}}{N_{b}+g_bP_b}\cdot\frac{h_{bm}P_{b}}{N_{m} + g_mP_m}.
\end{align*}

Since $r$ is strictly increasing in $\gamma$, to maximize $r$ it suffices to determine $P_{m}, P_{b}$ that maximize $\gamma$. 
The first and the second partial derivative of $\gamma$ with respect to $P_{m}$ are:
\begin{align}
\frac{\partial \gamma}{\partial P_{m}} 
=& \frac{h_{mb}}{N_{b} + g_b P_{b}} + \frac{h_{bm}P_{b}}{(N_{m} + g_mP_{m})^2}\left( \frac{h_{mb}N_m}{N_b + g_b P_{b}}-g_m \right),\label{eq:first-derivative-i}\\
\frac{\partial^2 \gamma}{\partial {P_{m}}^2} &= -2\frac{h_{bm}P_{b}g_m}{(N_{m} + g_mP_{m})^3}\Big( \frac{h_{mb}N_m}{N_b + g_b P_{b}}-g_m \Big).\label{eq:second-derivative-i}
\end{align}
From (\ref{eq:first-derivative-i}) and (\ref{eq:second-derivative-i}):
\begin{enumerate}
\item If $\frac{h_{mb}N_m}{N_b + g_b P_{b}}-g_m\geq 0$, then $\frac{\partial^2 \gamma}{\partial {P_{m}}^2}\leq 0$ and $\frac{\partial \gamma}{\partial P_{m}} >0$, i.e., $\gamma$ is concave and strictly increasing in $P_{m}$ {when $P_b$ is fixed}, and therefore maximized for $P_{m}=\xoverline{P_m}$.
\item If $\frac{h_{mb}N_m}{N_b + g_b P_{b}}-g_m < 0$, then $\frac{\partial^2 \gamma}{\partial {P_{m}}^2}> 0$, i.e., $\gamma$ is strictly convex in $P_{m}$ {when $P_b$ is fixed}. Therefore, $\gamma$ is maximized at either $P_{m} = 0$ or $P_{m}=\xoverline{P_m}$. Note that if $P_{m} = 0$, there is no signal on UL, in which case FD rate equals the maximum TDD UL rate.
\end{enumerate}

A similar results follows for $P_b$ by taking the first and the second partial derivative of $\gamma$ with respect to $P_{b}$.
\end{IEEEproof}

\subsubsection{{Mapping Gain over SINR Regions}}\label{section:bidirectional-SINR-region}

\begin{figure}
\centering
\vspace{-5pt}\subfloat[]
	{\label{fig:cap-region-gamma-1}
	\includegraphics[width = 0.45\linewidth]{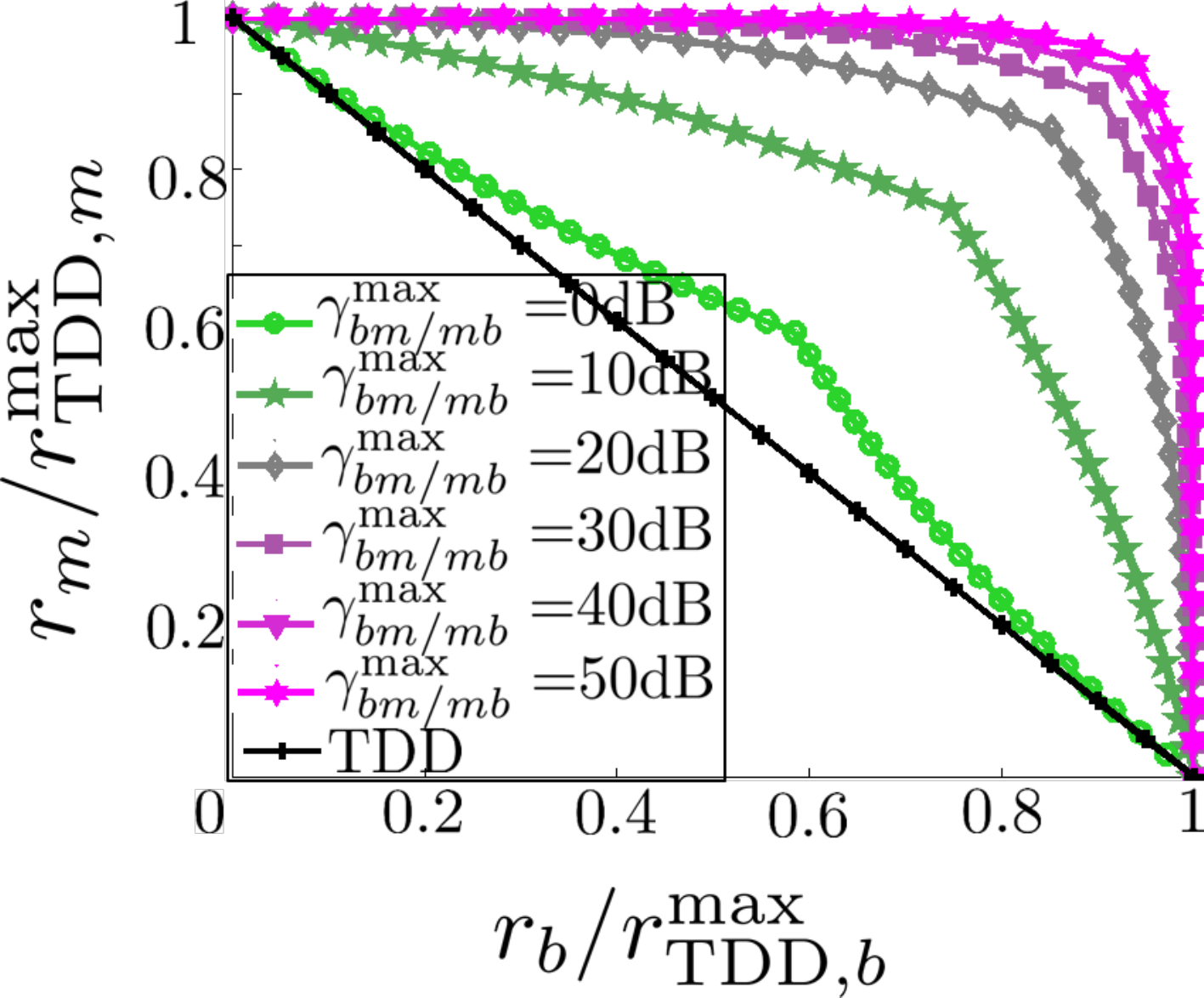}}\hspace{\fill}
    \vspace{-5pt}\subfloat[]
	{\label{fig:cap-region-gamma-10}
	\includegraphics[width = 0.43\linewidth]{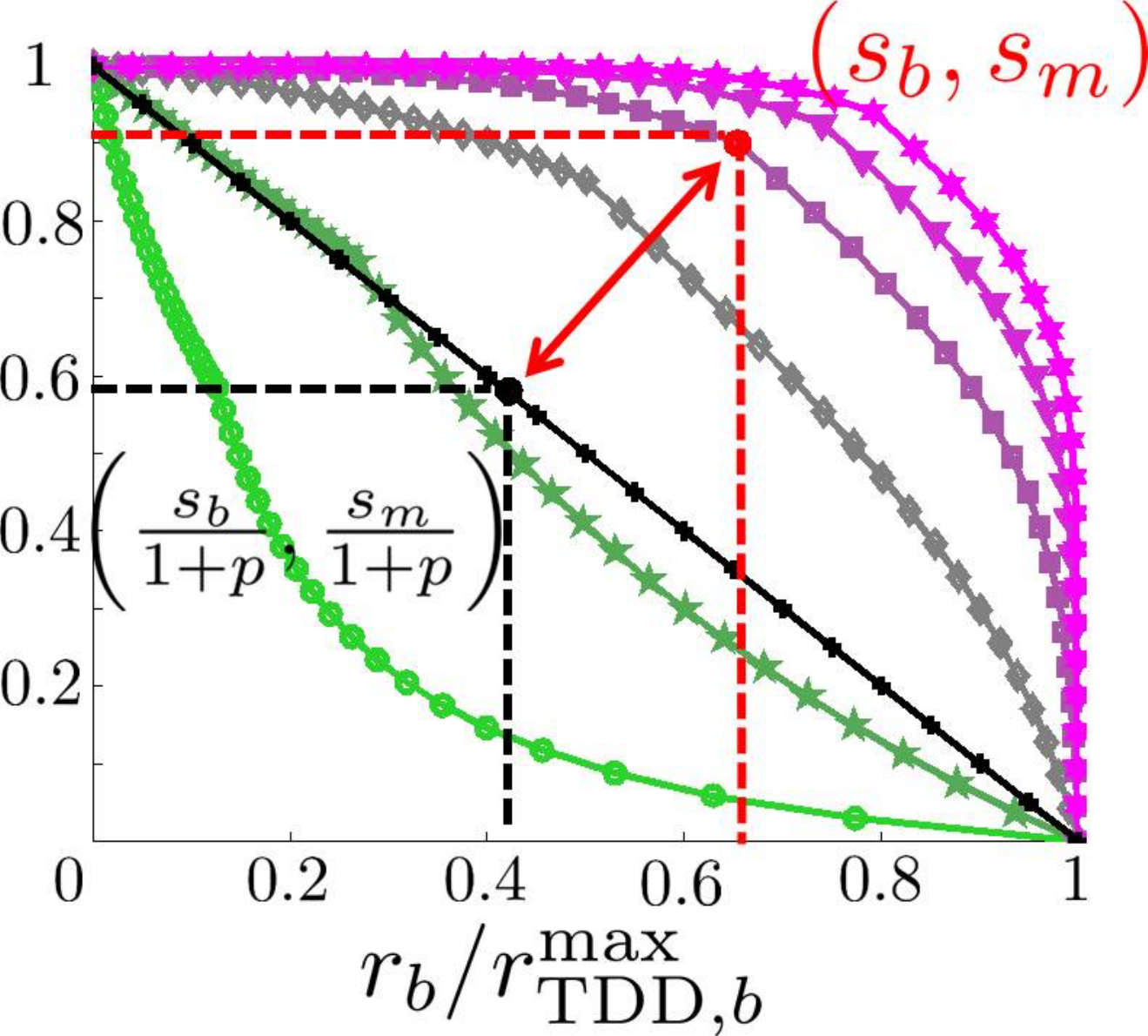}}
\caption{TDD and FD capacity regions, and FD extension. The capacity region is plotted for equal maximum SNRs: $\gamma_{mb}^{\max} = \gamma_{bm}^{\max}\equiv \gamma_{bm/mb}^{\max}$ and two cases of maximum XINRs: \protect\subref{fig:cap-region-gamma-1} $\gamma_{bb}^{\max} = 1$, $\gamma_{mm}^{\max} = 1$ and \protect\subref{fig:cap-region-gamma-10} $\gamma_{bb}^{\max} = 1$, $\gamma_{mm}^{\max} = 10$.}
\label{fig:cap-region}\vspace{-10pt}
\end{figure}

In this section we quantify the FD rate gains by comparing the FD capacity region to the corresponding TDD capacity region. Let $r_{b} = \log(1 + \frac{\gamma_{bm}}{1+\gamma_{mm}})$, $r_{m} = \log(1 + \frac{\gamma_{mb}}{1+\gamma_{bb}})$ denote DL and UL rates, respectively and let $r_{\text{TDD}, b}^{\max} = \log(1 + \gamma_{bm}^{\max})$, $r_{\text{TDD}, m}^{\max} = \log(1 + \gamma_{mb}^{\max})$ denote the maximum TDD rates. The FD capacity region is the set of all points $(r_b, r_m)$ such that $P_m\in[0, \xoverline{P_m}]$, $P_b\in[0, \xoverline{P_b}]$, while the TDD capacity region is the convex hull of the points $(0, 0), (r_{\text{TDD}, b}^{\max}, 0)$, and $(0, r_{\text{TDD}, m}^{\max})$. 
We also let $s_b =\log(1 + \frac{\gamma_{bm}^{\max}}{1+\gamma_{mm}^{\max}})$ and $s_m = \log(1 + \frac{\gamma_{mb}^{\max}}{1 + \gamma_{bb}^{\max}})$ be the FD DL and UL rates when both stations transmit at their maximum power levels $\xoverline{P_b}, \xoverline{P_m}$. 

Fig.~\ref{fig:cap-region} shows FD and TDD capacity regions for symmetric maximum SNRs $\gamma_{mb}^{\max} = \gamma_{bm}^{\max}$ and two cases of maximum XINRs: $\gamma_{bb}^{\max} = \gamma_{mm}^{\max} = 1$ and $\gamma_{bb}^{\max} = 1$, $\gamma_{mm}^{\max} = 10$. Here, the axes are normalized by $r_{\text{TDD}, b}^{\max}$ and $r_{\text{TDD}, m}^{\max}$, respectively. To determine the points at the boundary of the FD capacity region, we apply Lemma \ref{lemma:dependence-on-subch-power} as follows. 
For $r_b = \alpha s_b$, where $\alpha \in (0, 1)$, Lemma \ref{lemma:dependence-on-subch-power} implies that the UL rate $r_m$ is maximized for $P_{m} = \xoverline{P_m}$, regardless of the value of $P_b$. Therefore, the DL rate is lowered from $s_b$ to $r_b = \alpha s_b$ by lowering $P_b$. The point $(r_b, r_m)$ at the boundary of the FD capacity region is then determined by solving $r_b = \alpha s_b$ for $P_b$, and setting $r_m = r_m(P_b, \xoverline{P_m})$. An analogous procedure is carried out for $r_m = \alpha s_m$, where $\alpha \in (0, 1)$. 
We remark that FD capacity regions are not necessarily convex (e.g., Fig.~\ref{fig:cap-region}\subref{fig:cap-region-gamma-10} for $\gamma_{bm/mb}^{\max} = 0$dB and $\gamma_{bm/mb}^{\max} = 10$dB).

Lemma \ref{lemma:dependence-on-subch-power} states that the maximizer of the FD sum rate is either $(r_{\text{TDD}, b}^{\max}, 0)$, $(0, r_{\text{TDD}, m}^{\max})$ or $(s_b, s_m)$. In particular, to see whether FD operation increases the sum rate, it suffices to check whether $s_b+s_m>\max\{r_{\text{TDD}, b}^{\max}, r_{\text{TDD}, m}^{\max}\}$. This motivates us to focus on the pair $(s_b, s_m)$ when considering by how much the FD capacity region extends the corresponding TDD capacity region. We introduce the following definition (see Fig.~\ref{fig:cap-region}\subref{fig:cap-region-gamma-10} for a geometric interpretation).

\begin{definition}\label{def:cap-region-ext}
FD extends the corresponding TDD capacity region by $p\cdot100\%$ if 
$p \geq 0$ is the smallest number for which $\frac{s_b}{1 + p}, \frac{s_m}{1 + p}$ is inside the TDD capacity region.
\end{definition}

\begin{figure}[t]
\centering
	\subfloat[]
	{\label{fig:p11}
	\includegraphics[scale = 0.27]{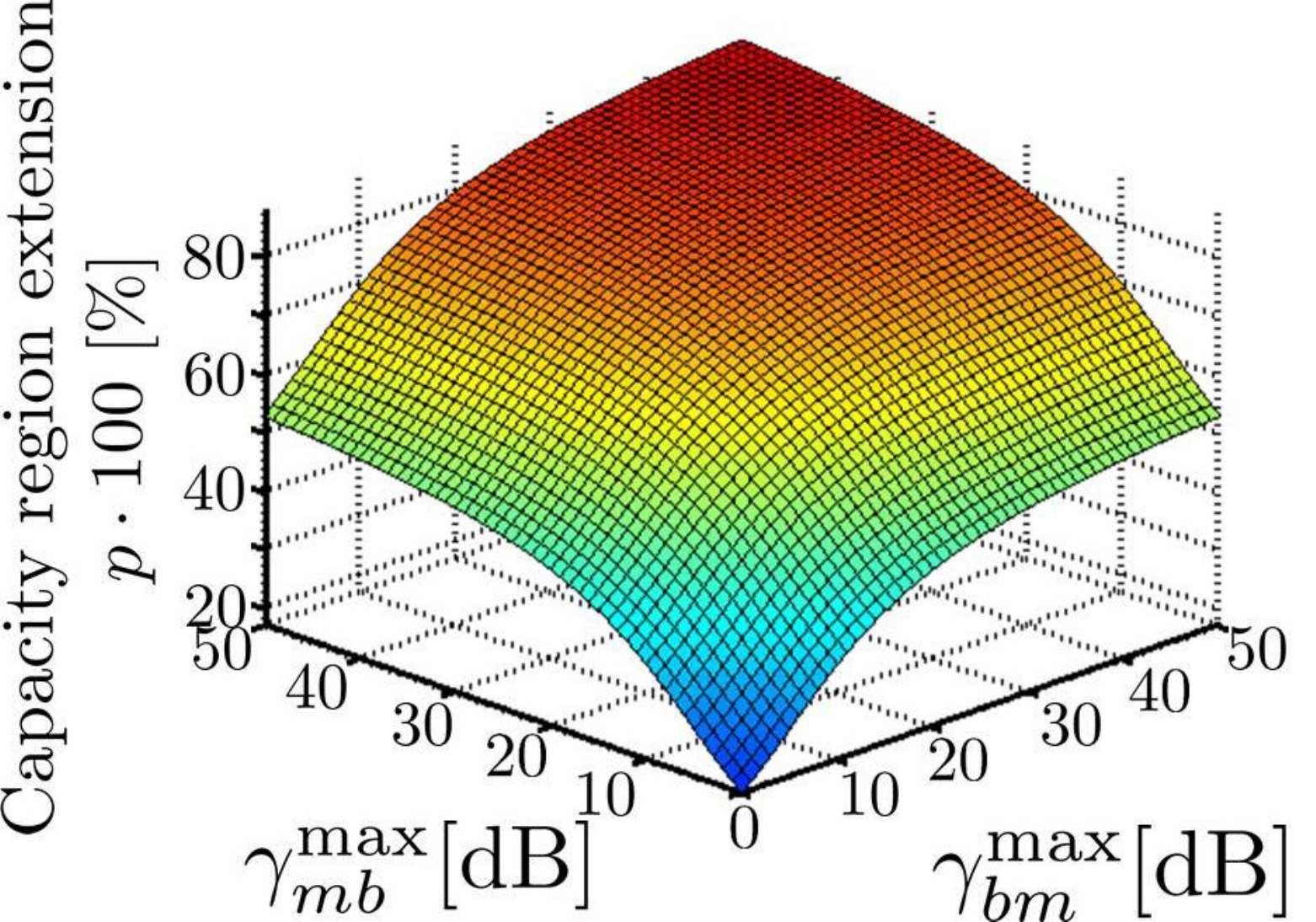}}\hspace{\fill}
	\subfloat[]
	{\label{fig:p110}
	\includegraphics[scale = 0.27]{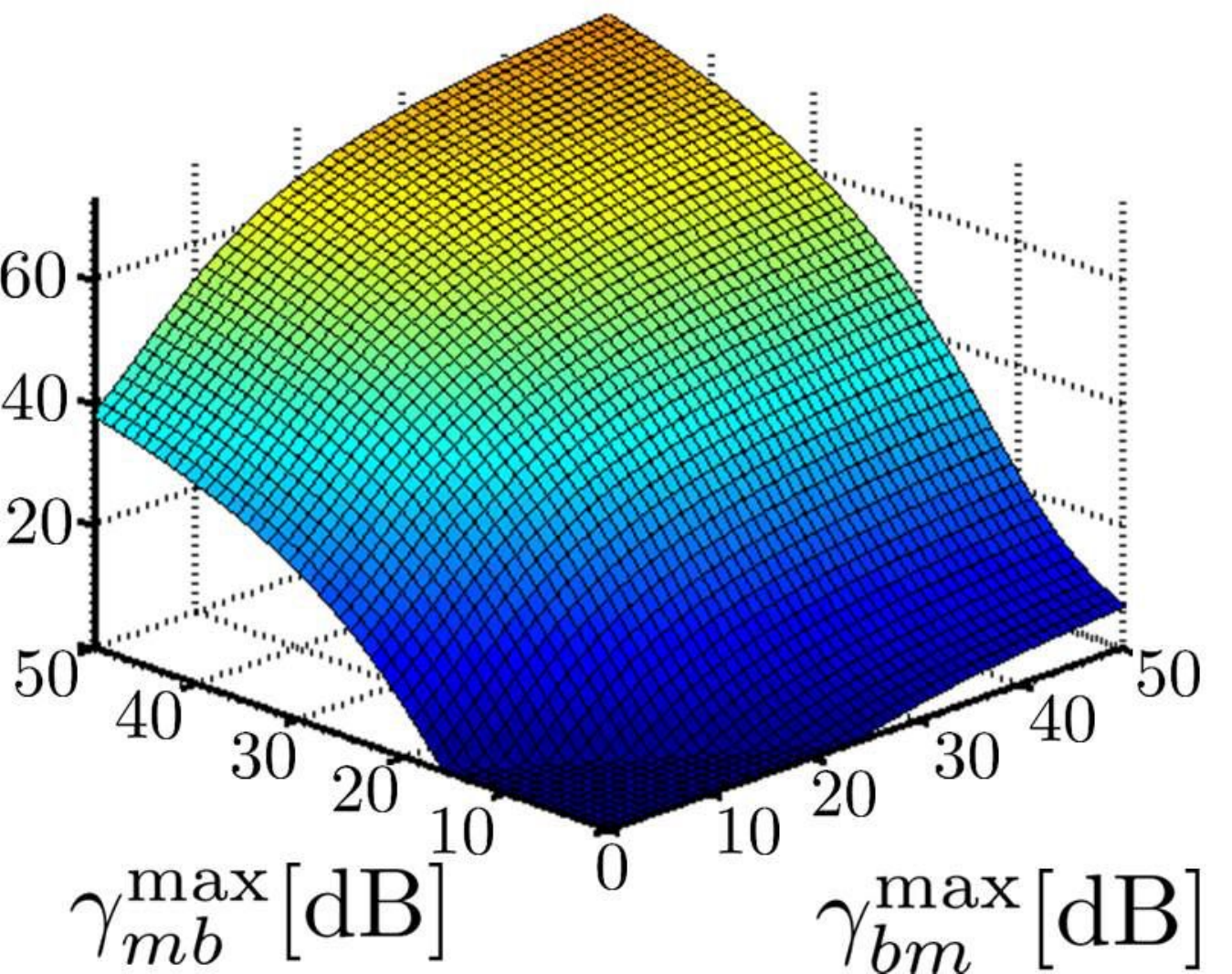}}
	\caption{TDD capacity region extension due to FD as a function of SNRs for \protect\subref{fig:p11} $\gamma_{bb}^{\max} = 1$, $\gamma_{mm}^{\max} = 1$ and \protect\subref{fig:p110} $\gamma_{bb}^{\max} = 1$, $\gamma_{mm}^{\max} = 10$.}
	\label{fig:p-in-terms-of-snrs}\vspace{-10pt}
\end{figure}

The following lemma provides a necessary and sufficient condition for the capacity region extension of $p\cdot100\%$.

\begin{lemma}\label{lemma:cap-region-extension-condition}
FD extends the TDD capacity region by \hbox{$p\cdot100\%$}, where $p\geq 0$, if and only if:
\begin{equation}
\frac{\log\Big(1 + \frac{\gamma_{bm}^{\max}}{1 +\gamma_{mm}^{\max}}\Big)}{\log(1 + \gamma_{bm}^{\max})} + \frac{\log\Big(1 +\frac{\gamma_{mb}^{\max}}{1 + \gamma_{bb}^{\max}}\Big)}{\log(1+\gamma_{mb}^{\max})} = 1+ p. \label{eq:cap-region-extension-condition}
\end{equation}
\end{lemma}

\begin{IEEEproof}[Proof Sketch]
The proof is based on the fact that since $p$ is the smallest number for which the point $(\frac{s_b}{1+p}, \frac{s_m}{1+p})$ is in the TDD capacity region, $(\frac{s_b}{1+p}, \frac{s_m}{1+p})$ must lie on the line connecting $r_{\text{TDD}, b}^{\max}$ and $r_{\text{TDD}, m}^{\max}$ (Fig.~\ref{fig:cap-region}\subref{fig:cap-region-gamma-10}), and therefore:
$\frac{s_m}{1+p} = r_{\text{TDD}, m}^{\max} - \frac{r_{\text{TDD}, m}^{\max}}{r_{\text{TDD}, b}^{\max}}\frac{s_b}{1 + p}$, which is equivalent to (\ref{eq:cap-region-extension-condition}).
\end{IEEEproof}

Fig.~\ref{fig:p-in-terms-of-snrs} shows the TDD capacity region extension due to FD operation, as a function of the received signals' SNR, for BS FD receiver that cancels SI to the noise level and MS FD receiver that cancels SI to (i) the noise level (Fig.~\ref{fig:p-in-terms-of-snrs}\protect\subref{fig:p11}) and (ii) one order of magnitude above noise (Fig.~\ref{fig:p-in-terms-of-snrs}\protect\subref{fig:p110}). Recall from Definition \ref{def:cap-region-ext} that the capacity region extension is computed for $P_m = \xoverline{P_m}$ and $P_b = \xoverline{P_b}$, and therefore the differences in the SNRs are due to signal propagation and not due to reduced transmission power levels. Fig.~\ref{fig:p-in-terms-of-snrs} suggests that to achieve non-negligible capacity region extension, SNRs at the MS and at the BS must be sufficiently high -- at least as high as to bring the resulting SINR to the level above 0dB. 
\subsubsection{{Sum Rate {Bi}concavity}}\label{section:bidirectional-concavity}

In this section, we establish a sufficient condition for the sum rate to be (strictly) {bi}concave and increasing in $P_m$ and in $P_b$ (Condition \ref{cond:concavity}). We also show that when the condition does not hold, 
using FD does not provide appreciable rate gains, 
as compared to the maximum rate achievable by TDD operation. {Intuitively, the condition states that a station's amount of SIC should be at least as high as the loss incurred due to wireless propagation on the path to the intended receiver.}

\begin{condition}\label{cond:concavity}
$\gamma_{mm}\leq \frac{\gamma_{mb}}{1 + \gamma_{bb}}$ and $\gamma_{bb}\leq \frac{\gamma_{bm}}{1 + \gamma_{mm}}$.
\end{condition}
\begin{proposition}\label{prop:suff-cond-concavity}
If $\gamma_{mm}\leq \frac{\gamma_{mb}}{1 + \gamma_{bb}}$,  
the sum rate $r$ is strictly concave and strictly increasing in $P_m$ {when $P_b$ is fixed}. Similarly, if $\gamma_{bb}\leq \frac{\gamma_{bm}}{1 + \gamma_{mm}}$, $r$ is strictly concave and strictly increasing in $P_b$ {when $P_m$ is fixed}. 
Thus, when Condition \ref{cond:concavity} holds, $r$ is strictly biconcave and strictly increasing 
 in $P_m$ and in $P_b$. Furthermore, when Condition \ref{cond:concavity} does not hold, 
$r - r_{\text{TDD}}^{\max}<1$b/s/Hz.
\end{proposition}
\begin{IEEEproof}
{Fix $P_b$. }From the proof of Lemma \ref{lemma:dependence-on-subch-power}, we can express $r$ as $r = \log(1 + \gamma)$, where $\gamma$ is strictly increasing and concave in $P_m$ whenever
\begin{equation}
\frac{h_{mb}N_m}{N_b + g_b P_{b}}-g_m\geq 0. \label{eq:gamma-concavity-condition}
\end{equation}
Multiplying both sides of (\ref{eq:gamma-concavity-condition}) by $\frac{P_m}{N_m}$ and reordering terms:
\begin{align*}
\frac{h_{mb}P_m}{N_b + g_b P_{b}}\geq \frac{g_mP_m}{N_m}\quad
\Leftrightarrow\quad \gamma_{mm}\leq \frac{\gamma_{mb}}{1 + \gamma_{bb}}.
\end{align*}
Whenever (\ref{eq:gamma-concavity-condition}), or equivalently, the inequality $\gamma_{mm}\leq \frac{\gamma_{mb}}{1 + \gamma_{bb}}$, holds, since $\gamma>0, \frac{\partial \gamma}{\partial P_m} > 0, \frac{\partial^2 \gamma}{\partial {P_m}^2} \leq 0$:
\begin{equation*}
\frac{\partial r}{\partial P_m} = \frac{1}{1 +\gamma}\cdot\frac{\partial \gamma}{\partial P_m} > 0,\text{ and,}
\end{equation*}\vspace{-10pt}
\begin{equation*}
\frac{\partial^2 r}{\partial {P_m}^2} = -\frac{1}{(1 + \gamma)^2}\cdot\Big(\frac{\partial \gamma}{\partial P_m}\Big)^2 + \frac{1}{1 + \gamma}\cdot\frac{\partial^2 \gamma}{\partial {P_m}^2} < 0,
\end{equation*}
and therefore $r$ is strictly increasing and strictly concave in $P_m$. 
{Similarly, whenever $\gamma_{bb}\leq \frac{\gamma_{bm}}{1 + \gamma_{mm}}$, $r$ is strictly increasing and strictly concave in $P_b$ {when $P_m$ is fixed}.}

{Now suppose that Condition \ref{cond:concavity} does not hold. 
Then, either $\gamma_{mm} > \frac{\gamma_{mb}}{1 + \gamma_{bb}}$ or $\gamma_{bb} > \frac{\gamma_{bm}}{1 + \gamma_{mm}}$. 
Suppose that $\gamma_{mm} > \frac{\gamma_{mb}}{1 + \gamma_{bb}}$. Then:
\begin{align*}
r &= \log\Big(1 + \frac{\gamma_{mb}}{1 + \gamma_{bb}}\Big) + \log\Big(1 + \frac{\gamma_{bm}}{1 + \gamma_{mm}}\Big)\\
&< \log\Big(1 + \frac{\gamma_{mb}}{1 + \gamma_{bb}}\Big) + \log\Big(1 + \frac{\gamma_{bm}}{1 + \frac{\gamma_{mb}}{1 + \gamma_{bb}}}\Big)\\
&= \log\Big(2\cdot\Big(1 + \frac{1}{2}\Big(\gamma_{bm} + \frac{\gamma_{mb}}{1 + \gamma_{bb}}-1\Big)\Big)\Big)\\
&= 1\mathrm{b/s/Hz} + \log\Big(1 + \frac{1}{2}\Big(\gamma_{bm} + \frac{\gamma_{mb}}{1 + \gamma_{bb}}-1\Big)\Big).
\end{align*}
Since $\frac{1}{2}\Big(\gamma_{bm} + \frac{\gamma_{mb}}{1 + \gamma_{bb}}-1\Big) <  \max\{\gamma_{mb}, \gamma_{bm}\}$, it follows that 
$
r < 1\mathrm{b/s/Hz} + r_{\text{TDD}}^{\max},
$
which completes the proof for $\gamma_{mm} > \frac{\gamma_{mb}}{1 + \gamma_{bb}}$. The proof for the case $\gamma_{bb} > \frac{\gamma_{bm}}{1 + \gamma_{mm}}$ follows the same line of argument and is omitted for brevity.}
\end{IEEEproof}

{\subsection{Two Unidirectional Links}\label{section:two-unidirectionals}}

\begin{figure*}[ht!]
\centering

	\subfloat
	{
	\includegraphics[scale = 0.26]{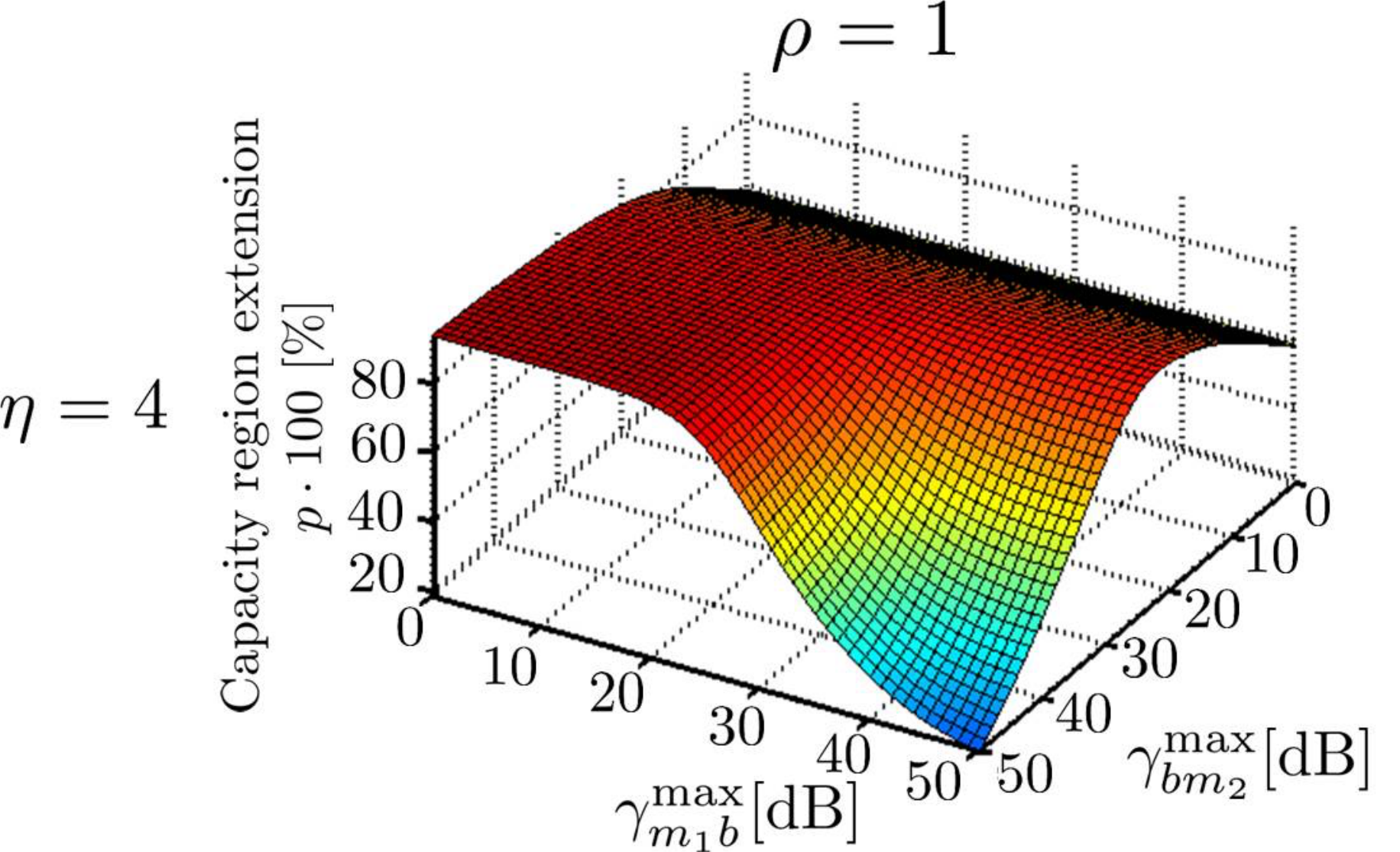}}\hspace{\fill}
	\subfloat
	{
	\includegraphics[scale = 0.26]{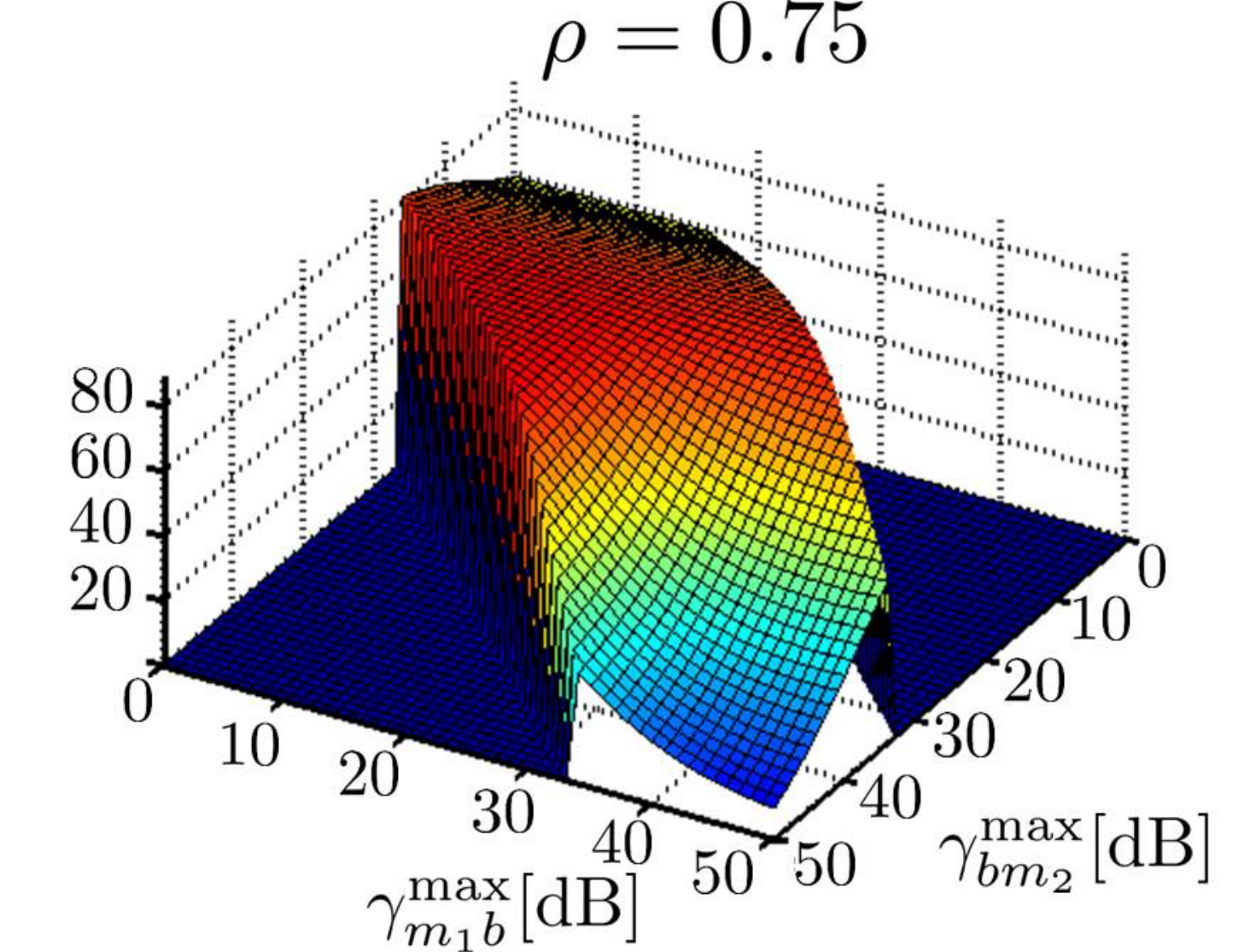}}\hspace{\fill}
    \subfloat
	{
	\includegraphics[scale = 0.26]{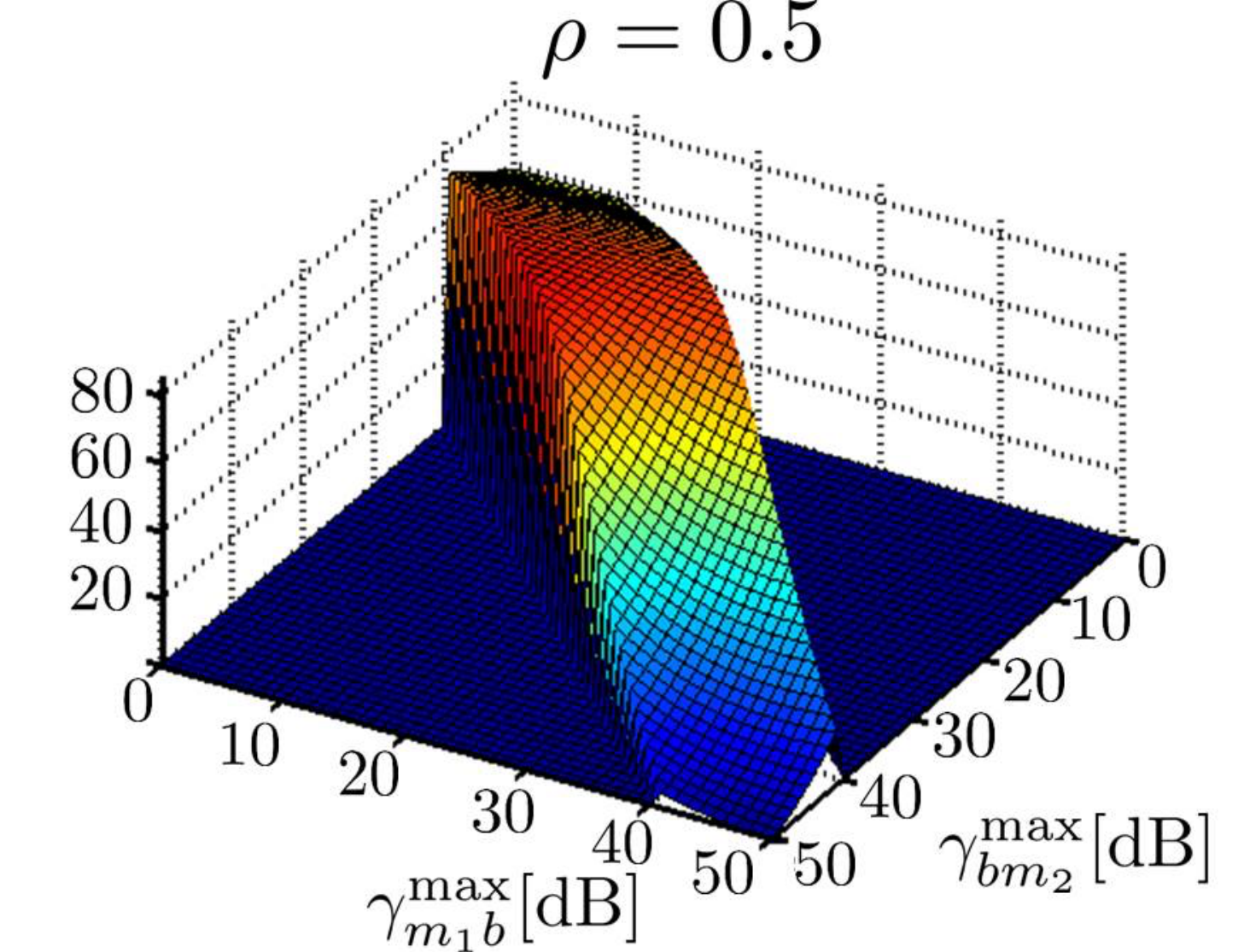}}\hspace{\fill}
    \subfloat
	{
	\includegraphics[scale = 0.26]{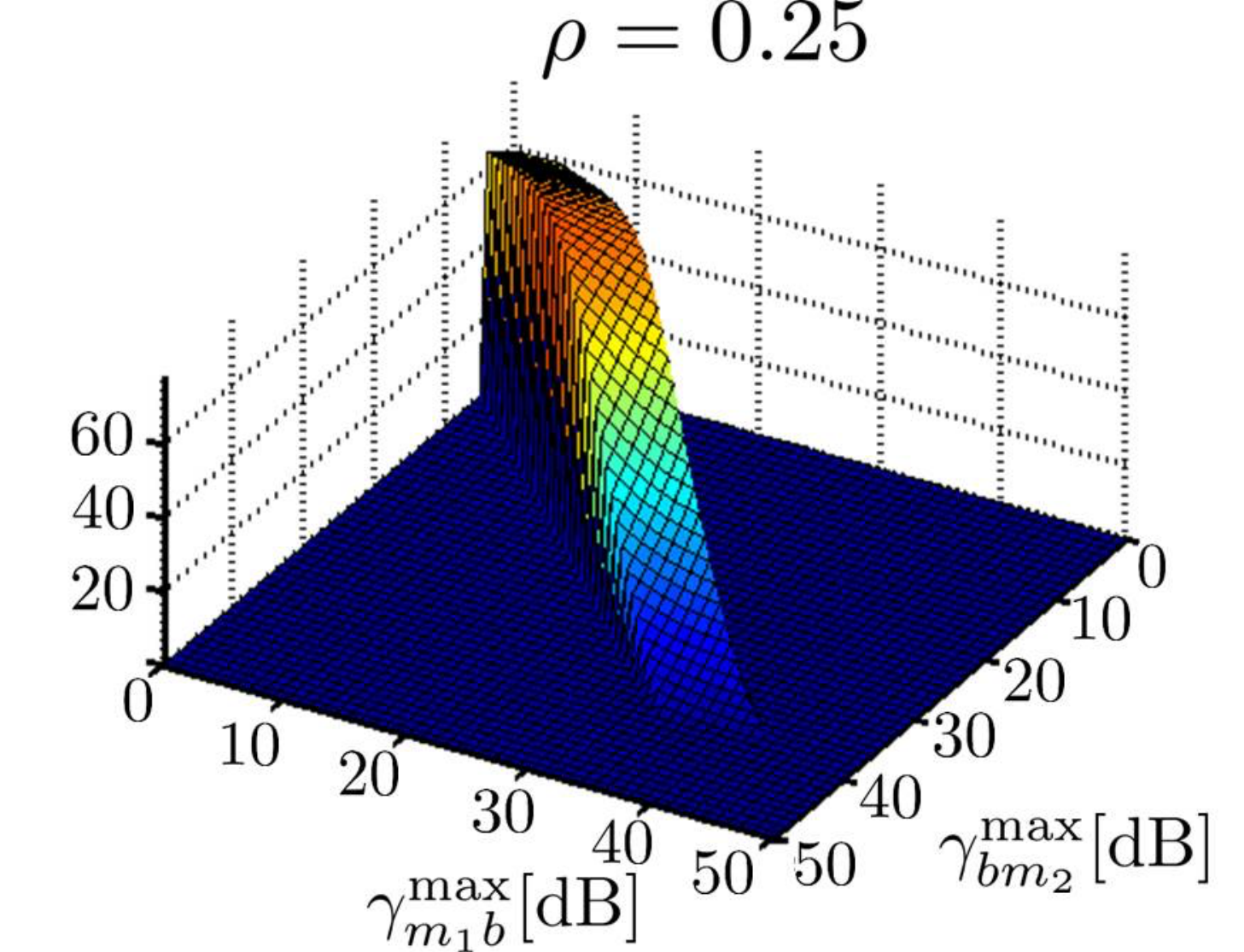}}\hspace*{\fill}\\[-10pt]
    \subfloat
	{
	\includegraphics[scale = 0.26]{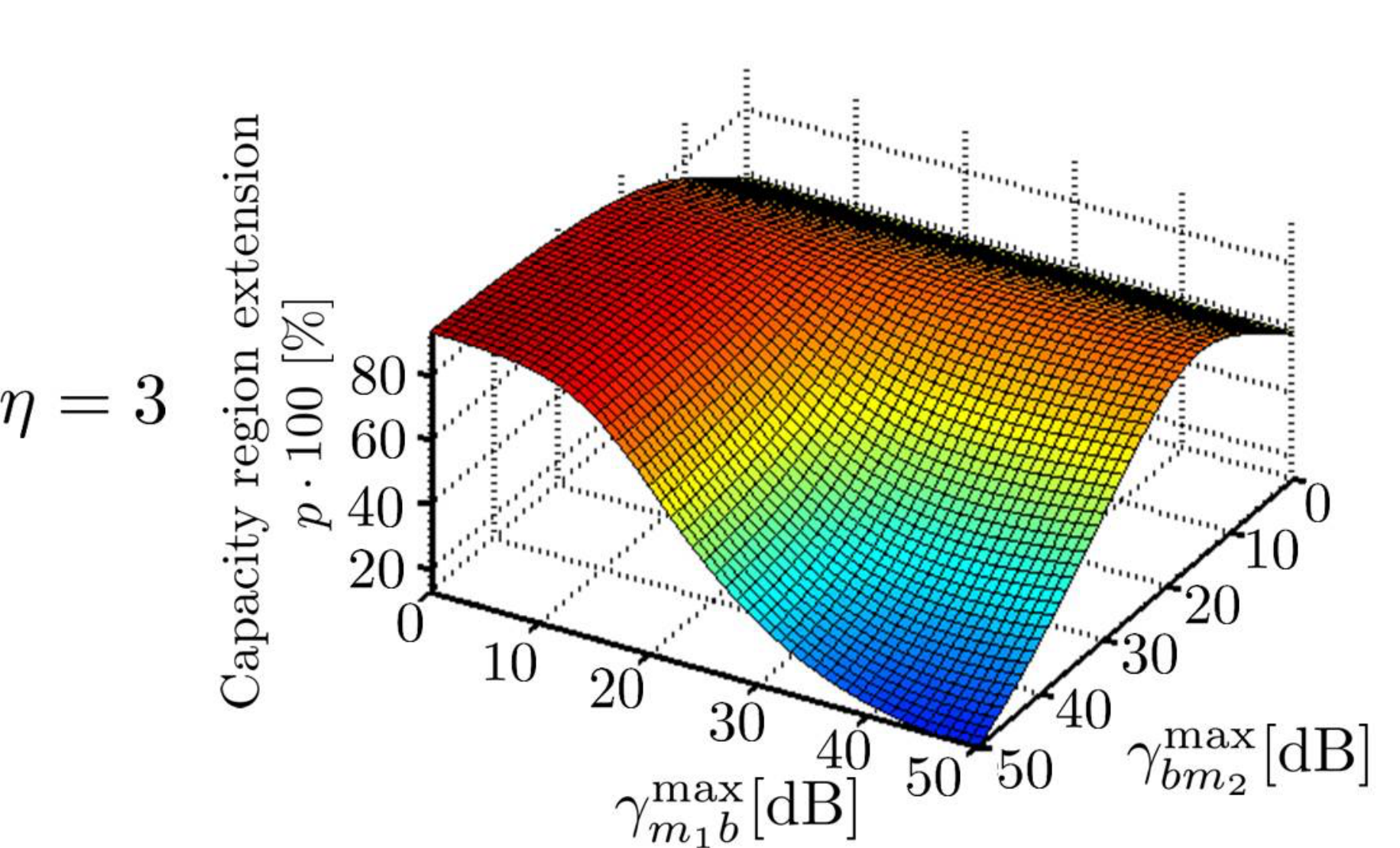}}\hspace{\fill}
	\subfloat
	{
	\includegraphics[scale = 0.26]{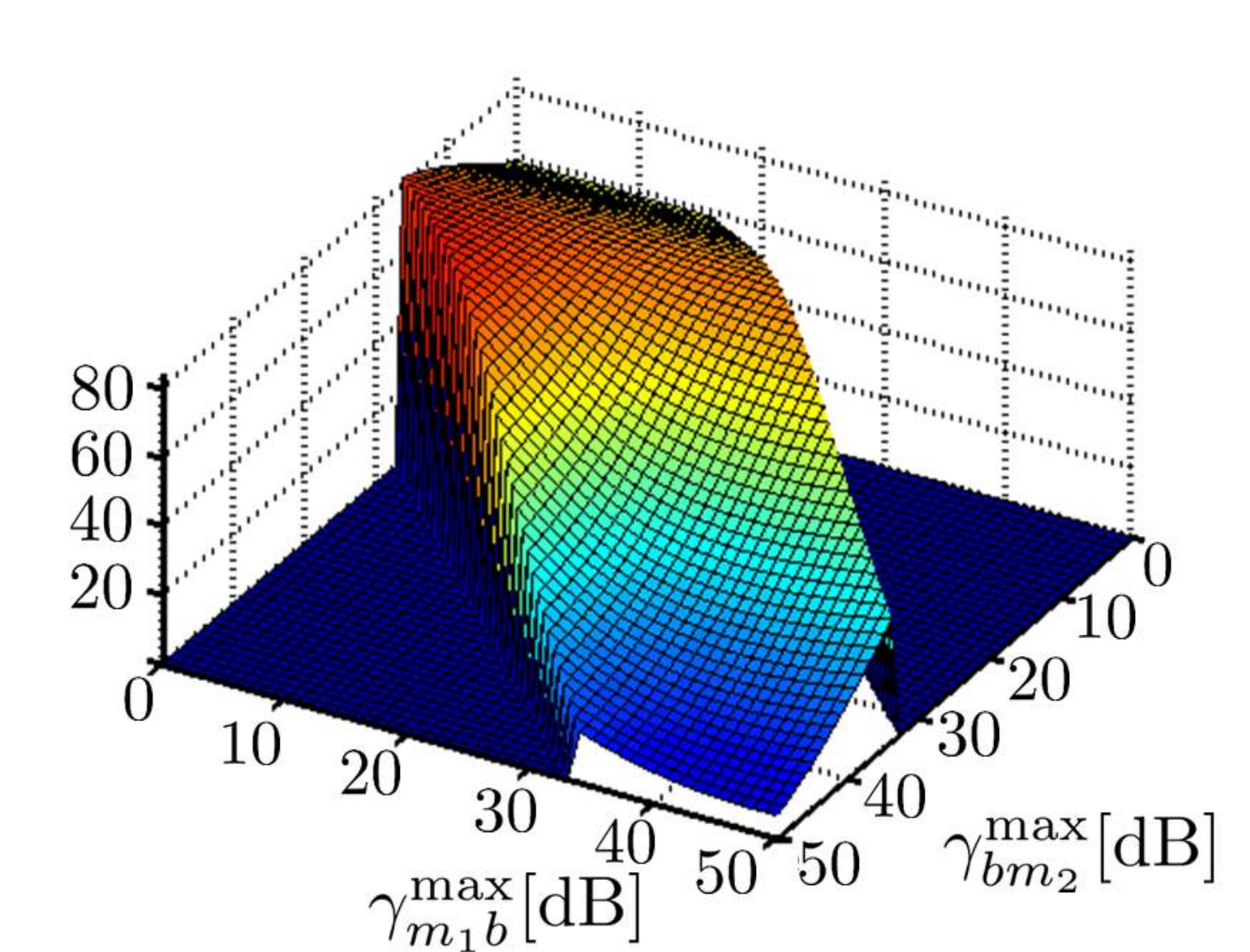}}\hspace{\fill}\subfloat
	{
	\includegraphics[scale = 0.26]{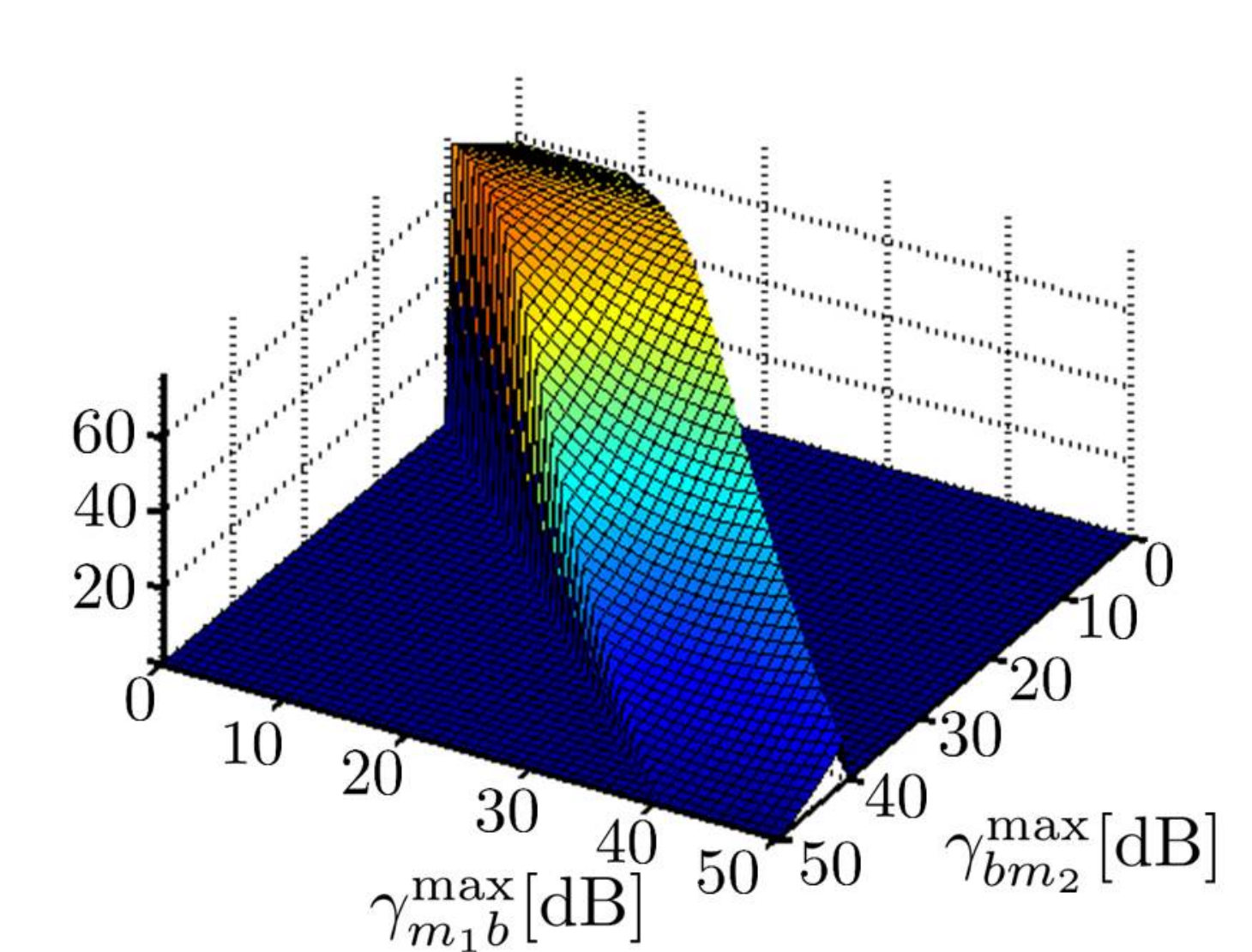}}\hspace{\fill}\subfloat
	{
	\includegraphics[scale = 0.26]{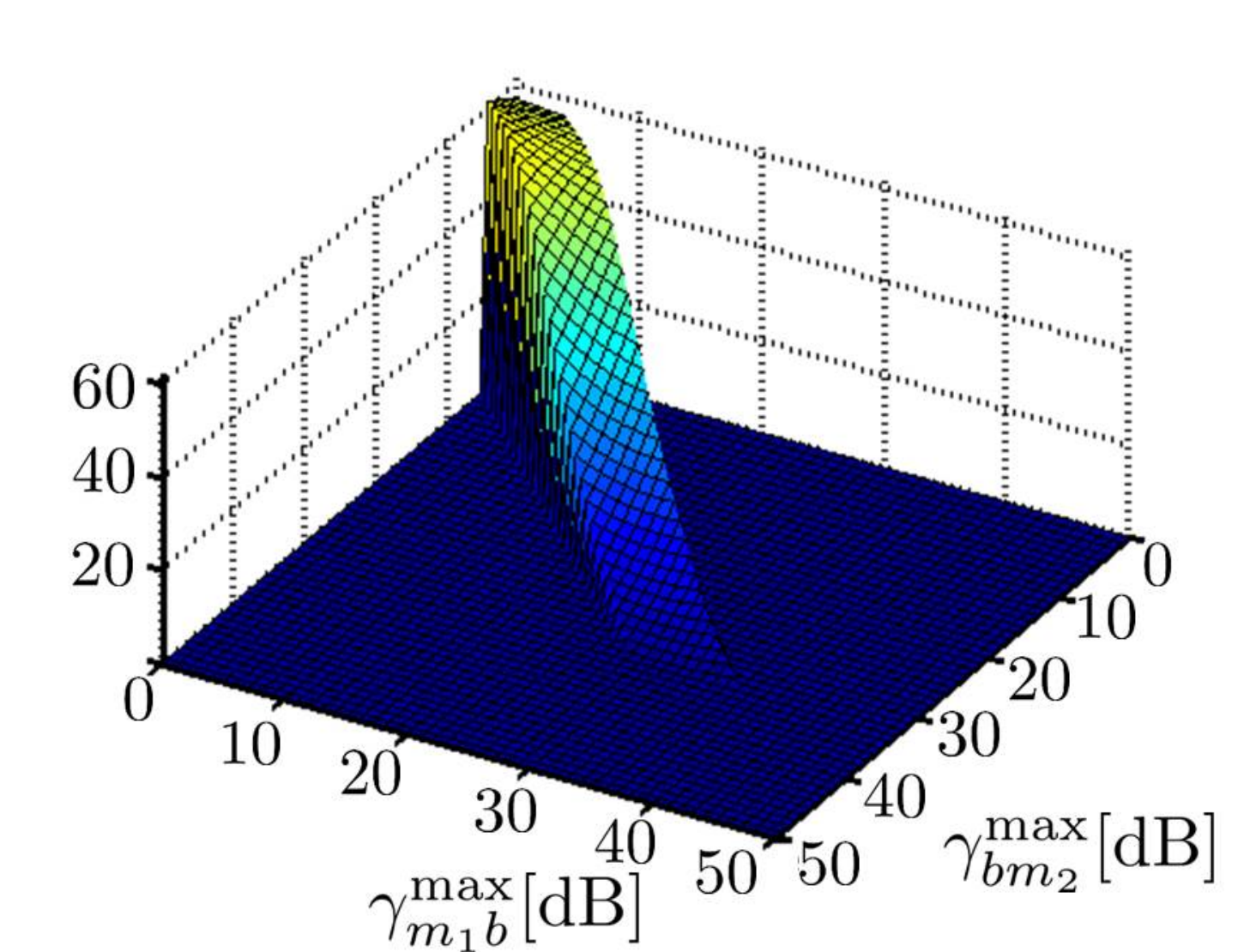}}\hspace*{\fill}\\[-10pt]
    \subfloat
	{
	\includegraphics[scale = 0.26]{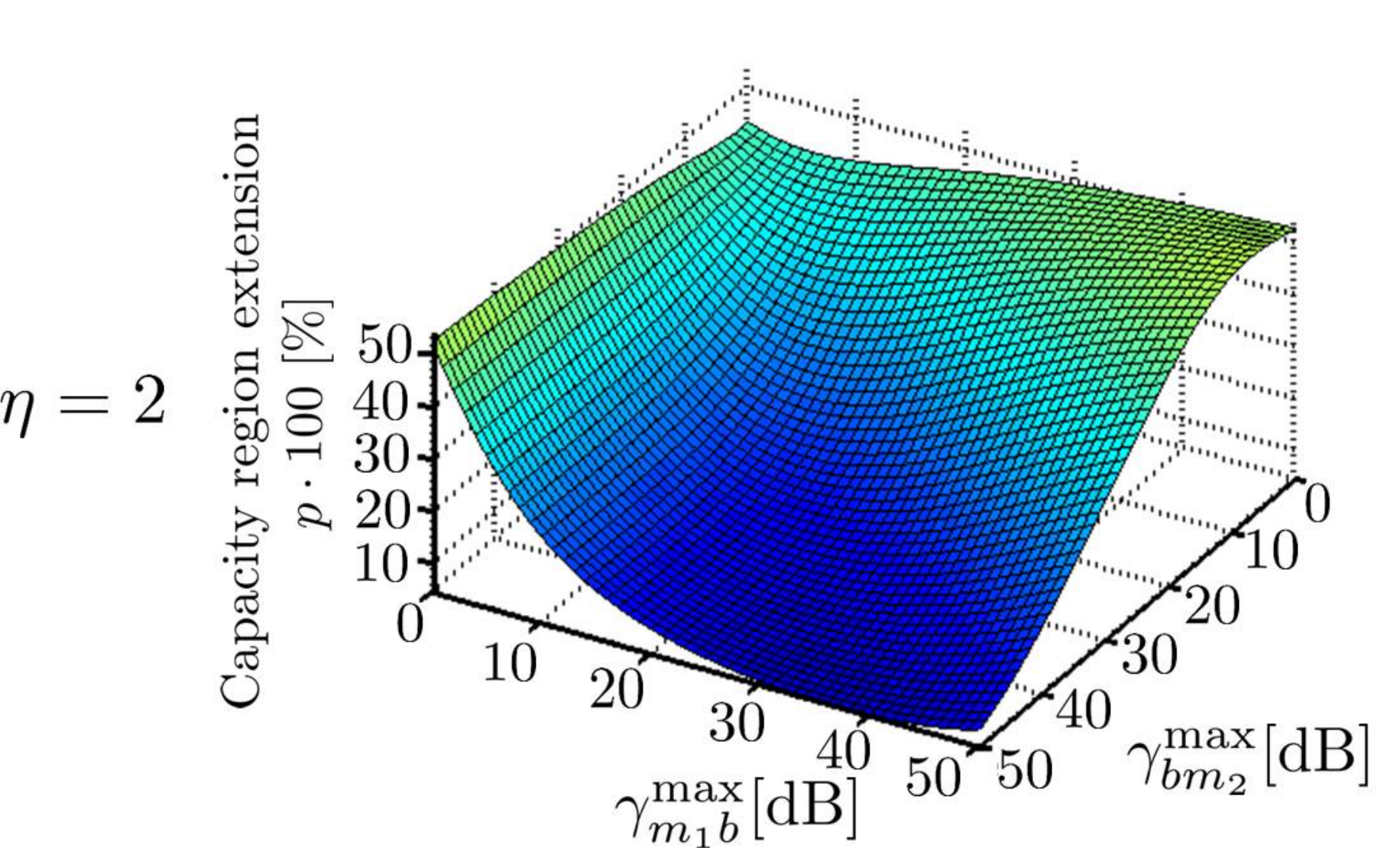}}\hspace{\fill}
	\subfloat
	{
	\includegraphics[scale = 0.26]{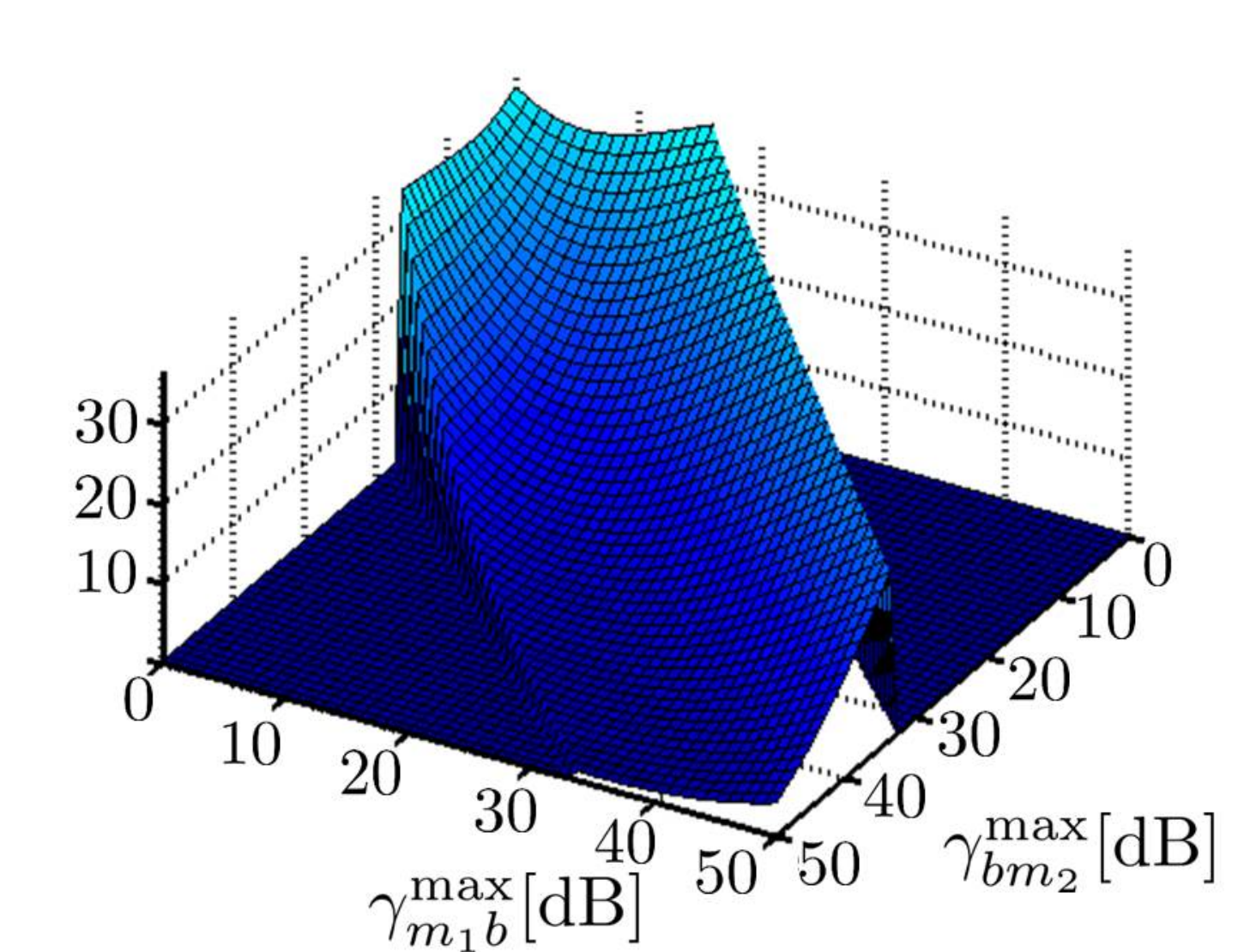}}\hspace{\fill}\subfloat
	{
	\includegraphics[scale = 0.26]{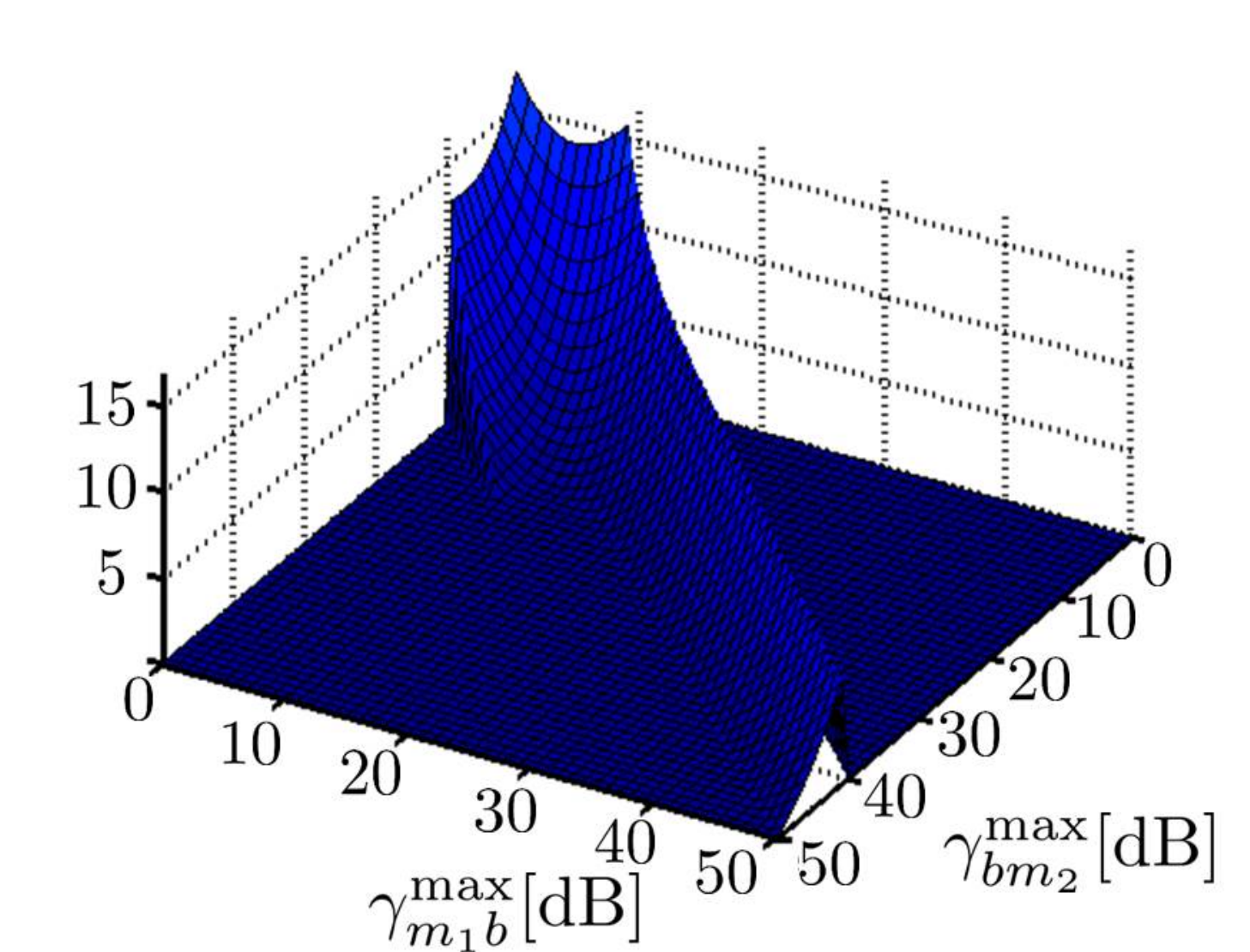}}\hspace{\fill}\subfloat
	{
	\includegraphics[scale = 0.26]{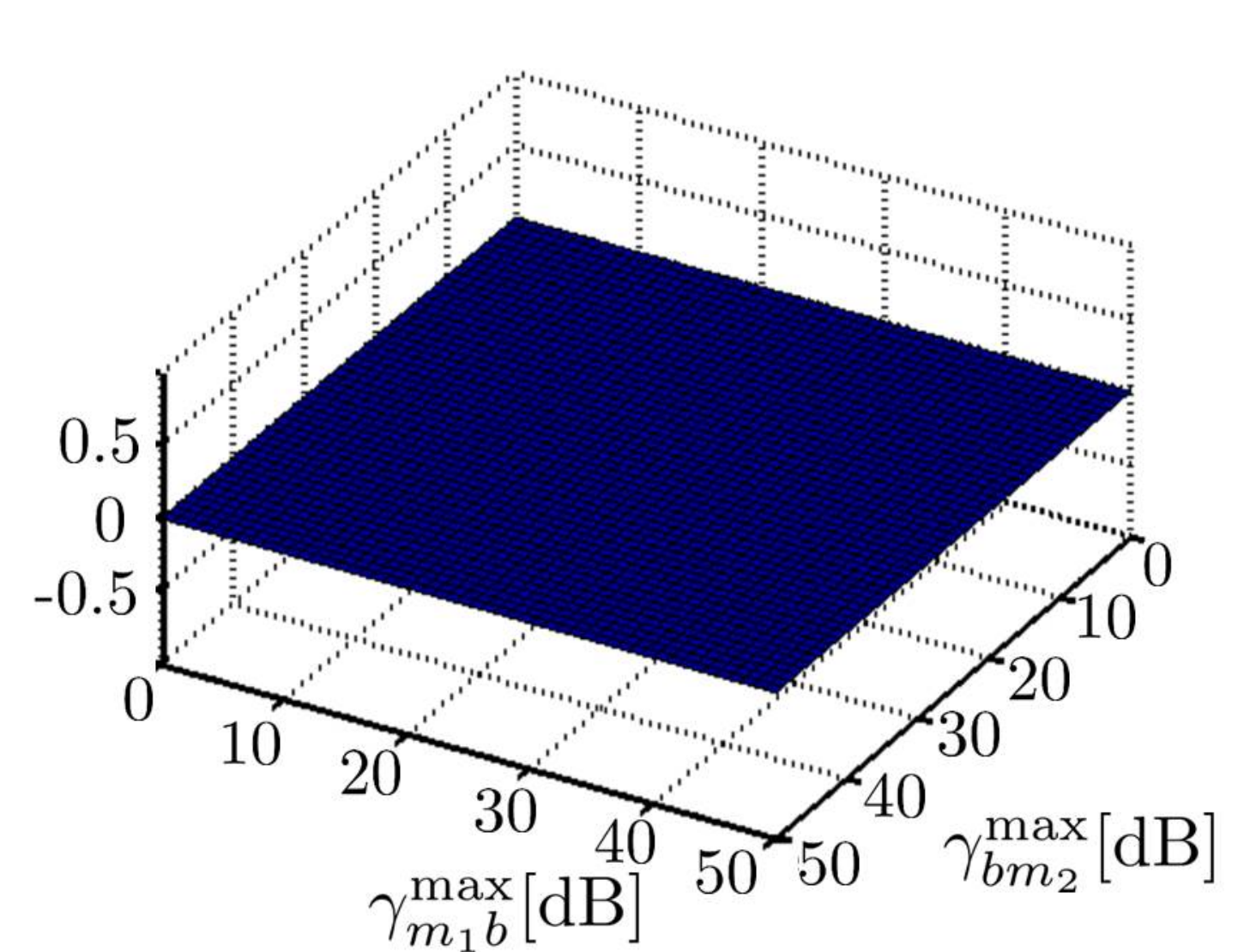}}
	\caption{TDD capacity region extension due to FD as a function of SNRs, where SNRs change due to path loss with exponent $\eta$, and distance between MS 1 and MS 2 is $d_{m_1m_2} = \rho(d_{m_1b} + d_{bm_2})$. Transmission power levels are set to maximum. In SNR regions where the triangle inequality of the distances is not satisfied, $p$ is set to 0. }
	\label{fig:p-in-terms-of-snrs-plus-ini}\vspace{-10pt}
\end{figure*}

Much of the analysis for use case (i) (Section~\ref{section:bidirectional}) extends to use case (ii) (Fig.~\ref{fig:full-duplex-links}\subref{fig:cross}), due to the similarity between the sum rate as a function of transmission power levels for these two use cases (see Eqs.~(\ref{eq:bidirectional-single-SE}) and (\ref{eq:two-uni-SE})). However, there are also important differences. First, the interfering signal at MS 2 in use case (ii), unlike the self-interfering signal at the MS in the bidirectional link case, is not known at the receiver, and therefore, cannot be cancelled (unless an additional channel is used, which we do not consider). Second, in use case (ii), the channel gains between MSs cannot take arbitrary values. This is because the channel gains typically conform to a path loss model of propagation, where the SNR depends on distances between MSs, which in turn need to satisfy the triangle inequality. 
The following two Lemmas are similar to Lemmas \ref{lemma:dependence-on-subch-power} and  \ref{lemma:cap-region-extension-condition}. We state them without proofs.

\begin{lemma}\label{lemma:two-uni-power levels}
If there exists an FD sum rate that is higher than the maximum TDD rate, then the FD sum rate is maximized at $P_{m_1} = \xoverline{P_{m}}$ for MS 1, and $P_b = \xoverline{P_b}$ for the BS. 
\end{lemma}

\begin{lemma}\label{lemma:two-uni-cap-region-extension-condition}
FD extends the TDD capacity region by \hbox{$p\cdot100\%$} if and only if:
\begin{equation}
\frac{\log\left(1 + \frac{\gamma_{bm_2}^{\max}}{1 +\gamma_{m_1m_2}^{\max}}\right)}{\log(1 + \gamma_{bm_2}^{\max})} + \frac{\log\left(1 +\frac{\gamma_{m_1b}^{\max}}{1 + \gamma_{bb}^{\max}}\right)}{\log(1+\gamma_{m_1b}^{\max})} = 1+ p. \label{eq:two-uni-cap-region-extension-condition}
\end{equation}
\end{lemma}

In a path loss model of propagation, the wireless channel gain between two stations is a function of the distance between the stations: $h_{uv} = \left(\frac{L}{d_{uv}}\right)^{\eta}$, where $u, v \in \{b, m_1, m_2\}$, $u\neq v$, $\eta$ is the path loss exponent, and $L$ is a constant. Therefore, as distances $d_{m_1b}$, $d_{bm_2}$, and $d_{m_1m_2}$ need to satisfy the triangle inequality, SNRs $\gamma_{m_1b}$, $\gamma_{bm_2}$ and INR $\gamma_{m_1m_2}$ cannot take arbitrary values. 
To evaluate rate gains in use case (ii), we consider path loss exponents $\eta \in \{2, 3, 4\}$, since typical range for the path loss exponent is between 2 and 4 \cite{rappaport2002wireless}. We assume fixed maximum power levels at the BS and the MS 1, equal noise levels $N$ at the BS and the MS 2, and we vary SNRs and the INR as the function of distance, as follows:
\begin{align*}
\gamma_{m_1b} &= \frac{h_{m_1b}\xoverline{P_{m_1}}}{N} = \frac{h_{m_1b}}{h_{m_1b}^{\max}}\cdot \gamma_{m_1b}^{\max}
=\Big(\frac{d_{m_1b}}{d_{m_1b}^{\min}}\Big)^{\eta}\gamma_{m_1b}^{\max},\\
\gamma_{m_1m_2} &= \frac{h_{m_1m_2}\xoverline{P_{m_1}}}{N} = \frac{h_{m_2m_2}}{h_{m_1m_2}^{\max}}\cdot \gamma_{m_1m_2}^{\max}
=\Big(\frac{d_{m_1m_2}}{d_{m_1m_2}^{\min}}\Big)^{\eta}\gamma_{m_1m_2}^{\max},\\
\gamma_{bm_2} &= \frac{h_{bm_2}\xoverline{P_b}}{N} = \frac{h_{bm_2}}{h_{bm_2}^{\max}}\cdot \gamma_{bm_2}^{\max}
=\Big(\frac{d_{bm_2}}{d_{bm_2}^{\min}}\Big)^{\eta}\gamma_{bm_2}^{\max},
\end{align*}
where $d_{uv}^{\min}$ is a reference distance at which $\gamma_{uv} = \gamma_{uv}^{\max}$ for $u, v \in \{b, m_1, m_2\}$, $x \neq y$. 

For the purpose of comparison, we will assume that $d_{bm_2}^{\min} = d_{m_1b}^{\min} = d_{m_1m_2}^{\min}\equiv d^{\min}$, which would correspond to $\xoverline{P_b} = \xoverline{P_m}$, and normalize all distances to $d^{\min}$. 

Capacity region extension as a function of SNRs is shown in Fig.~\ref{fig:p-in-terms-of-snrs-plus-ini}, for different values of the path loss exponent and $d_{m_1m_2} = \rho (d_{m_1b} + d_{bm_2})$, for $\rho \in \{0.25, 0.5, 0.75, 1\}$. For all combinations of SNRs at which the triangle inequality is not satisfied, we set the capacity region extension $p$ to 0.

Fig.~\ref{fig:p-in-terms-of-snrs-plus-ini} suggests that to achieve over 50\% capacity region extension, the environment needs to be sufficiently lossy, i.e., with the path loss exponent $\eta > 2$. Moreover, to achieve high capacity region extension, the SNRs at the BS and at the MS~2 need to be low enough, meaning that the corresponding distances $d_{m_1b}$ and $d_{bm_2}$ need to be large, since the differences in the SNR shown in all the graphs are due to different distances (and consequently different path loss).

\section{OFDM Bidirectional Links}\label{section:OFDM}

In this section, we focus on the rate maximization for use case (iii) (Fig.~\ref{fig:full-duplex-links}\subref{fig:multi}). Recall that in this use case the FD receiver at the MS has a frequency-selective SIC profile (Fig.~\ref{fig:SIC_Results}\subref{fig:SIC_BW_Circ}). Requiring two technical conditions (Conditions \ref{eq:multi-channel-concavity-condition} and \ref{item:assumption-1}), we derive {an} algorithm {(Algorithm 1, \textsc{MaximumRate})} for the sum rate maximization. {The algorithm is guaranteed to converge to a stationary point, which in practice is typically a global maximum.}  
While the derived algorithm {runs in polynomial time}, its running time is high because it requires invoking a large number of biconvex programming methods. We therefore consider a high SINR approximation of the sum rate, and develop an efficient power allocation algorithm for the sum rate maximization. We also prove that in the high SINR regime it is always optimal to set the maximum SIC frequency in the middle of the used frequency band. 
\vspace{-10pt}
\subsection{Analysis of Sum Rate}\label{section:OFDM-sum-rate}
\subsubsection{Dependence on Channel Power Levels}\label{section:OFDM-sum-rate-power levels}

The analysis of the sum rate in terms of transmission power levels extends from the single-channel case (Section \ref{section:bidirectional}). In particular:
\begin{observation} \label{observ:}
If
\begin{equation}
\frac{g_m(k-c)^2}{N_m}\leq \frac{h_{mb, k}}{N_b + g_bP_{b, k}} \text{ and } {\frac{g_b}{N_b} \leq \frac{h_{bm, k}}{N_m + g_mP_{m, k}(k-c)^2}}\label{eq:multi-channel-concavity-condition-pre}
\end{equation}
hold, then the sum rate is {bi}concave in $P_{m, k}$ and $P_{b, k}$.
\end{observation}
This result is simple to show, since $(k-c)^2$ term is independent of the transmission power levels, and $P_{b, k}$ and $P_{m, k}$ only appear in one summation term ($r_k$). Therefore, we get the same form of partial derivatives in $P_{b, k}$ and $P_{m, k}$ as in the case of a single channel (proof of Lemma \ref{lemma:dependence-on-subch-power}). Similar to the case of a single channel, if condition (\ref{eq:multi-channel-concavity-condition-pre}) is not satisfied, then the achievable rate improvement is low.

The first inequality in (\ref{eq:multi-channel-concavity-condition-pre}) guarantees concavity in $P_{m, k}$ {when $P_{b, k}$ is fixed}, while the second one guarantees concavity in $P_{b, k}$ {when $P_{m, k}$ is fixed}. The condition (\ref{eq:multi-channel-concavity-condition-pre}) cannot be satisfied for any $P_{b, k}\geq 0$, $P_{m, k}\geq 0$ (e.g., the first inequality cannot be satisfied if $\frac{g_m(k-c)^2}{N_m}>\frac{h_{mb, k}}{N_b}$). However, since the role of condition (\ref{eq:multi-channel-concavity-condition-pre}) is to guarantee {bi}concavity in the power levels, we can replace this condition by either $P_{m, k} = 0$ or $P_{b, k} = 0$, which implies rate concavity in $P_{m, k}, P_{b, k}$. Specifically, to guarantee that the sum rate is {bi}concave in all $P_{m, k}, P_{b, k}$, we require the following condition: 
\begin{condition}\label{eq:multi-channel-concavity-condition}
(a) $\frac{g_m(k-c)^2}{N_m}\leq \frac{h_{mb, k}}{N_b + g_bP_{b, k}}$ if $\frac{g_m(k-c)^2}{N_m}< \frac{h_{mb, k}}{N_b}$, otherwise $P_{m, k} = 0$, and \\
(b) $\frac{g_b}{N_b} \leq \frac{h_{bm, k}}{N_m + g_mP_{m, k}(k-c)^2}$ if $\frac{g_b}{N_b}<\frac{h_{bm, k}}{N_m}$; otherwise $P_{b, k} = 0$ if $P_{m, k}$ was not set to 0 by (a).
\end{condition}
Note that Condition \ref{eq:multi-channel-concavity-condition} forces a channel $k$ to be used in half-duplex (only one of $P_{m, k}, P_{b, k}$ is non-zero) whenever it is not possible to satisfy the sufficient condition (\ref{eq:multi-channel-concavity-condition-pre}) for the sum rate {bi}concavity in $P_{m, k}, P_{b, k}$ for any $P_{m, k}\geq 0$ and $P_{b, k}\geq 0$.
\subsubsection{Dependence on Maximum SIC Frequency}

The following lemma shows that choosing optimal $c$ for a given power allocation $\{P_{b, k}, P_{m, k}\}$ is hard in general, since the sum rate $r$ as a function of $c$ is neither convex nor concave, and can have $\Omega(K)$ local maxima. Proof is provided in Appendix.

\begin{lemma}\label{lemma:ci-maxima-localization}
The sum rate $r$ is neither convex nor concave in $c$. All (local) maxima of $r(c)$ lie in the interval $(1, K)$. In general, the number of local maxima is $\Omega(K)$. 
\end{lemma}

Even though $r(c)$ can have multiple maxima in $c$, if we restrict the analysis to the values of $\gamma_{mb, k}$ and $\gamma_{mm, k}$ that are relevant in practice, the selection of $c$, together with the power allocation, are tractable if the following inequalities hold:
\begin{equation} \frac{g_m}{N_m}\leq \frac{h_{mb, k}}{N_b + g_b P_{b,k}}, \forall k\in\{1,...,K\}.  \label{eq:bounded-derivative-ineq} 
\end{equation}
Note that these inequalities are implied by Condition \ref{eq:multi-channel-concavity-condition} for $|k-c|\geq 1$, and that there can be at most 2 channels with $|k-c|<1$. For $|k-c|<1$, the corresponding inequality limits SI on channel $k$. 
The following lemma bounds the first partial derivative of $r$ with respect to $c$. This bound will  prove useful in maximizing $r$ as a function of $c$ and $\{P_{b, k}, P_{m, k}\}$ (Section \ref{section:OFDM-OPS-general}).

\begin{lemma}\label{lemma:bounded-derivative-ci}
If inequalities (\ref{eq:bounded-derivative-ineq}) hold, then: 
\begin{equation*}
\left|\frac{\partial r}{\partial c}\right| \leq \frac{2}{\ln2}(\ln(K)+1+2\sqrt{3})\quad \forall c\in (1, K).
\end{equation*}
\end{lemma}

Similarly as for Condition \ref{eq:multi-channel-concavity-condition}, since (\ref{eq:bounded-derivative-ineq}) cannot be satisfied for $P_b\geq 0$ when $\frac{g_m}{N_m} > \frac{h_{mb, k}}{N_b}$, we require the following:
\begin{condition} \label{item:assumption-1} $\forall k\in\{1,...,K\}$: $\frac{g_m}{N_m}\leq \frac{h_{mb, k}}{N_b + g_b P_{b, k}}$ if $\frac{g_m}{N_m} < \frac{h_{mb, k}}{N_b}$, and $P_{m, k} = 0$ otherwise.   
\end{condition}
Proof of Lemma \ref{lemma:bounded-derivative-ci} can be found in the appendix.

\subsection{Parameter Selection Algorithms}\label{section:OFDM-OPS}

\subsubsection{General SINR Regime}\label{section:OFDM-OPS-general}

The pseudocode of the algorithm for maximizing the sum rate in the general SINR regime is provided in Algorithm 1 -- \textsc{MaximumRate}. We claim the following:
\begin{lemma}\label{lemma:maximum-rate}
{Under Conditions \ref{eq:multi-channel-concavity-condition} and \ref{item:assumption-1}, the sum rate maximization problem is biconvex. If biconvex programming subroutine in \textsc{MaximumRate} finds a global optimum for $\{P_{b, k}, P_{m, k}\}$, then \textsc{MaximumRate} determines $c$ and the power allocation $\{P_{b, k}, P_{m, k}\}$ that maximize sum rate up to an absolute error $\epsilon$, for any $\epsilon > 0$.}
\end{lemma}
\begin{algorithm}
\small
\caption{\textsc{MaximumRate}($\epsilon$)}
\begin{algorithmic}[1]
\Statex Input: $K, \xoverline{P_b}, \xoverline{P_m}, g_b, g_m, N_m, N_b$
\State $c_{1} = 1$, $c_{2} = K$, $\Delta c = \frac{\epsilon}{\frac{2}{\ln2}(\ln(K)+1+2\sqrt{3})}$
\State $c^{\max} = r^{\max} = 0$, $\{P_{b, k}^{\max}\} = \{P_{m, k}^{\max}\} = \{0\}$
\For{$c = c_1$, $c < c_2$, $c = c + \Delta c$}
	\State \label{step:grad-ascent} Solve via {biconvex programming}:
    \Statex$\hspace{0.3in}\max\quad r = \sum_{k=1}^K r_k$, where $r_k$ is given by (\ref{eq:bidirectional-SE})
    \Statex\hspace{0.3in}\textbf{s.t.} $\>$ Conditions \ref{eq:multi-channel-concavity-condition} and \ref{item:assumption-1} hold
    \Statex\hspace{0.6in}$\sum_{k=1}^K P_{m, k} \leq \xoverline{P_m},\quad \sum_{k=1}^K P_{b, k}\leq \xoverline{P_b}$
    \Statex\hspace{0.6in}$ P_{b, k} \geq 0,\quad P_{m, k}\geq 0\quad, \forall k\in \{1,...,K\}$.
   \If{$r > r^{\max}$}
   \State $r^{\max} = r,\quad c^{\max} = c,$
   \State $\{P_{b, k}^{\max}\} = \{P_{b, k}\},\quad \{P_{m, k}^{\max}\} = \{P_{m, k}\}$
   \EndIf
\EndFor
\State \textbf{return} $c^{\max}, \{P_{b, k}^{\max}\}, \{P_{m, k}^{\max}\}, r^{\max}$.
\end{algorithmic}
\end{algorithm}

{Note that without Condition \ref{eq:multi-channel-concavity-condition}, the biconvex programming subroutine in \textsc{MaximumRate} would not be guaranteed to converge to a stationary point (see, e.g., \cite{gorski2007biconvex}). Moreover, since the sum rate is highly nonlinear in the parameter $c$ (Lemma \ref{lemma:ci-maxima-localization}), $c$ cannot be used as a variable in the biconvex programming routine (or a convex programming method). Nevertheless, as a result of Lemma \ref{lemma:bounded-derivative-ci} that bounds the first derivative of $r$ with respect to $c$ when condition \ref{item:assumption-1} is applied, we can restrict our attention to $c$'s from a discrete subset of the interval $(1, K)$.}

\begin{IEEEproof}[Proof of {Lemma \ref{lemma:maximum-rate}}]
Consider the optimization problem in Step \ref{step:grad-ascent} of the algorithm. Since  Condition \ref{eq:multi-channel-concavity-condition} is required by the constraints, the objective $r$ is concave in $P_{b, k}$ {whenever $P_{m, k}$'s are fixed, and, similarly}, {concave in} $P_{m, k}$ {whenever $P_{b, k}$'s are fixed}. {Therefore, $r$ is biconcave in $P_{b, k}, P_{m, k}$.} The feasible region of the problem from Step \ref{step:grad-ascent} is determined by linear inequalities and Conditions \ref{eq:multi-channel-concavity-condition} and \ref{item:assumption-1}. 

Condition \ref{eq:multi-channel-concavity-condition} is either an inequality or an equality for each $P_{m, k}, P_{b, k}$ that (possibly rearranging the terms) is linear in $P_{m, k}, P_{b, k}$. Condition \ref{item:assumption-1} is a linear inequality in $P_{m, k}$. Therefore, the feasible region in the problem of Step \ref{step:grad-ascent} is a polyhedron and therefore convex. It follows immediately that this problem {is biconvex}.

{Suppose that the biconvex programming method from Step 4 of \textsc{MaximumRate} finds a global optimum.} The{n the} algorithm finds an optimal power allocation for each $c$ from the set of $\frac{(K-1)(\frac{2}{\ln2}(\ln(K)+1+2\sqrt{3}))}{\epsilon} - 2$ equally spaced points from the interval $(1, K)$, and chooses $c$ and power allocation that provide maximum sum rate $r$. 

What remains to prove is that by choosing any alternative $c \neq c^{\max}$ and accompanying optimal power allocation the sum rate cannot be improved by more than an additive $\epsilon$. 

Recall from Lemma \ref{lemma:ci-maxima-localization} that optimal $c$ must lie in $(1, K)$. 
Suppose that there exist $c^*, \{P_{b, k}^*, P_{m, k}^*\}$ such that $c^*\in(1, K)$, $c^*\neq c^{\max}$ and $r(c^*, \{P_{b, k}^*, P_{m, k}^*\}) > r^{\max} + \epsilon$.

From the choice of points $c$ in the algorithm, there must exist at least one point $c^a$ that the algorithm considers such that $|c^a-c^*|<\Delta c = \frac{\epsilon}{\frac{2}{\ln2}(\ln(K)+1+2\sqrt{3} )}$. From Lemma \ref{lemma:bounded-derivative-ci}, 
\begin{align*}
r&(c^*, \{P_{b, k}^*, P_{m, k}^*\}) - r(c^a, \{P_{b, k}^*, P_{m, k}^*\}) < \\
&\frac{\epsilon}{\frac{2}{\ln2}(\ln(K)+1+2\sqrt{3})}\cdot \Big(\frac{2}{\ln2}(\ln(K)+1+2\sqrt{3}) \Big)= \epsilon,
\end{align*}
since in any finite interval $I$ any continuous and differentiable function $f(x)$ cannot change by more than the length of the interval $I$ times the maximum value of its first derivative $f'(x)$ (a simple corollary of the Mean-Value Theorem). 

Since the algorithm finds an optimal power allocation for each $c$, we have that $r(c^a, \{P_{b, k}^*, P_{m, k}^*\})\leq r(c^a, \{P_{b, k}^a, P_{m, k}^a\})\leq r^{\max}$. Therefore: $r(c^*, \{P_{b, k}^*, P_{m, k}^*\}) - r^{\max} < \epsilon$, which is a contradiction.
\end{IEEEproof}

\subsubsection{High SINR Regime}\label{section:OFDM-OPS-high-SINR}

A high SINR approximation of the sum rate is:
\begin{equation}
r \approx \sum_{k=1}^K \Big(\log\Big(\frac{\gamma_{mb, k}}{1 + \gamma_{bb, k}}\Big) + \log\Big(\frac{\gamma_{bm, k}}{1 + \gamma_{mm, k}}\Big)\Big). \label{eq:se-high-sinr-approx}
\end{equation}

While in the high SINR regime the dependence of sum rate on each power level $P_{b, k}, P_{m, k}$ for $k\in\{1,..., K\}$ becomes concave (regardless of whether Condition \ref{eq:multi-channel-concavity-condition} holds or not), the dependence on the parameter $c$ remains neither convex nor concave as long as we consider a general power allocation. Therefore, we cannot derive a closed form expression for $c$ in terms of an arbitrary power allocation. However, as we show in Lemma~\ref{lemma:high-sinr-ci-at-middle}, when power allocation and the choice of parameter $c$ are considered jointly, it is always optimal to place $c$ in the middle of the interval $(1, K)$: $c = \frac{K+1}{2}$. 
The following proposition and lemma characterize the optimal power allocation for a given $c$. 

\begin{proposition}\label{prop:high-sinr-equal-power}
Under high-SINR approximation and any power allocation $\{P_{m, k}\}$ at the MS and any choice of $c$, it is always optimal to allocate BS power levels as $P_{b, k} = \frac{\xoverline{P_b}}{K}$.
\end{proposition}

\begin{IEEEproof}
Let $P_b$ denote the total irradiated power by the BS. Write power levels on individual subchannels as $P_{b, k} = \beta_k P_{b}$, where $\beta_k \geq 0$, $\forall k \in \{1,..., K\}$, and $\sum_{k=1}^K \beta_k = 1$. Then the sum rate can be written as:
\begin{align}
r = \sum_{k=1}^K &\Big(\log\Big( \frac{h_{mb, k}P_{m, k}}{N_b + g_b \beta _k P_b} \Big)\notag\\ &
+ \log\Big( \frac{h_{bm, k}\beta_k P_b}{N_m + g_m(k-c)^2P_{m, k}} \Big)\Big).\notag
\end{align}

First, observe that 
\begin{align*}
\frac{\partial r}{\partial P_b} &= \sum_{k=1}^K \mathds{1}_{\{\beta_k>0\}}\left(\frac{1}{P_b} - \frac{g_b\beta_k}{N_b + g_b\beta_k P_b}\right)\\
&= \sum_{k=1}^K \mathds{1}_{\{\beta_k>0\}}\Big({P_b}^{-1} - ({N_b}/({g_b\beta_k}) +  P_b)^{-1}\Big),
\end{align*}
where $\mathds{1}_{\{.\}}$ is an indicator function. Since $\beta_k \geq 0$ $\forall k \in \{1,..., K\}$ and $\sum_{k=1}^K \beta_k = 1$, it follows that there exists at least one strictly positive $\beta_k$. For each such $\beta_k$, 
$\mathds{1}_{\beta_k>0}\Big(\frac{1}{P_b} - \frac{1}{\frac{N_b}{g_b\beta_k} +  P_b}\Big)>0, 
$
since $P_b < \frac{N_b}{g_b\beta_k} +  P_b$. Therefore,  $\frac{\partial r}{\partial P_b}>0$, which implies that it is optimal to choose $P_b = \xoverline{P_b}$.

Taking the first and the second partial derivative of $r$ with respect to each $\beta_k$, it is simple to show that $r$ has the same dependence on each $\beta_k$, and, moreover, is strictly concave in each $\beta_k$, as 
$\frac{\partial^2 r}{\partial {\beta_k}^2} = -\frac{1}{{\beta_k}^2} +\frac{1}{\left(\frac{N_b}{g_bP_b}+\beta_k\right)^2} <0,
$ 
where the inequality follows from $\beta_k < \frac{N_b}{g_bP_b} + \beta_k$. Therefore, $r$ is maximized for $\beta_k = \frac{1}{K}$.
\end{IEEEproof}

\begin{lemma}\label{lemma:high-sinr-pik}
Under high-SINR approximation and for a given, fixed, $c$ the optimal power allocation at the MS satisfies $P_{m, k} = \alpha_k\cdot \xoverline{P_{m}}$, where $\alpha_k \geq 0$, $\sum\alpha_{k} = 1$, and for $k\neq K$:\\
(i) $\alpha_k = \left(\frac{1}{\alpha_K} - \frac{1}{N_m/R_{K} + \alpha_K}\right)^{-1}$ if $k=c$,\\
(ii) $\alpha_k = \frac{-N_m+\sqrt{{N_m}^2+4\alpha_K(N_m + R_{K}\alpha_K)R_{k}}}{2R_{k}}$ if $k\neq c$,\\
where $R_{k} = g_m(k-c)^2P_m$ for $k\in\{1,...,K\}$.
\end{lemma}

\begin{IEEEproof}
Let $P_{m, k} = \alpha_k\cdot P_m$, where $\alpha_k > 0$, $\forall k$, and $\sum_{k=1}^K \alpha_k =1$. The sum rate can then be written as:
\begin{align}
r = \sum_{k=1}^K &\Big(\log\Big( \frac{h_{mb, k}\alpha_kP_{m}}{N_b + g_b P_{b, k}} \Big)\notag\\ 
&+ \log\Big( \frac{h_{bm, k}P_{b,k}}{N_m + g_m(k-c)^2\alpha_kP_{m}} \Big)\Big)\notag.
\end{align}
Proving that at the optimal solution that maximizes $r$ we necessarily have $P_m = \xoverline{P_m}$ is analogous to the proof given in Proposition~\ref{prop:high-sinr-equal-power} for $P_b = \xoverline{P_b}$, and it is therefore omitted.

As $\sum_{k=1}^K \alpha_k =1$, only $K-1$ $\alpha_k$'s can be chosen independently, while the value of the remaining one is implied by their sum being equal to 1. Choose $\alpha_K = 1 - \sum_{k=1}^{K-1}\alpha_k$, and observe that that $K-c \neq 0$ (and therefore $R_K\neq 0$) is always true since, similar as in the proof of Lemma \ref{lemma:ci-maxima-localization}, at the optimum it must be $c\in(1, K)$.

If $R_{k} = g_m(k-c)^2P_m = 0$, then the first and the second derivative of $r$ with respect to $\alpha_k$ are given as:
\begin{align*}
\frac{\partial r}{\partial \alpha_k} &= \frac{1}{\alpha_k} + \frac{1}{\alpha_K}\frac{\partial \alpha_K}{\partial\alpha_k} - \frac{R_{K}}{N_m + R_{K} \alpha_K}\frac{\partial \alpha_K}{\partial\alpha_k}\\
&= \frac{1}{\alpha_k} - \frac{1}{\alpha_K} + \frac{R_{K}}{N_m + R_{K} \alpha_K},\\
\frac{\partial^2 r}{\partial {\alpha_k}^2} =& - \frac{1}{\alpha_k^2} - \left(\frac{1}{\alpha_K^2} - \frac{{R_{K}}^2}{(N_m + R_{K} \alpha_K)^2}\right)\\
=& - \frac{1}{\alpha_k^2} - \left(\frac{1}{\alpha_K^2} - \frac{1}{(N_m/R_{K} + \alpha_K)^2}\right)<0.
\end{align*}
It follows that $r$ is concave in $\alpha_k$ and maximized for 
\begin{align}
\alpha_k &= \big({\alpha_K}^{-1} - ({N_m/R_{K} + \alpha_K})^{-1}\big)^{-1},\label{eq:alphak-for-Rk-eq0}
\end{align}
where $R_K = g_m(K-c)^2P_m$.

If $R_{k}\neq 0$, then the first and the second derivative are:
\begin{align*}
\frac{\partial r}{\partial \alpha_k} &= \frac{1}{\alpha_k} - \frac{R_{k}}{N_i + R_{k} \alpha_k} - \frac{1}{\alpha_K} + \frac{R_{K}}{N_m + R_{K} \alpha_K},\\
\frac{\partial^2 r}{\partial {\alpha_k}^2} =& - \frac{1}{\alpha_k^2} + \frac{{R_{k}}^2}{(N_m + R_{k} \alpha_k)^2} - \frac{1}{\alpha_K^2} + \frac{{R_{K}}^2}{(N_m + R_{K} \alpha_K)^2}\\
=& - \left({\alpha_k^{-2}} -  {\left({N_m}/{R_{k}} + \alpha_k\right)^{-2}}\right) \\&- \left({\alpha_K^{-2}} - {\left({N_m}/{R_{k}} + \alpha_K\right)^{-2}}\right)
<0.
\end{align*}
It follows that $r$ is concave in $\alpha_k$ and maximized for:
\begin{equation}
\frac{\partial r}{\partial \alpha_k} = \frac{1}{\alpha_k} - \frac{R_{k}}{N_i + R_{k} \alpha_k} - \frac{1}{\alpha_K} + \frac{R_{K}}{N_m + R_{K} \alpha_K} = 0. \label{eq:dr-dalphak}
\end{equation}
After simplifying (\ref{eq:dr-dalphak}), we get:
\begin{equation}
\alpha_k(N_m + R_{k}\alpha_k) = \alpha_K(N_m + R_{K}\alpha_K). \label{eq:quadratic-alphak}
\end{equation}
Solving the quadratic equation (\ref{eq:quadratic-alphak}) for $\alpha_k$ and using that $\alpha_k>0$, it follows that $r$ is maximized when $\alpha_k$ satisfies
\begin{equation}
\alpha_k = \frac{-N_m+\sqrt{{N_m}^2+4\alpha_K(N_m + R_{K}\alpha_K)R_{k}}}{2R_{k}},\label{eq:alphak-for-Rk-neq0}
\end{equation}
where $R_k = g_m(k-c)^2P_m$.
\end{IEEEproof}

It is relatively simple to show (using similar approach as in the proof of Lemma \ref{lemma:ci-maxima-localization}) that under general power allocation $r$ can have up to $K$ local maxima with respect to $c$. However, if $c$ is considered with respect to the optimal power allocation corresponding to $c$ (Proposition \ref{prop:high-sinr-equal-power} and Lemma \ref{lemma:high-sinr-pik}), it is always optimal to place $c$ in the middle of the interval $(1, K)$, as the following lemma states. 
 
\begin{lemma} \label{lemma:high-sinr-ci-at-middle}
If $\left(c, \{P_{b, k}, P_{m, k}\}\right)$ maximizes the sum rate under high SINR approximation, then $c = \frac{K+1}{2}$.
\end{lemma}
Even though this result may seem intuitive because the optimal power allocation is always symmetric around $c$ (Proposition \ref{prop:high-sinr-equal-power} and Lemma \ref{lemma:high-sinr-pik}), the proof does not follow directly from this property and requires many technical details. For this reason, the proof is deferred to the appendix. 

A simple corollary of Lemma \ref{lemma:high-sinr-ci-at-middle} is that:
\begin{corollary}\label{cor:high-sinr-symmetric-pik}
If $(c^*, \{P_{m, k}^{\max}, P_{b, k}^{\max}\})$ maximizes $r$ under high SINR approximation, then the power allocation $\{P_{m, k}^{\max}\}$ is symmetric around $\frac{K+1}{2}$ and decreasing in $|k-c|$.
\end{corollary}

\begin{IEEEproof}
The first part follows directly from $c^{\max} = \frac{K+1}{2}$. The second part is proved in Lemma \ref{lemma:high-sinr-ci-at-middle}.
\end{IEEEproof}

\begin{lemma}
A solution $(c^{\max}, \{P_{m, k}^{\max}, P_{b, k}^{\max}\})$ that maximizes $r$ under high SINR approximation up to an absolute error $\epsilon$ can be computed in $O\left(K\log\left(\frac{1}{\epsilon}\right)\right)$ time.
\end{lemma}

\begin{IEEEproof}
From Proposition \ref{prop:high-sinr-equal-power}, at the optimum $P_{b, k}^{\max} = \frac{P_{b,\max}}{K}$, $\forall k \in \{1,..., K\}$. This can be computed in constant time, and requires $\Theta(K)$ time to assign the values to all the $P_{b, k}$'s. 
From Lemma \ref{lemma:high-sinr-ci-at-middle}, $c^{\max} = \frac{K+1}{2}$.

From Lemma \ref{lemma:high-sinr-pik}, $P_{m, k}^{\max} = \alpha_k \xoverline{P_m}$, where $\{\alpha_k\}$ are positive coefficients given by (\ref{eq:alphak-for-Rk-eq0}), (\ref{eq:alphak-for-Rk-neq0}) and $\sum_{k=1}^K \alpha_k = 1$. Recall that all the $\alpha_k$'s are given in terms of $\alpha_K$, so we can find the allocation $\{\alpha_k\}$ by performing a binary search for $\alpha_K$ until $\sum_{k=1}^K \alpha_k \in [1 - \epsilon', 1]$. Corollary \ref{cor:high-sinr-symmetric-pik} implies that $\alpha_K \leq \frac{1}{K}$, so it is sufficient to perform the binary search for $\alpha_K\in \big[0, \frac{1}{K}\big]$. Such a binary search requires $O\big(\log\big(\frac{1}{K\epsilon'}\big)\big)$ iterations, with each iteration requiring $O(K)$ time to compute $\{\alpha_k\}$ and evaluate $\sum_{k=1}^K \alpha_k$, for the total time $O\big(K\log\big(\frac{1}{K\epsilon'}\big)\big)$.

The last part of the proof is to determine an appropriate $\epsilon'$ so that $r(c^{\max}, \{P_{m, k}^{\max}, P_{b, k}^{\max}\})\geq \max r - \epsilon$, where the maximum is taken over all feasible points $(c, \{P_{m, k}, P_{b, k}\})$. Notice that we are only deviating from the optimal solution in that $\sum_{k=1}^K P_{m, k}^{\max} = \xoverline{P_m}\cdot \sum_{k=1}^K \alpha_k \in [\xoverline{P_m} (1-\epsilon'), \xoverline{P_m}]$ instead of $\sum_{k=1}^K P_{m, k}^{\max} = \xoverline{P_m}$. Therefore, $(c^{\max}, \{P_{m, k}^{\max}, P_{b, k}^{\max}\})$ is the optimal solution to the problem that is equivalent to the original problem, with maximum total power at the MS equal to $\xoverline{P_m} \cdot \sum_{k=1}^K \alpha_k $. Observe that:
\begin{align*}
\frac{\partial r}{\partial P_m} &= \sum_{k=1}^K \Big( \frac{1}{P_m} - \frac{1}{\frac{N_m}{g_m(k-c)^2} + P_i}\cdot \mathds{1}_{\{k \neq c\}} \Big)
\leq \frac{K}{P_m}.
\end{align*}
As $\frac{\partial r}{\partial P_m} (P_m) \leq \frac{K}{\xoverline{P_m}(1-\epsilon')}$ for $P_m \in [\xoverline{P_m} (1-\epsilon'), \xoverline{P_m}]$, it follows that: 
$\max r - r(c^{\max}, \{P_{m, k}^{\max}, P_{b, k}^{\max}\}) \leq \frac{K}{\xoverline{P_m}(1-\epsilon')}\cdot \xoverline{P_m} \epsilon'$.
Setting: 
$\frac{K}{\xoverline{P_m}(1-\epsilon')}\cdot \xoverline{P_m} \epsilon' = \epsilon \quad \Leftrightarrow \quad \epsilon' = \frac{\epsilon}{K + \epsilon}$,
we yield the total running time of: 
$O\left(K\log\left(\frac{K+\epsilon}{K\epsilon}\right)\right) = O\left(K\log\left(\frac{1}{\epsilon}\right)\right)$.
\end{IEEEproof}

We summarize the results from this section in Algorithm 2 -- \textsc{HSINR-MaximumRate}.
\begin{algorithm}
\small
\caption{\textsc{HSINR-MaximumRate}($\epsilon$)}
\begin{algorithmic}[1]
\Statex Input: $K, \xoverline{P_b}, \xoverline{P_m}, g_b, g_m, N_m, N_b$
\State $c^{\max} = (K+1)/{2}$
\State $\{P_{b, k}^{\max}\} = \xoverline{P_b}/K$, $\forall k \in \{1,...,K\}$
\For{$\alpha_K \in [0, 1/K]$, via a binary search}
\State Compute $\alpha_k$ for $1\leq k\leq K-1$ using (\ref{eq:alphak-for-Rk-eq0}) and (\ref{eq:alphak-for-Rk-neq0})
\State End binary search when $\sum_{k=1}^K \alpha_k \in [1- \epsilon/(K+\epsilon), 1]$
\EndFor
\State \textbf{return} $c^{\max}, \{P_{b, k}^{\max}\}, \{P_{m, k}^{\max}\}$.
\end{algorithmic}
\end{algorithm}
\vspace{-10pt}

\begin{figure*}[t]
\centering
\subfloat[BS, $\gamma^{\mathrm{avg}}=0$dB]{\includegraphics[scale = 0.22]{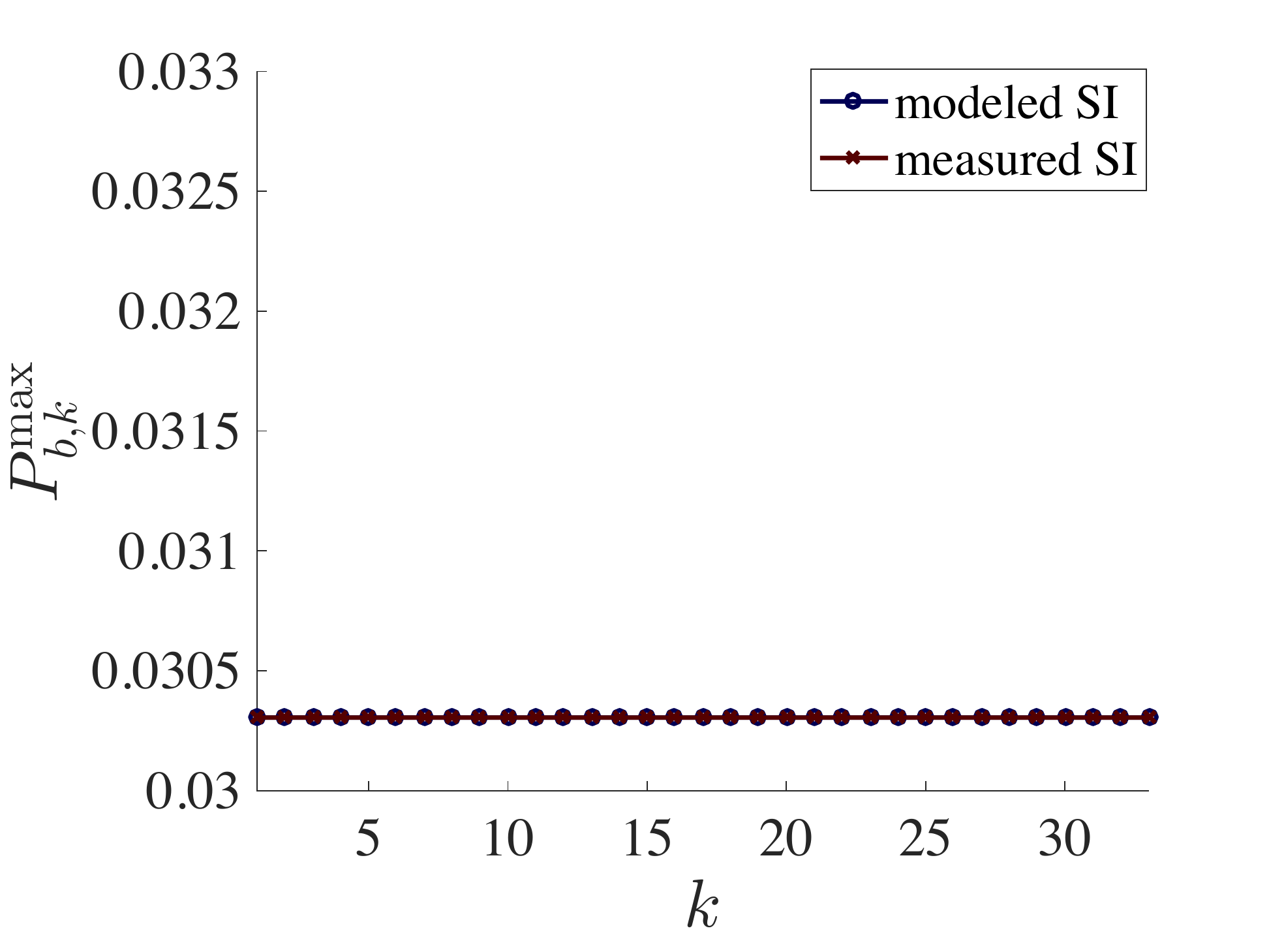}\label{fig:Pb_0dB}}\hspace{0.05in}
\subfloat[BS, $\gamma^{\mathrm{avg}}=10$dB]{\includegraphics[scale = 0.22]{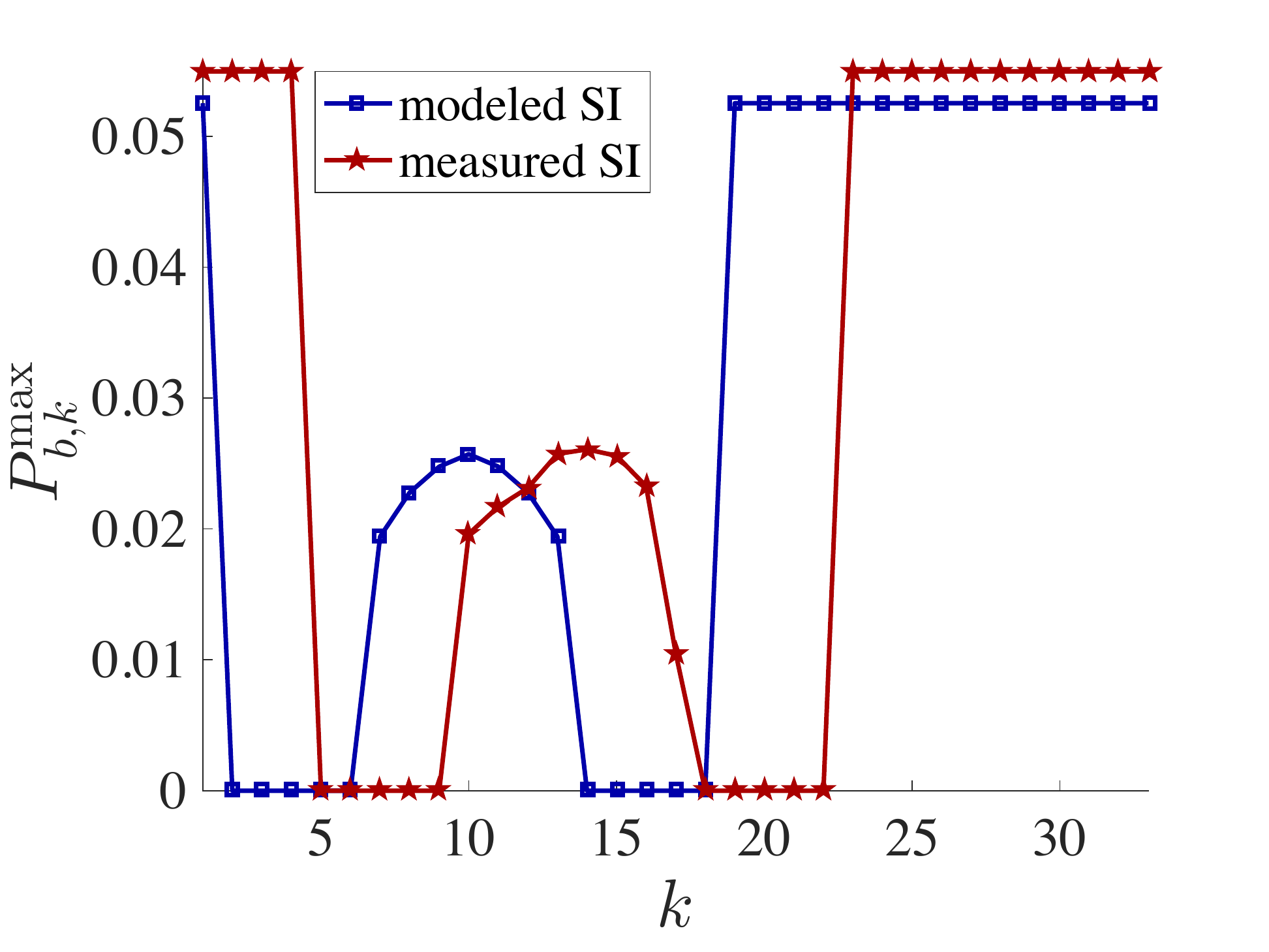}\label{fig:Pb_10dB}}
\subfloat[BS, $\gamma^{\mathrm{avg}}=20$dB]{\includegraphics[scale = 0.22]{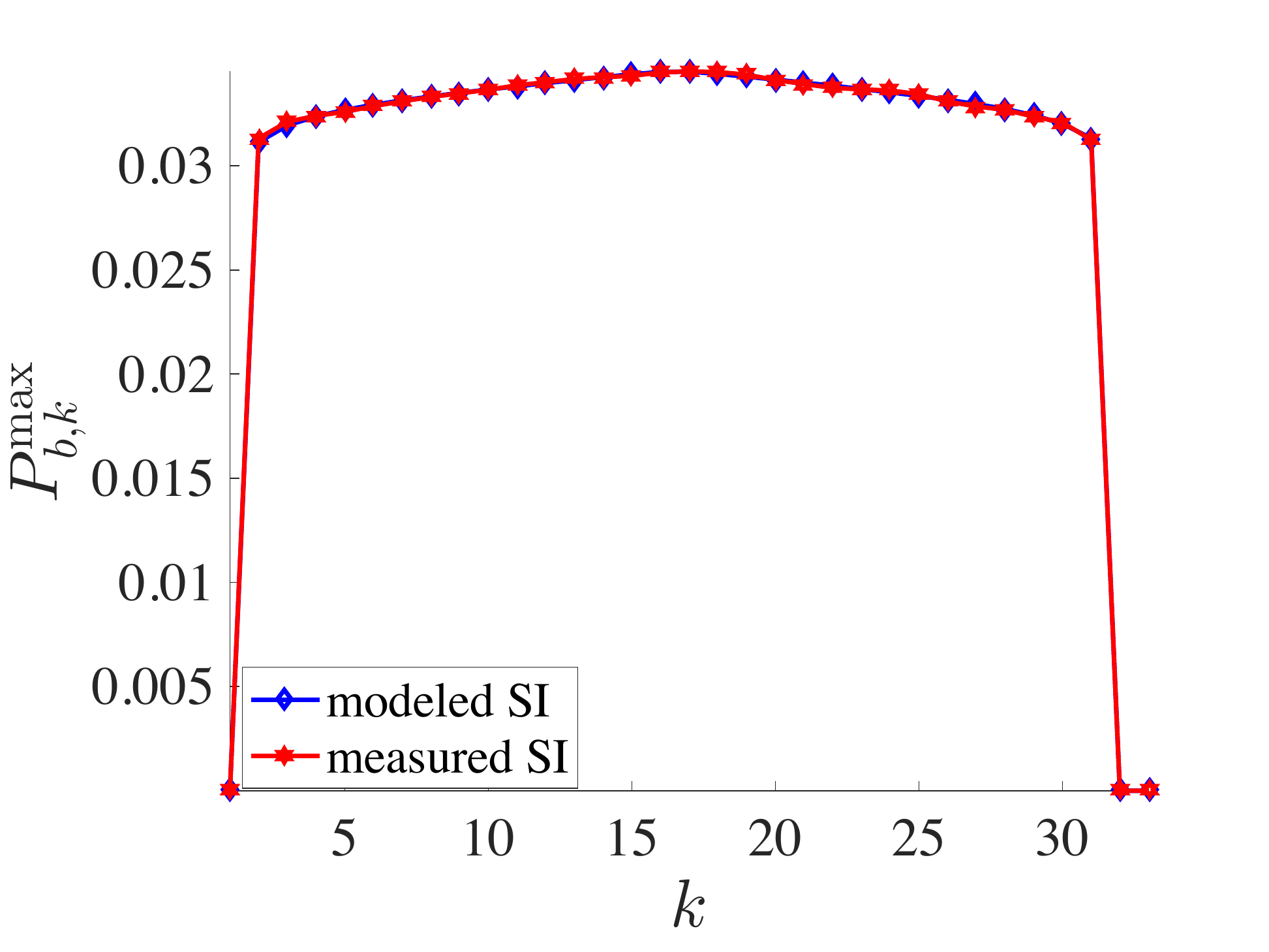}\label{fig:Pb_20dB}}
\subfloat[BS, $\gamma^{\mathrm{avg}}\in\{30, 40, 50\}$dB]{\includegraphics[scale = 0.22]{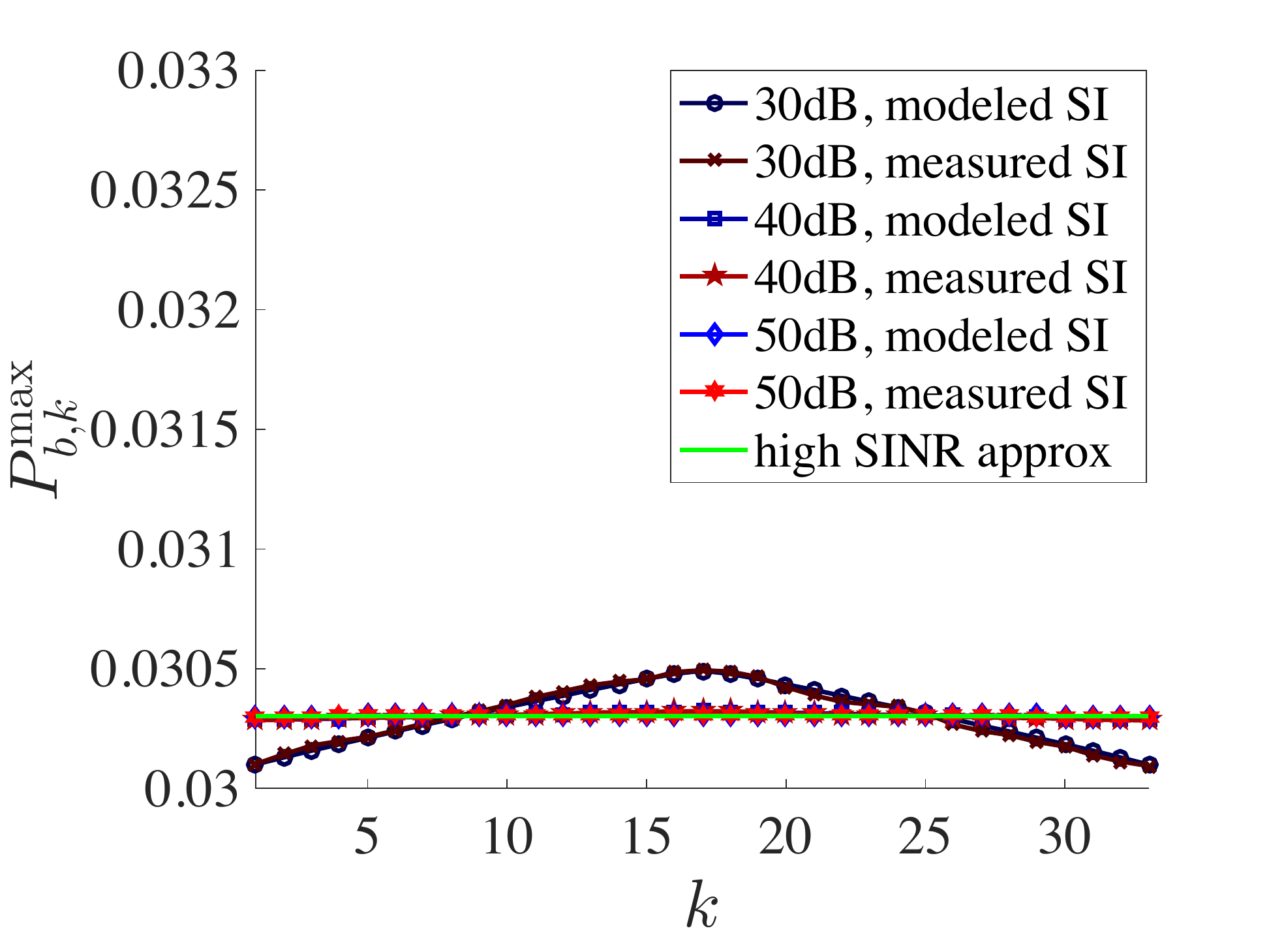}\label{fig:Pb_hsinr}}\\\vspace{-10pt}
\hspace{0.15in}\subfloat[MS, $\gamma^{\mathrm{avg}}=0$dB]{\includegraphics[scale = 0.22]{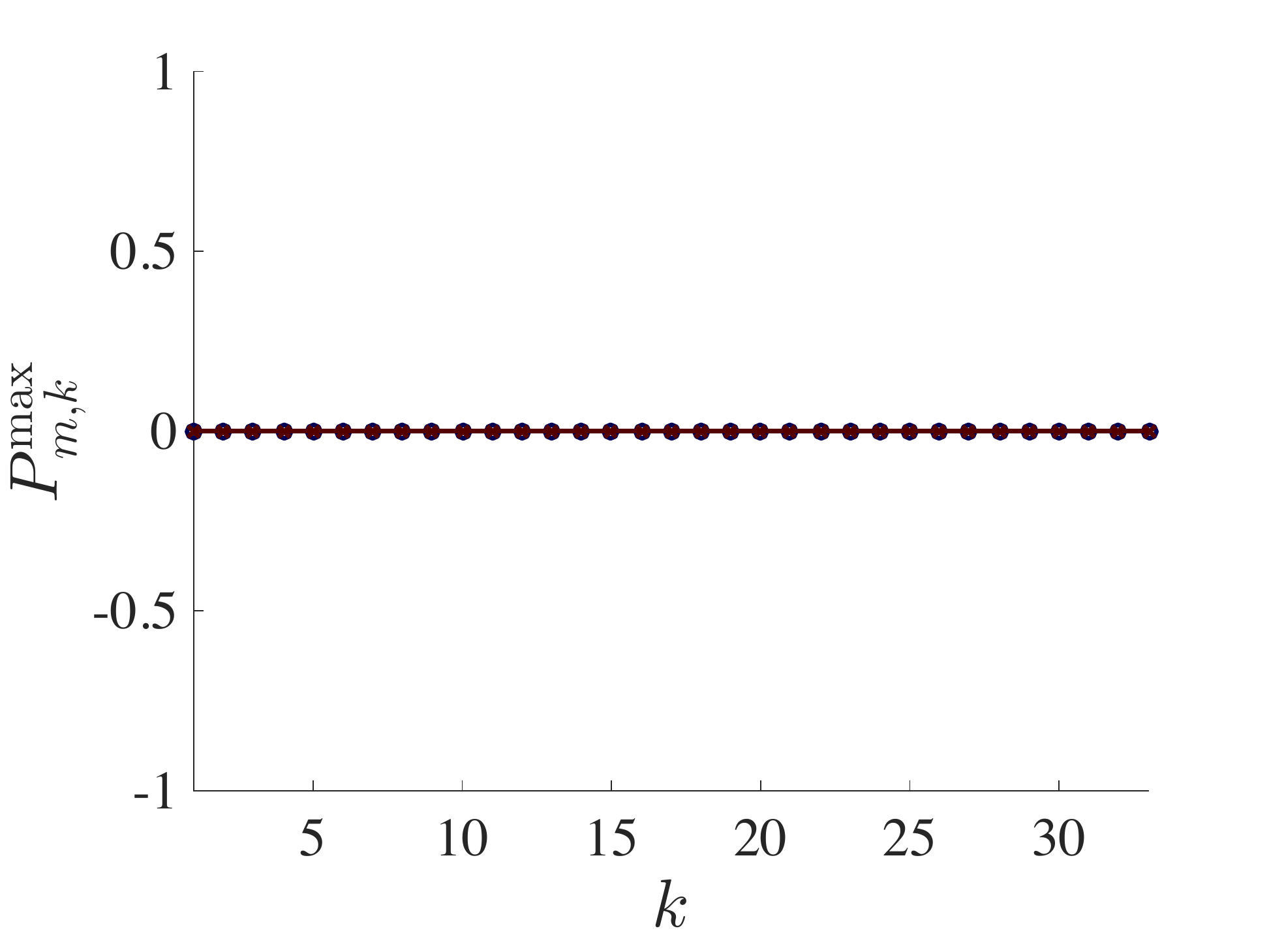}\label{fig:Pm_0dB}}
\hspace{0in}\subfloat[MS, $\gamma^{\mathrm{avg}}=10$dB]{\includegraphics[scale = 0.22]{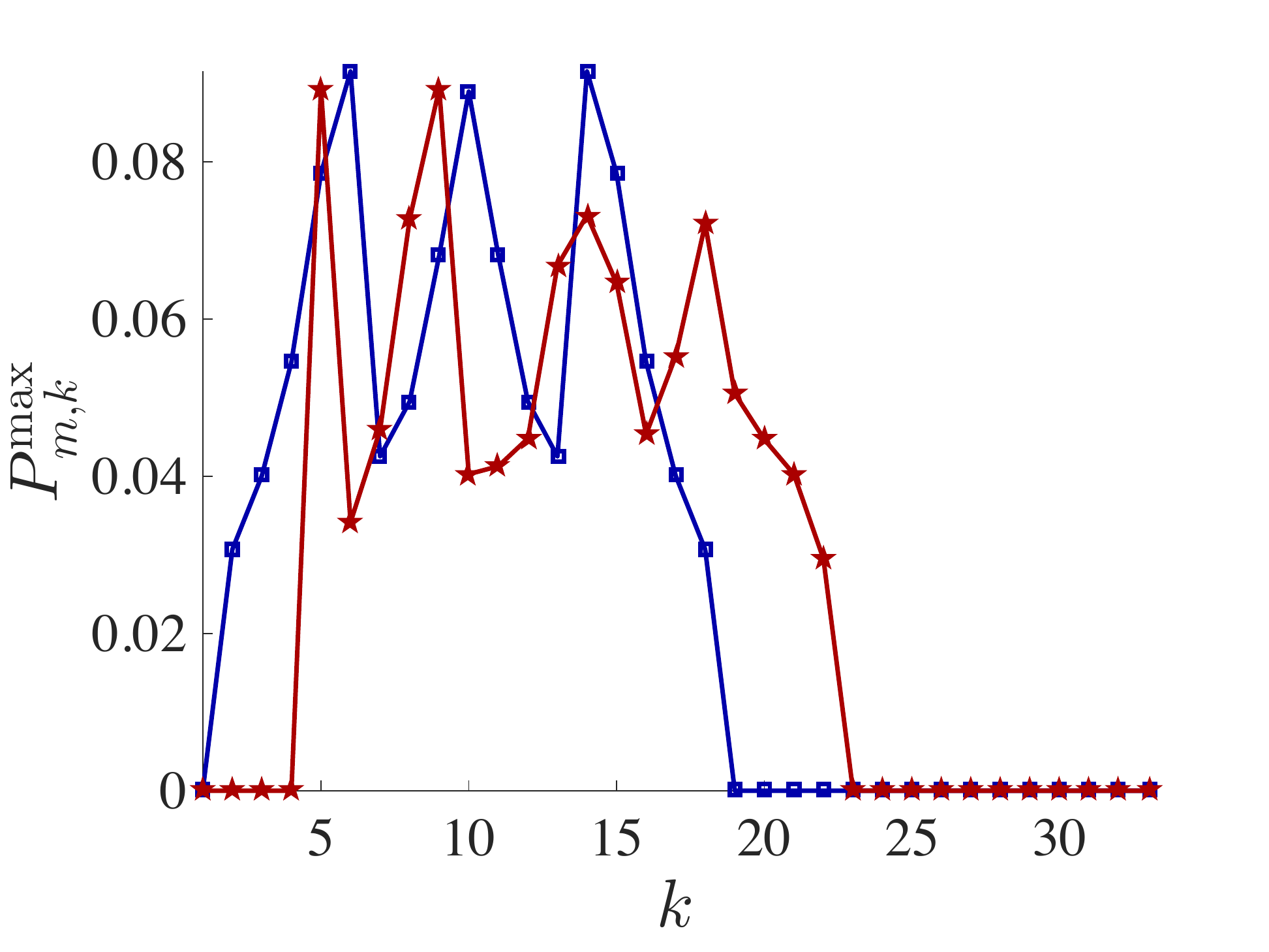}\label{fig:Pm_10dB}}
\hspace{0.05in}\subfloat[MS, $\gamma^{\mathrm{avg}}=20$dB]{\includegraphics[scale = 0.22]{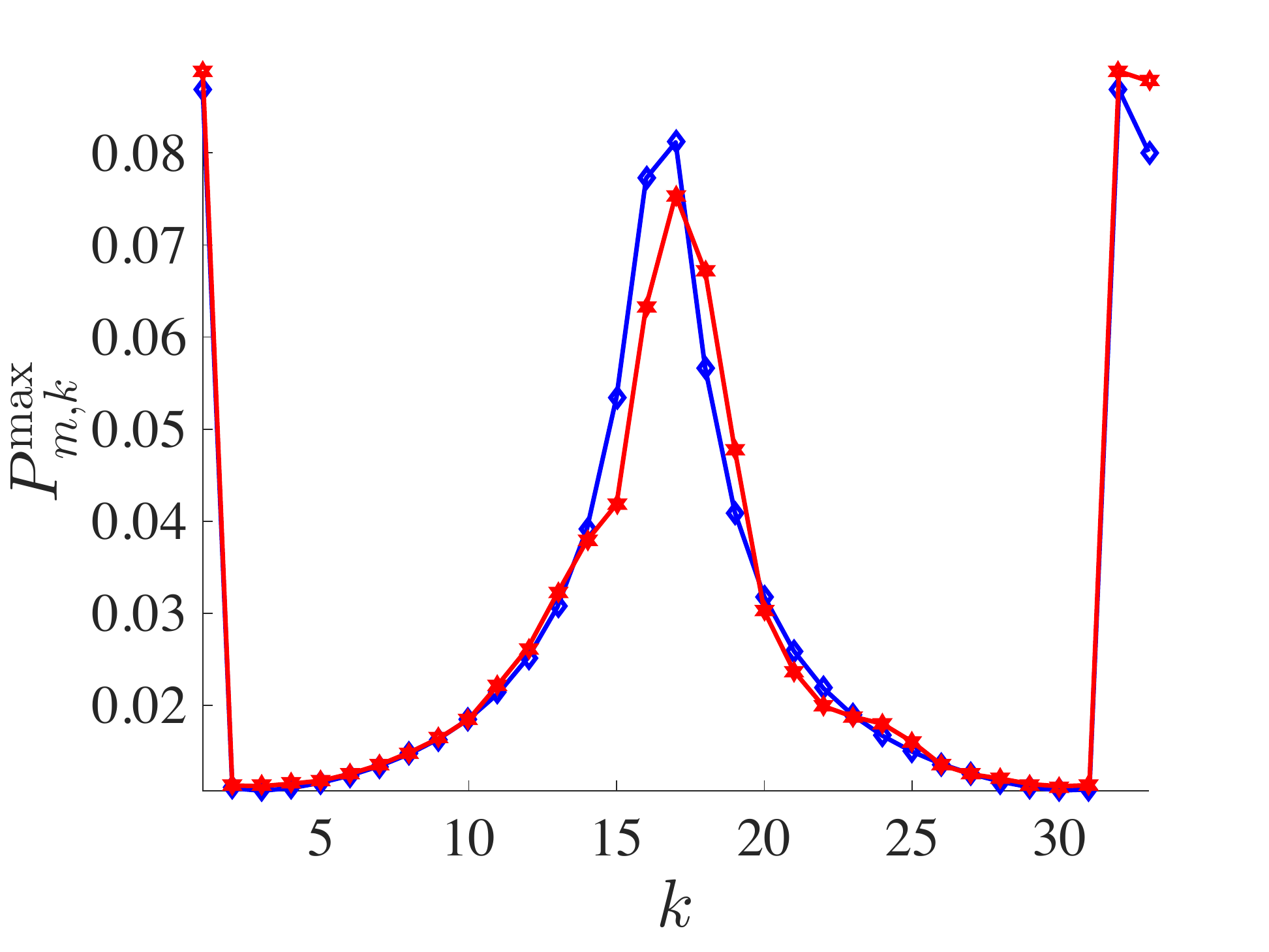}\label{fig:Pm_20dB}}
\hspace{.05in}\subfloat[MS, $\gamma^{\mathrm{avg}}\in\{30, 40, 50\}$dB]{\includegraphics[scale = 0.22]{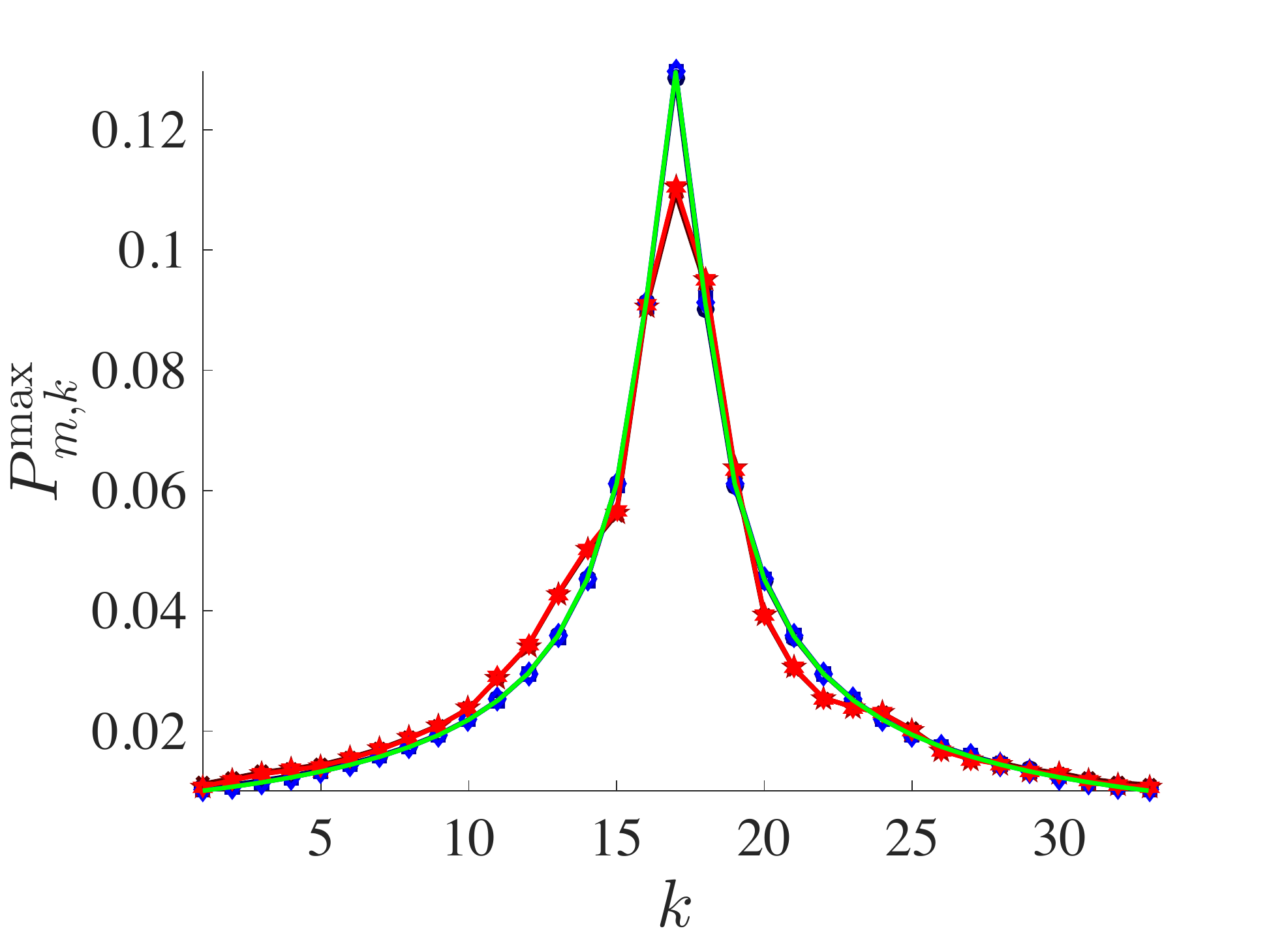}\label{fig:Pm_hsinr}}
     \caption{ 
     {Power allocation over $K=33$ channels (20MHz bandwidth) at the BS and MS for different values of average SNR ($\gamma^{\mathrm{avg}}$). The higher the $\gamma^{\mathrm{avg}}$, the more channels are used in full-duplex, and the closer the power allocation gets to the high SINR approximation one (computed by \textsc{HSINR-MaximumRate}}).
}\vspace{-10pt}
     \label{fig:multich-PA}
\end{figure*}

\section{{Measurement-based Numerical Evaluation}}\label{section:OFDM-num-results}

This section presents numerical evaluations for use case (iii). Numerical evaluations for use cases (i) and (ii) were already provided in Sections \ref{section:bidirectional} and \ref{section:two-unidirectionals}, respectively. 
We focus on the impact of a frequency-selective SIC profile in a small form factor hardware at the MS (Fig.~\ref{fig:SIC_Results}\subref{fig:SIC_BW_Circ}), and evaluate achievable rate gains from FD. 

\begin{figure*}
\centering
\subfloat[$K=33$]{\includegraphics[scale = 0.22]{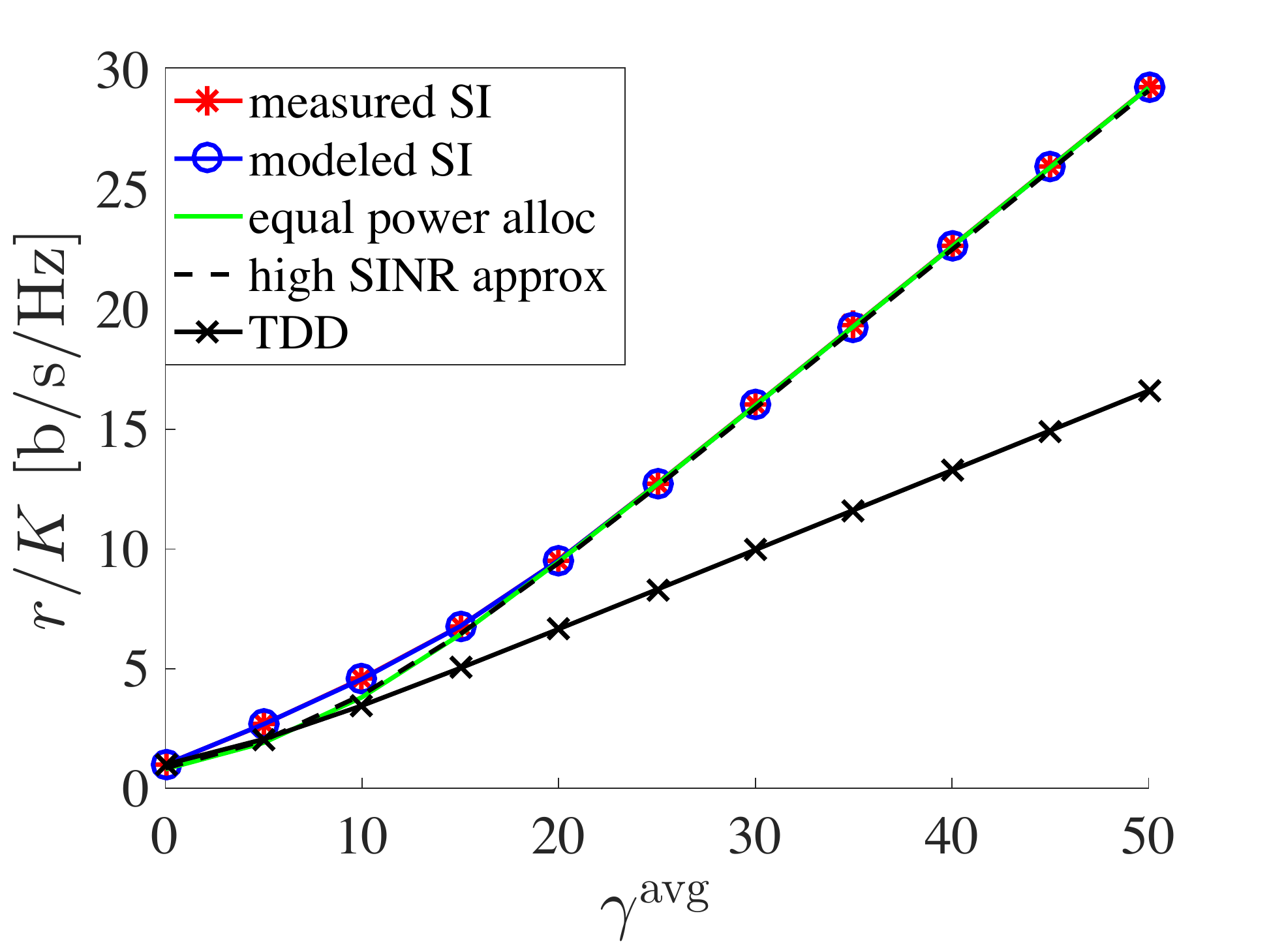}\label{fig:rate_multich}}\hspace{\fill}
\subfloat[$K=33$]{\includegraphics[scale = 0.22]{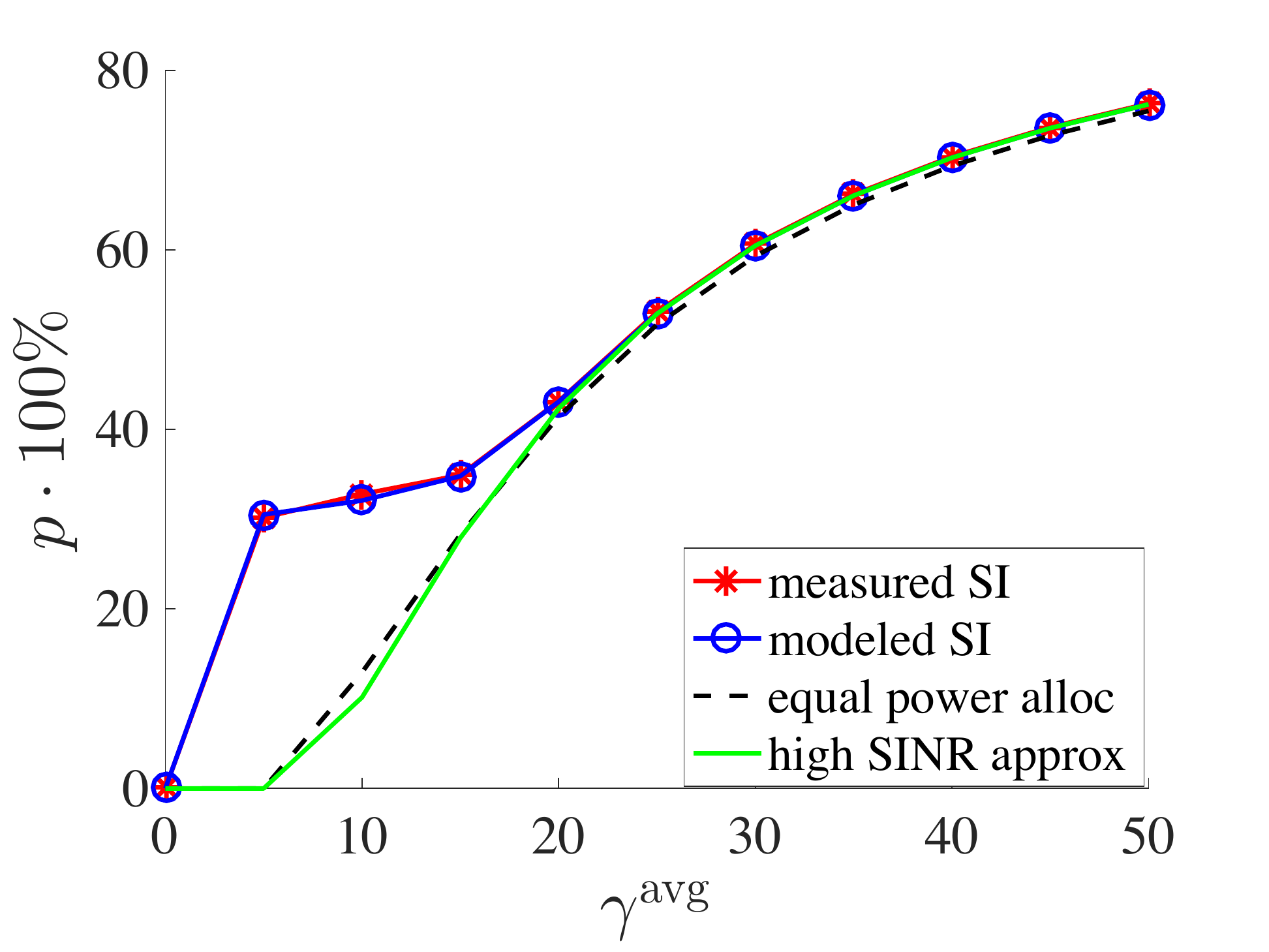}\label{fig:extension_multich_K_33}}\hspace{\fill}
\subfloat[$K=17$]{\includegraphics[scale = 0.22]{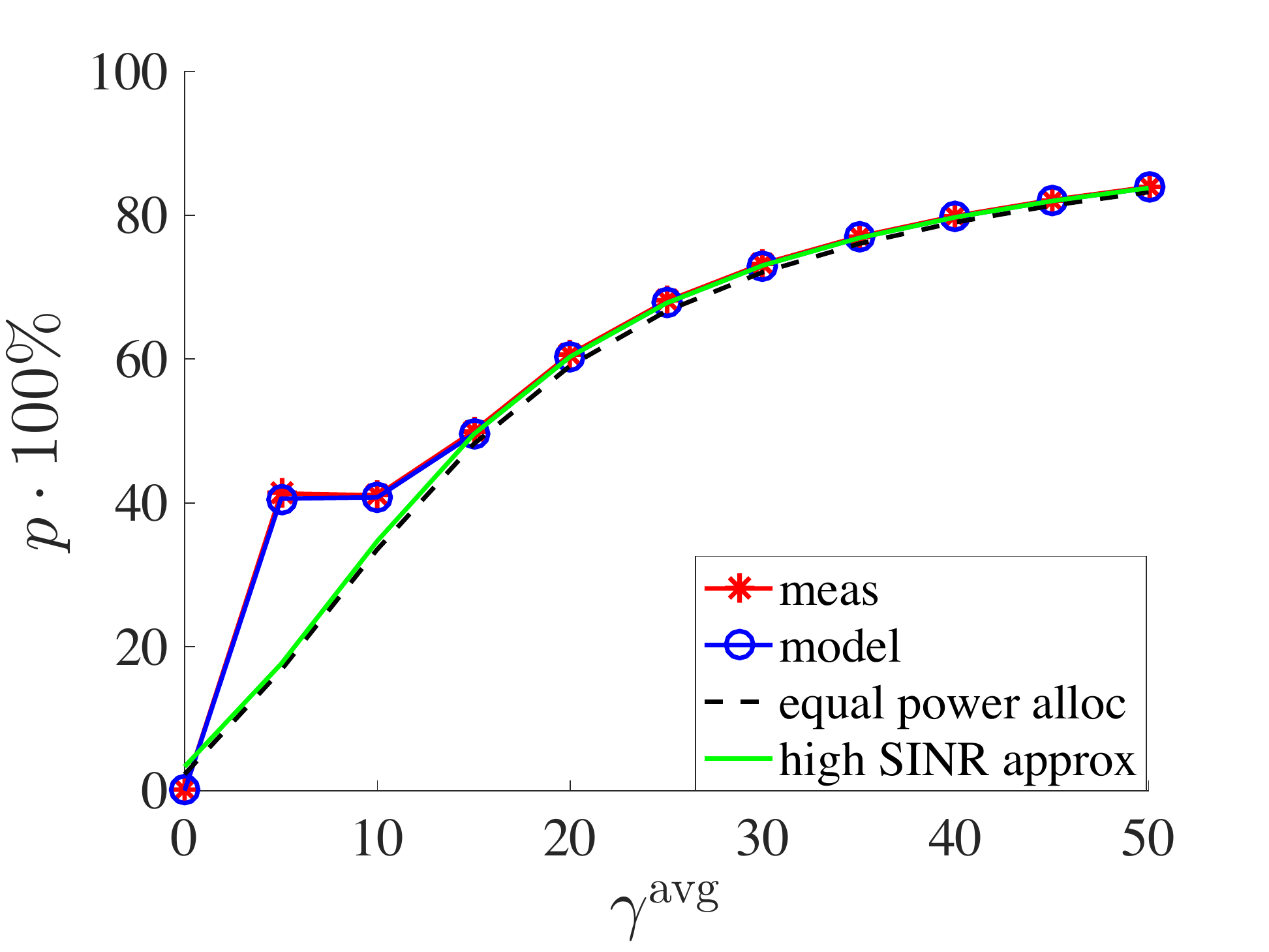}\label{fig:extension_multich_K_17}}\hspace{\fill}
\subfloat[$K=9$]{\includegraphics[scale = 0.22]{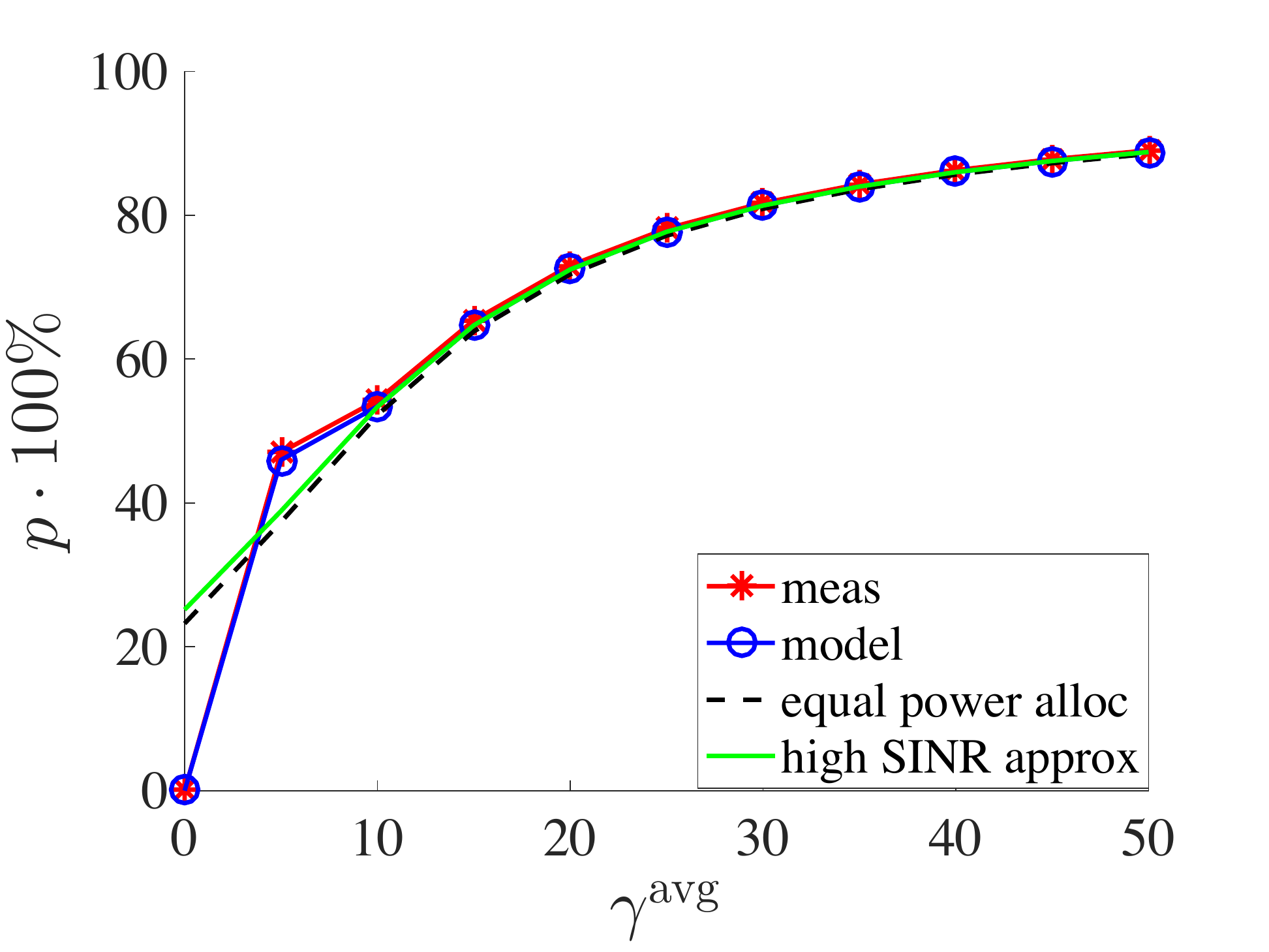}\label{fig:extension_multich_K_9}}
\caption{{Evaluated \protect\subref{fig:rate_multich} sum rate for $K=33$, normalized to the number of channels 
$K$, and \protect\subref{fig:extension_multich_K_33}--\protect\subref{fig:extension_multich_K_9} capacity region extension for \protect\subref{fig:extension_multich_K_33} $K=33$, \protect\subref{fig:extension_multich_K_17} $K=17$, and \protect\subref{fig:extension_multich_K_9} $K=9$. The graphs suggest that higher average SNR ($\gamma^{\mathrm{avg}}$) and better cancellation (lower bandwidth -- fewer frequency channels $K$) lead to higher rate gains.   }
}\vspace{-10pt}
\label{fig:rate-and-extension}
\end{figure*}
\begin{figure}
\centering
\subfloat[$K=33$]{\includegraphics[scale = 0.22]{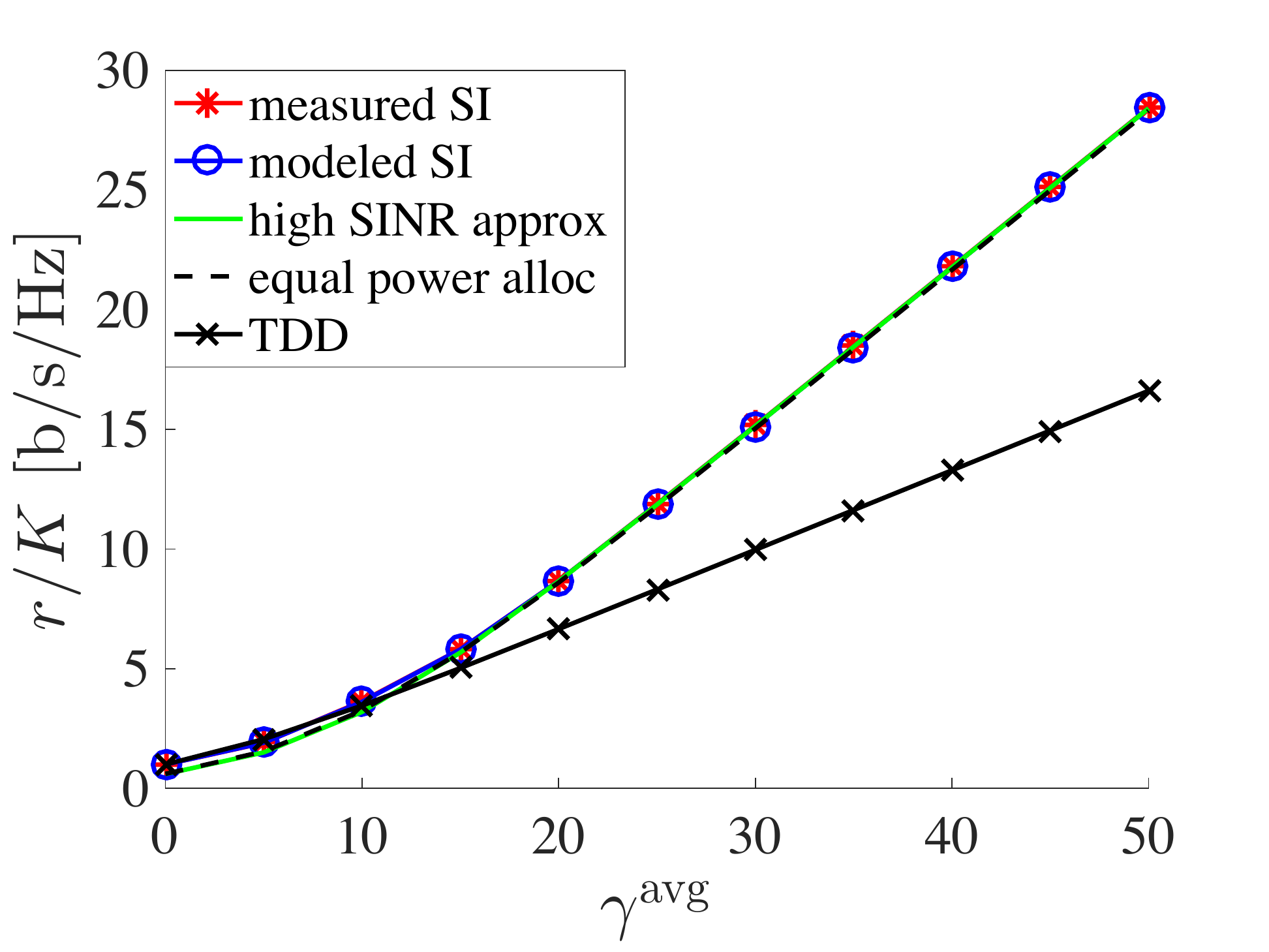}\label{fig:rate_scaled_33}}\hspace{\fill}
\subfloat[$K=\{9, 17, 33\}$]{\includegraphics[scale = 0.22]{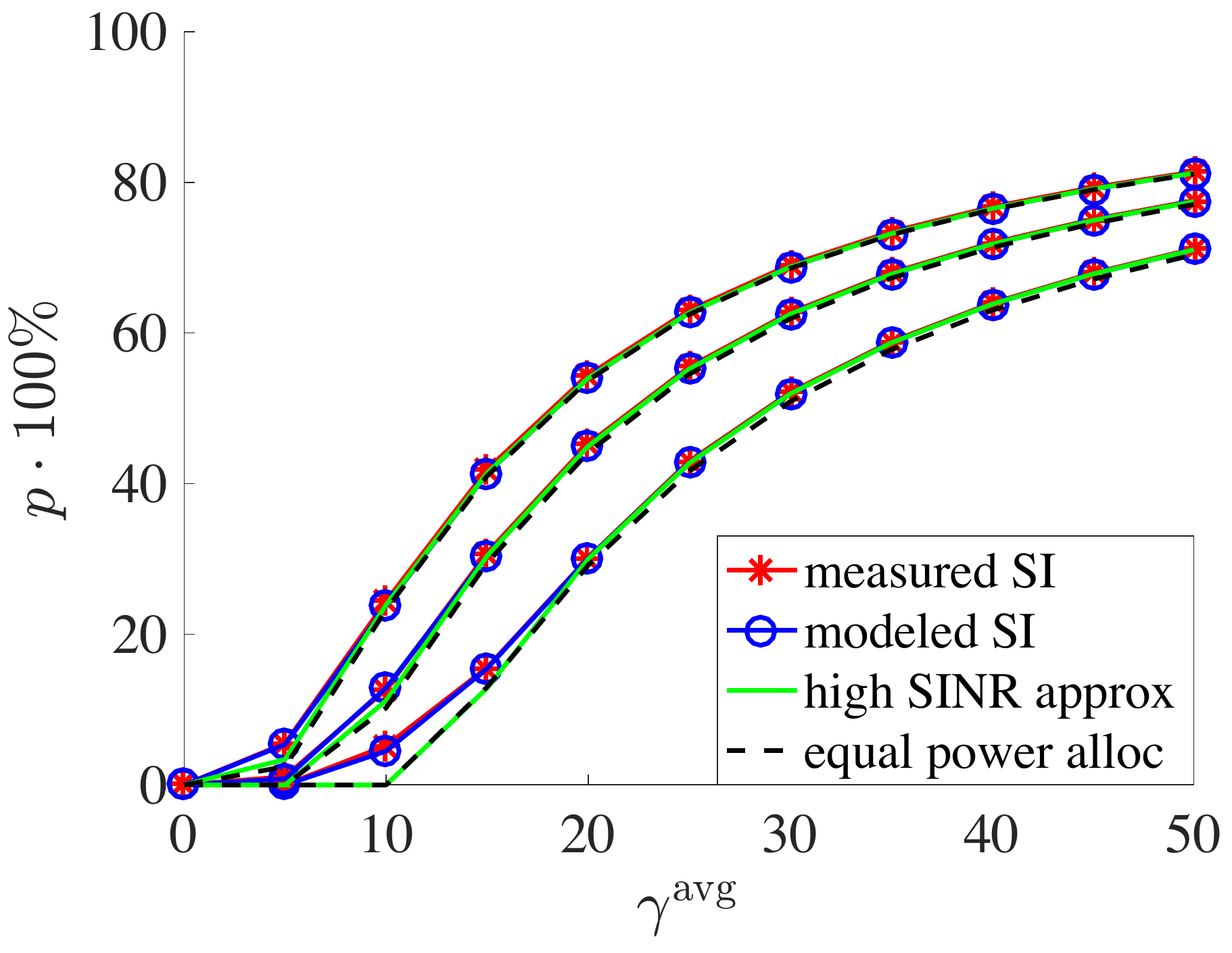}\label{fig:p_scaled_combined}}
\caption{{Evaluated \protect\subref{fig:rate_scaled_33} sum rate for $K=33$, normalized to 
$K$, and \protect\subref{fig:p_scaled_combined} capacity region extension for $K\in\{9, 17, 33\}$, for the sum of the total transmission power levels at the MS and at the BS scaled so that it is the same as in the TDD case.   }
}\vspace{-20pt}
\label{fig:rate-and-extension-scaled}
\end{figure}

\noindent\textbf{Evaluation Setup.} To determine the position $c^{\max}$ of maximum SIC and the power allocation $\{P_{m, k}^{\max}, P_{b, k}^{\max}\}$ that maximize the sum rate, we run an implementation of the \textsc{MaximumRate} algorithm separately for measured (~\cite{ZhouActiveTXCancISSCC14, Zhou_NCSIC_JSSC14} and Fig.~\ref{fig:SIC_Results}\subref{fig:SIC_BW_Circ}) and modeled (Eq. (\ref{eq:RSI-MS})) SIC profiles of the MS FD receiver. Additionally, we determine $c^{\max}, \{P_{m, k}^{\max}, P_{b, k}^{\max}\}$ for the high SINR approximation of the sum rate using the \textsc{HSINR-MaximumRate} algorithm. {We also compare the results to the case when the total transmission power is allocated equally among the frequency channels (we refer to this case as equal power allocation).}

Since the measurements were performed only for the analog part of the FD receiver, we assume additional 50dB of cancellation from the digital domain.\footnote{Fig.~\ref{fig:SIC_Results}\protect\subref{fig:SIC_BW_Circ} only shows isolation from the SIC in the analog domain.} Similar to \cite{bharadia2013full}, we assume that when either station transmits at maximum total power that is equally allocated across channels (so that $P_{m, k} = {\xoverline{P_m}}/{K}, P_{b, k} = {\xoverline{P_b}}/{K}$), the noise on each channel is 110dB below the transmitted power level.

We consider a total bandwidth of: (i) 20MHz in the range 2.13--2.15GHz, (ii) 10MHz in the range 2.135--2.145GHz, and (iii) 5MHz in the range 2.1375--2.1425GHz. We adopt the distance between the measurement points as the OFDM channel width ($\approx600$kHz), so that there are $K=33$, $K=17$, and $K=9$ channels, respectively, in the considered bands.  
For the SIC at the BS, we take $g_b \xoverline{P_b}/K = N_b$ \cite{bharadia2013full}.

We scale all the power variables so that $\xoverline{P_m} = \xoverline{P_b} = 1$. 
We consider flat frequency fading (so that $h_{mb, k}$ and $h_{bm, k}$ are constant across channels $k$), and perform numerical evaluations for ${h_{mb, k}}\xoverline{P_m}/{N_b} = {h_{bm, k}}\xoverline{P_b}/{N_m}\equiv\gamma^{\text{avg}}\cdot K$, $\forall k$, where $\gamma^{\mathrm{avg}}\in\{0, 10, 20, 30, 40, 50\}$ [dB]. 

 We run \textsc{MaximumRate} for $\Delta c = 0.01$, which corresponds to an absolute error of up to $\epsilon\approx 0.2$ for $r$. We evaluate the sum rate and the capacity region extension using the measurement data for the remaining SI and $c^{\max}, \{P_{m, k}^{\max}, P_{b, k}^{\max}\}$ returned by the algorithm.  
We assume that the amount of SIC around $f_c$ does not change as $f_c$ (and correspondingly $c$) is varied. 
To run the algorithm for $c$ positioned at any point between two neighboring channels, we interpolate the measurement data. 

\noindent\textbf{Results. }
Due to space constraints, we provide detailed results for the power allocation only for the 20MHz bandwidth ($K=33$) case, in Fig.~\ref{fig:multich-PA}. For the {10MHz ($K=17$) and } 5MHz ($K=9$) {cases}, we only provide the results for the capacity region extension, in Fig.~\ref{fig:rate-and-extension}.

{Fig. \ref{fig:multich-PA} shows the power allocations at the BS (Fig.~ \ref{fig:multich-PA}\subref{fig:Pb_0dB}--\subref{fig:Pb_hsinr}) and at the MS (Fig.~ \ref{fig:multich-PA}\subref{fig:Pm_0dB}--\subref{fig:Pm_hsinr}) computed by \textsc{MaximumRate} for both measured and modeled SI and for different values of average SNR $\gamma^{\mathrm{avg}}$. Additionally, Figs.~\ref{fig:multich-PA}\subref{fig:Pb_hsinr} and \ref{fig:multich-PA}\subref{fig:Pm_hsinr} compare the power allocation computed by \textsc{MaximumRate} to the one computed by \textsc{HSINR-MaximumRate}. As Fig.~\ref{fig:multich-PA} suggests, when $\gamma^{\mathrm{avg}}$ is too low, most channels are used as half-duplex -- i.e., only one of the stations transmits on a channel. As $\gamma^{\mathrm{avg}}$ increases, the number of channels used as full-duplex increases: at $\gamma^{\mathrm{avg}}=10$dB about seven channels are used as full-duplex, while for $\gamma^{\mathrm{avg}}=20$dB all but two channels are used as full-duplex, and when $\gamma^{\mathrm{avg}}\geq 30$dB, we reach the high SINR approximation for the FD power allocation.}

{Fig.~\ref{fig:rate-and-extension} shows \subref{fig:rate_multich} sum rate normalized to the number of channels for $K=33$ (20MHz bandwidth) and \subref{fig:extension_multich_K_33}--\subref{fig:extension_multich_K_9} capacity region extension for $K=33$ (20MHz bandwidth), $K=17$ (10MHz bandwidth), and $K=9$ (5MHz bandwidth). As Fig.~\ref{fig:rate-and-extension} suggests, the FD rate gains increase as $\gamma^{\mathrm{avg}}$ increases and the SIC becomes better across the channels (i.e., as we consider lower bandwidth -- lower $K$).}

{We observe in Fig.~\ref{fig:rate-and-extension}\subref{fig:extension_multich_K_33}--\subref{fig:extension_multich_K_9} that there is a ``jump'' in the capacity region extension as $\gamma^{\mathrm{avg}}$ increases from 0dB to 5dB. This happens because at $\gamma^{\mathrm{avg}}=0$dB Conditions \ref{eq:multi-channel-concavity-condition} and \ref{item:assumption-1} force all the power levels at the MS to zero, and we have the HD case where only the BS is transmitting. At $\gamma^{\mathrm{avg}}=5$dB Conditions \ref{eq:multi-channel-concavity-condition} and \ref{item:assumption-1} become less restrictive and some of the channels are used as FD. At the same time, the total irradiated power (considering both MS and BS) is doubled compared to the case when $\gamma^{\mathrm{avg}}=0$dB (and to the TDD operation), so a large portion of the rate improvement comes from this increase in the total irradiated power. To isolate the rate gains caused by FD operation from those caused by the increase in the total irradiated power, we normalize the total irradiated power so that it is the same as in the TDD regime and compute the sum rate for $K=33$ and the capacity region extension for $K=\{33, 17, 9\}$, as shown in Fig.~\ref{fig:rate-and-extension-scaled}. The results suggest that the rate gains that are solely due to FD operation increase smoothly with $\gamma^{\mathrm{avg}}$ and the rate gains are almost indistinguishable for different power allocation policies (\textsc{MaximumRate} for measured and modeled SI, \textsc{HSINR-MaximumRate}, and equal power allocation).}

Since for the transmitted power of $1/K$ and $c$ placed in the middle of the frequency band XINR at the first and the last channel is about 35 ($\approx15$dB) for $K=33$, about 8.5 ($\approx 9$dB) for $K=17$, and about 2.5 ($\approx 4$dB) for $K=9$, our numerical results suggest, as expected (see e.g., Figs.~\ref{fig:cap-region} and Fig.~\ref{fig:p-in-terms-of-snrs}), that to achieve high rate gain{s}, $\gamma^{\text{avg}}$ needs to be sufficiently high. This is demonstrated by the results shown in Fig.~\ref{fig:rate-and-extension} {and \ref{fig:rate-and-extension-scaled}}. {In particular, the rate gains obtained solely from FD operation are non-negligible when on most channels $\mathrm{XINR}\geq 0$dB. Moreover, simple power allocation policies, such as equal power allocation and high SINR approximation power allocation are near-optimal when the rate gains are non-negligible, as demonstrated by Fig.~\ref{fig:rate-and-extension-scaled}.}

\section{Conclusion and Future Work}\label{section:conclusion}

In this paper we considered three basic use cases of FD, including single- and multi-channel scenarios. In order to analyze the multi-channel scenario, we developed a new model that is grounded in realistic FD receiver implementations for small form factor devices. We characterized the rate gains in different scenarios and solved power allocation and frequency selection problems either analytically or algorithmically. Our numerical results demonstrate the gains from FD in scenarios and for receiver models that have not been studied before. 

This is one of the first steps towards understanding the benefits and the complexities associated with FD. Hence, there are still many open problems to consider. In particular, {generalizing our results to the MIMO settings is of high relevance and interest. Additionally,} SIC that has different impacts on different channels calls for the design of algorithms for OFDM networks with multiple access and MSs modeled as small form-factor devices. Moreover, we plan to develop scheduling algorithms that support the co-existence of half- and full-duplex users. While significant attention has been given to scheduling and resource allocation in half duplex OFDM networks (see, e.g., \cite{HSAB2009} and references therein), as demonstrated in this paper, the special characteristics of FD pose new challenges that have not been addressed.

\section*{Acknowledgements}
This research was supported in part by the NSF grant ECCS-1547406, DARPA RF-FPGA program HR0011-12-1-0006, Qualcomm Innovation Fellowship, and the People Programme (Marie Curie Actions) of the European Union's Seventh Framework Programme (FP7/2007-2013) under REA grant agreement n${^{\text{o}}} $[PIIF-GA-2013-629740].11.

\bibliographystyle{IEEETran}
{
\bibliography{references_FD}}

\appendix

\begin{IEEEproof}[Proof of {Lemma \ref{lemma:ci-maxima-localization} }]
Since $r = \sum_{k=1}^K r_k$, we will first observe partial derivatives of $r_k$ with respect to $c$. 

Observe that in the expression (\ref{eq:bidirectional-SE}) for $r_k$ only $\gamma_{mm, k}$ depends on $c$. Moreover, since $\gamma_{mm, k} = \frac{g_m (k-c)^2 P_{m, k}}{N_m}$, we have that $(k-c)\frac{\partial \gamma_{mm, k}}{\partial c} = -2 \gamma_{mm, k}$.  

Observe partial derivatives of $r_k$ with respect to $c$:
\begin{myalign}
\frac{\partial r_{k}}{\partial c} &= \frac{2}{\ln2}\cdot\frac{g_m P_{m, k}}{N_m}\cdot\gamma_{mb,k}\notag\\ &\cdot\frac{k-c}{\big(1 + \gamma_{mb, k} + \gamma_{mm, k}(c)\big)\big(1+\gamma_{mm, k}(c)\big)},\label{eq:first-derivative-ci}\\
\frac{\partial^2 r_{k}}{\partial {c}^2} &= \frac{2}{\ln2}\cdot\frac{g_m P_{m, k}}{N_m}\cdot\gamma_{mb,k} \notag \\ &\cdot\frac{\gamma_{mm, k}(c)\big(2 + \gamma_{mb, k} + 3\gamma_{mm, k}(c)\big) - \big(1 + \gamma_{mb, k}\big)}{\big(1 + \gamma_{mb, k} + \gamma_{mm, k}(c)\big)^2\big(1+\gamma_{mm, k}(c)\big)^2}.\label{eq:second-derivative-ci}
\end{myalign}

From (\ref{eq:first-derivative-ci}), $\frac{\partial r_k}{\partial c}$ equals zero for $c = k$, it is positive for $c<k$ and negative for $c>k$. Therefore, $r_k$ is a has a unique maximum in $c$, with the maximum attained at $k = c$. Since this is true for every $k \in \{1,..., K\}$, it follows that for $c\leq 1$ $\forall k\in\{1,..., K\}$: $\frac{\partial r_k}{\partial c}\geq 0$ (with equality only for $k = c$), and therefore $\frac{\partial r}{\partial c}>0$. Similarly, $\frac{\partial r}{\partial c}<0$ for $c\geq K$. Therefore, all (local) maxima of $r(c)$ must lie in the interval $(1, K)$.

As $\gamma_{mm, k} = \frac{g_m (k-c)^2 P_{m, k}}{N_m}$, $r_k$ is symmetric around $c = k$. From (\ref{eq:second-derivative-ci}), $\frac{\partial^2 r_k}{\partial {c}^2}$ is negative for $k - c = 0$, and there exits a unique $c_0$ at which $\frac{\partial^2 r_{k}}{\partial {c}^2} = 0$ (this part can be shown by solving $\gamma_{mm, k}(c)\big(2 + \gamma_{mb, k} + 3\gamma_{mm, k}(c)\big) - \big(1 + \gamma_{mb, k}\big) = 0$, which is a quadratic equation in terms of $(k-c)^2$ with a unique zero; see the proof of Lemma \ref{lemma:bounded-derivative-ci}). For $|k-c|>|k-c_0|$, $\frac{\partial^2 r_k}{\partial {c}^2}$ is positive. This is true, e.g., for $\gamma_{mm, k}(c) \geq 1$.  

Visually, each $r_k$ as a function of $c$ is a symmetric bell-shaped curve centered at $k$. Therefore, $r$ can be seen as a sum of shifted and equally spaced symmetric bell-shaped curves. This sum, in general, can have linear in $K$ number of local maxima. Examples with $K$ local maxima can be constructed by choosing sufficiently large $\frac{g_m P_{m, k}}{N_m}$ (sufficiently ``narrow" bell-shaped curves). 
\end{IEEEproof}

\begin{IEEEproof}[of {Lemma \ref{lemma:bounded-derivative-ci}}]
Assume that $\gamma_{mm, k}>0$ and $\gamma_{mb, k} >0$ $\forall k\in\{1,..., K\}$, as otherwise $\left|\frac{\partial r_k}{\partial c}\right| = 0$ and can be ignored.

\noindent\textbf{Case 1.} Assume first that $c = k^*$ for some $k^*\in \{1,..., K\}$. Then, using (\ref{eq:first-derivative-i}), $\frac{\partial r_k^*}{\partial c} = 0$, and for every $k\neq k^*$:
\begin{myalign}
\left|\frac{\partial r_k}{\partial c}\right| &\leq \frac{2}{\ln2}\gamma_{mb, k}\frac{g_mP_m}{N_m}\frac{|k-c|}{(1 + \gamma_{mm, k}(c))(1 + \gamma_{mb, k})}\notag\\
&\leq \frac{2}{\ln2}\frac{\frac{g_mP_m}{N_m}|k-c|}{1 + \frac{g_mP_m}{N_m}(k-c)^2}\cdot \frac{\gamma_{mb, k}}{1 + \gamma_{mb, k}}
\leq \frac{2}{\ln2} \frac{1}{|k-c|},\notag
\end{myalign}
since $k-c \geq 1$. Observe that since $c = k^* \in \{1,..., K\}$, every $c-k$ is a positive integer. Therefore:
\begin{myalign}
\left|\frac{\partial r}{\partial c}\right| &= \left|\sum_{k=1}^K \frac{\partial r_k}{\partial c}\right|
\leq \frac{2}{\ln2}\left|-\sum_{j=1}^{k^*-1}\frac{1}{|j-k^*|} + \sum_{k=k^* + 1}^{K}\frac{1}{|k-k^*|}\right|\notag\\
&\leq \frac{2}{\ln2}\sum_{k=1}^{K-1}\frac{1}{k} = \frac{2}{\ln2}H_{K-1}\notag,
\end{myalign}
where $H_{K-1}$ is the $(K-1)^{\text{th}}$ harmonic number. Using the known inequality $H_n < \ln(n) + 0.58 + \frac{1}{2n}$ for $n\in\mathbb{N}$ \cite{young199175} and assuming $K\geq 4$, we get: 
$\left|\frac{\partial r}{\partial c}\right| < \frac{2}{\ln2}(\ln(K)+1)$. 
For $K<4$, by inspection: ${\sum_{k=1}^{K-1}\frac{1}{k}< \ln(K)+1}$.

\noindent\textbf{Case 2.} Assume that $c\notin\{1, ..., K\}$, and observe that for $|k-c| \geq 1$: $\left|\frac{\partial r_k}{\partial c}\right|\leq \frac{2}{\ln2}\frac{1}{|k-c|}\leq \frac{2}{\ln2}\frac{1}{\lfloor|k-c|\rfloor}$. 

There can be at most two $k$'s with $|k-c|<1$. For such $k$, we bound $\left|\frac{\partial r_k}{\partial c}\right|$ as follows. First, observe from (\ref{eq:first-derivative-ci}) and (\ref{eq:second-derivative-ci}) that $\frac{\partial}{\partial |k-c|}\left|\frac{\partial r_k}{\partial c}\right| = - \frac{\partial^2 r_k}{\partial c^2}$. From (\ref{eq:second-derivative-ci}), $\frac{\partial^2 r_k}{\partial c^2} = 0$ if and only if for some $c_0$:
\begin{myalign}
&\gamma_{mm, k}(c_0)\big(2 + \gamma_{mb, k} + 3\gamma_{mm, k}(c_0)\big) - \big(1 + \gamma_{mb, k}\big) = 0\notag\\
\Leftrightarrow &\gamma_{mm, k}(c_0) = \frac{(2 + \gamma_{mb, k}) + \sqrt{(2 + \gamma_{mb, k})^2 + 12(1 +\gamma_{mb_k})}}{6}.\notag
\end{myalign}
Note we have used that $\gamma_{mm, k}>0$ to get a unique solution for $\gamma_{mm, k}$. 
Since $\gamma_{mm, k}(c_0) = \frac{g_mP_{m,k}}{N_m}(k-c_0)^2$:
\begin{myalign}
(k-c_0)^2 &= \frac{N_m}{g_mP_{m, k}}\gamma_{mm, k}(c_0)\notag\\
&> \frac{N_m}{g_mP_{m, k}} \frac{2\cdot (2 + \gamma_{mb, k})}{6}
> \frac{N_m}{g_mP_{m, k}} \frac{\gamma_{mb, k}}{3}.\notag
\end{myalign}
From condition \ref{item:assumption-1} we have that $\frac{N_m}{g_mP_{m, k}}\cdot\gamma_{mb, k} \geq 1$, which gives $|k-c_0|> \frac{1}{\sqrt{3}}$. It is clear from (\ref{eq:first-derivative-i}) and $\gamma_{mm, k} = \frac{g_mP_{m, k}}{N_m}(k-c)^2$ that $\frac{\partial^2 r_k}{\partial c^2}$ is negative for $|k-c|<|k-c_0|$ and positive for $|k-c|>|k-c_0|$. Since $\frac{\partial}{\partial |k-c|}\left|\frac{\partial r_k}{\partial c}\right| = - \frac{\partial^2 r_k}{\partial c^2}$, it follows directly that $\left|\frac{\partial r_k}{\partial c}\right|$ is maximized at $c = c_0$. Therefore, for $|k-c|<1$, we have that $\left|\frac{\partial r_k}{\partial c}\right|<\frac{2}{\ln2}\frac{1}{|k-c_0|}<\frac{2}{\ln2}\sqrt{3}$.

Combining the results for $|k-c|\geq 1$ and $|k-c|<1$:
\begin{myalign}
\left|\frac{\partial r}{\partial c}\right| &\leq \sum_{k=1}^K \Big|\frac{\partial r_k}{\partial c}\Big|
\leq \frac{2}{\ln2}\Big(\Big|-\sum_{j=1}^{\lfloor c\rfloor -1}\frac{1}{|j-c|} + \sum_{k=\lceil c\rceil + 1}^{K}\frac{1}{|k-c|}\Big| + {2}{\sqrt{3}}\Big)\notag\notag\\
&\leq \frac{2}{\ln2}\Big(\sum_{k=1}^{K-1}\frac{1}{k}+ {2}{\sqrt{3}}\Big)
< \frac{2}{\ln2}(\ln(K)+1+2\sqrt{3}). \notag
\end{myalign}
\end{IEEEproof}

\begin{IEEEproof}[of {Lemma \ref{lemma:high-sinr-ci-at-middle}}]
From Lemma \ref{lemma:high-sinr-pik}, $P_{m, k}^* = \alpha_k P_{m}$, where $P_{m} = \xoverline{P_m}$, and recalling that $R_{k} = g_m \xoverline{P_m} (k-c)^2$:
\begin{itemize}
\item $\alpha_k =
\frac{\alpha_K \cdot \left(N_m + \alpha_K g_m P_{m} (K-c)^2 \right)}{N_m}$ if $k = c$;
\item $\alpha_k =\frac{-N_m + \sqrt{{N_m}^2 + 4 \alpha_K(N_m + \alpha_K g_m P_{m} (K-c)^2)g_m P_m (k-c)^2}}{2g_m P_{m} (k-c)^2}$ if $k \neq c$;
\end{itemize}
and $\alpha_K>0$ is chosen so that $\sum_{k=1}^K \alpha_k = 1$. To simplify the notation, we will let $\gamma_{mm} = \frac{g_m P_m}{N_m}$, and write $\alpha_k$ as:
\begin{myalign}
\alpha_k = \begin{cases}
{\alpha_K \cdot \left(1 + \alpha_K \gamma_{mm} (K-c)^2 \right)}, & \text{if } k = c,\\
\frac{-1 + \sqrt{1 + 4 \alpha_K(1+ \alpha_K \gamma_{mm}(K-c)^2)\gamma_{mm}(k-c)^2}}{2\gamma_{mm} (k-c)^2}, & \text{if } k \neq c
\end{cases}.\label{eq:alpha-k}
\end{myalign}
\begin{figure}[t]
     \centering\hspace{\fill}
     \subfloat[t][$c - 5 < \frac{1}{2}$]{\includegraphics[scale = 0.26]{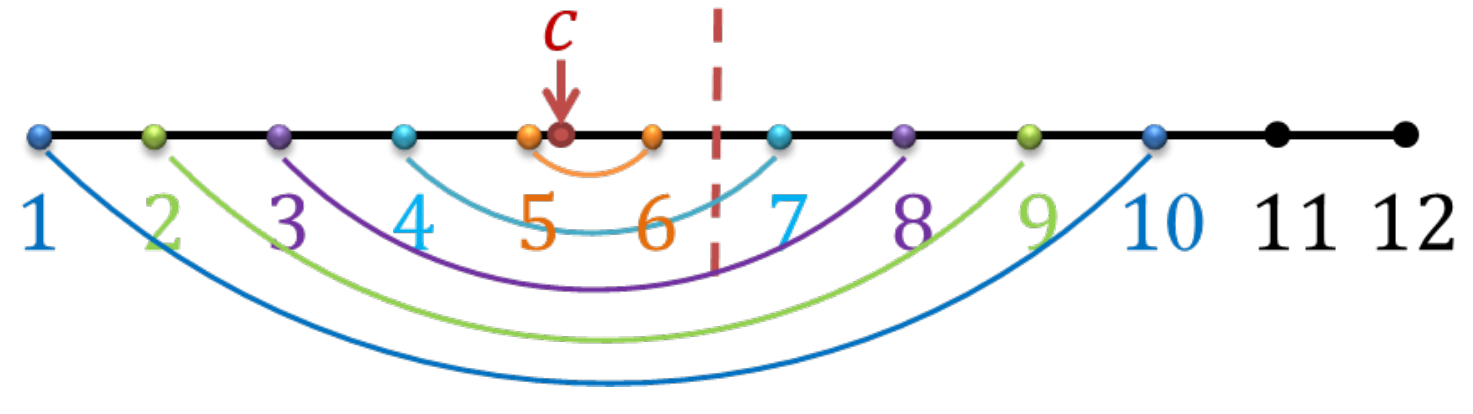}\label{fig:ci-positioning-left}}\hspace{\fill}
     \subfloat[t][$c - 5 > \frac{1}{2}$]{\includegraphics[scale = 0.26]{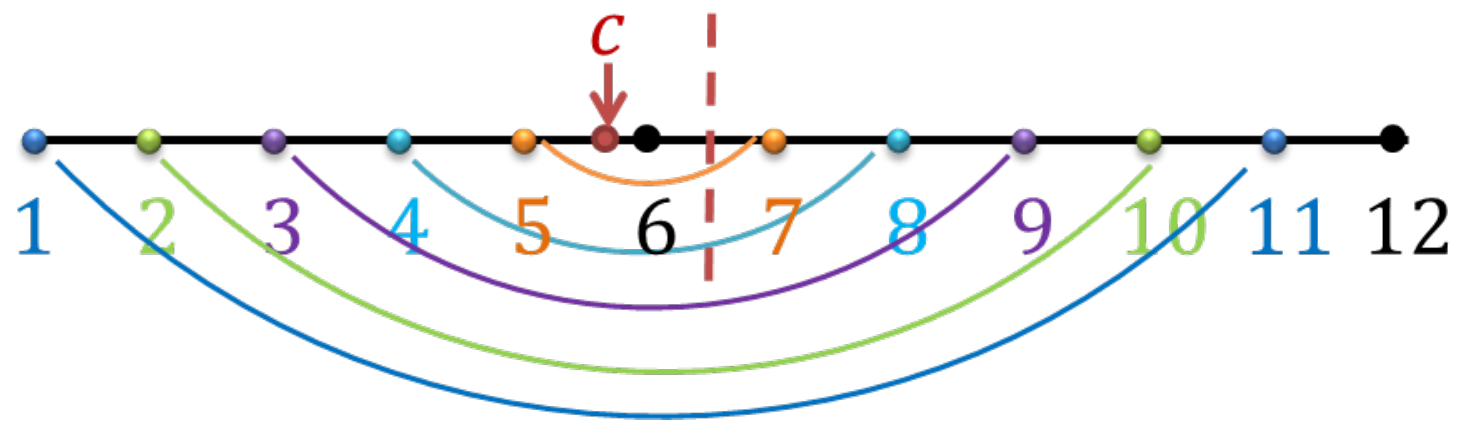}\label{fig:ci-positioning-right}}
     \hspace*{\fill}
     \caption{Pairing of points that are left and right from $c$ for \protect\subref{fig:ci-positioning-left} $c\in (5, 5.5)$ and \protect\subref{fig:ci-positioning-right} $c\in (5.5, 6)$.}
     \label{fig:pairings}\vspace{-10pt}
\end{figure}
Notice that for $c = 1 + l\cdot \frac{1}{2}$, $l \in \{1, 2, ..., 2K -3\}$, the power allocation is symmetric around $c$, that is : $\alpha_{\lfloor \frac{c}{2}\rfloor} = \alpha_{\lceil \frac{c}{2} \rceil}$, $\alpha_{\lfloor \frac{c}{2}\rfloor - 1} = \alpha_{\lceil \frac{c}{2} \rceil + 1}$, etc.

The first partial derivative of $r$ with respect to $c$ is:
\begin{myalign}
\frac{\partial r}{\partial c} 
&= \sum_{k=1}^K \frac{\partial}{\partial c} \left(\log\left( \frac{1}{N_m (1 + \frac{g_m P_m}{N_m} \alpha_k(k-c)^2)}\right)\right)\notag\\
&= \sum_{k=1}^K \frac{\partial}{\partial c} \left(\log\left( \frac{1}{1 + \gamma_{mm} \alpha_k(k-c)^2}\right)\right)\notag\\
&= \sum_{k=1}^K \frac{2 \gamma_{mm} \alpha_k (k-c)}{1 + \gamma_{mm} \alpha_k(k-c)^2}\label{eq:high-sinr-ci-deriv}
\end{myalign}
Observe that given the optimal power allocation (\ref{eq:alpha-k}):
\begin{itemize}
\item If $c = \frac{K+1}{2}$, then (from (\ref{eq:alpha-k})) $\alpha_1 = \alpha_K$, $\alpha_2 = \alpha_{K-1}$,..., $\alpha_{\lfloor\frac{K+1}{2}\rfloor} = \alpha_{\lceil\frac{K+1}{2}\rceil}$, and it follows that $\frac{\partial r}{\partial c} = 0$.

\item If $c = 1 + l\cdot \frac{1}{2}$, for $l\in\{0, 1, ..., K - 2\}$, then, as $\{\alpha_k\}$
is symmetric around $c$: $\frac{\partial r}{\partial c} = \sum_{i=1}^{\lfloor c \rfloor} \frac{2 \gamma_{mm} \alpha_i (i-c)}{1 + \gamma_{mm} \alpha_i(i-c)^2} + \sum_{j=\lfloor c \rfloor + 1}^{2 \lfloor c \rfloor} \frac{2 \gamma_{mm} \alpha_j (j-c)}{1 + \gamma_{mm} \alpha_j(j-c)^2} + \sum_{k=2\lfloor c \rfloor + 1}^{K} \frac{2 \gamma_{mm} \alpha_k (k-c)}{1 + \gamma_{mm} \alpha_k(k-c)^2} = \sum_{k=2\lfloor c \rfloor + 1}^{K} \frac{2 \gamma_{mm} \alpha_k (k-c)}{1 + \gamma_{mm} \alpha_k(k-c)^2} > 0$.

\item If $c = 1 + l\cdot \frac{1}{2}$, for $l\in\{K, ..., K - 2\}$, then, as $\{\alpha_k\}$
is symmetric around $c$: $\frac{\partial r}{\partial c} = \sum_{i=1}^{2 c - K - 1} \frac{2 \gamma_{mm} \alpha_i (i-c)}{1 + \gamma_{mm} \alpha_i(i-c)^2} + \sum_{j=2 c - K}^{ \lfloor c \rfloor} \frac{2 \gamma_{mm} \alpha_j (j-c)}{1 + \gamma_{mm}\alpha_j(j-c)^2} + \sum_{k=\lfloor c \rfloor + 1}^{K} \frac{2 \gamma_{mm} \alpha_k (k-c)}{1 + \gamma_{mm} \alpha_k(k-c)^2} = \sum_{i=1}^{2 c - K - 1} \frac{2 \gamma_{mm} \alpha_i (i-c)}{1 + \gamma_{mm} \alpha_i(i-c)^2} < 0$.
\end{itemize}
In other words, if we restrict our attention only to those $\{\alpha_k\}$ that determine the optimal power allocation, then considering $c$'s from the set $1 + l\cdot \frac{1}{2}$, where $l \in \{0, 1, ..., 2K - 2\}$, we get that the first derivative of $r$ with respect to $c$ is positive for $c < \frac{K+1}{2}$,l equal to zero for $c = \frac{K+1}{2}$, and negative for $c > \frac{K+1}{2}$. 
To conclude that at the global maximum for $r$ we have $c = \frac{K+1}{2}$ by considering $c\in(1, K)$ it remains to show that for $c \in (1 + l \cdot\frac{1}{2}, 1 + (l+1) \cdot\frac{1}{2})$, where $l \in \{0, 1, ..., 2K - 2\}$, we have that $\frac{\partial r}{\partial c}>0$ if $l \leq K-2$ and $\frac{\partial r}{\partial c}<0$ if $l \geq K-1$.

Fix any $l \in \{0, 1, ..., K - 2\}$ (on the left half of the interval $[1, K]$) and let $c \in (1 + l \cdot\frac{1}{2}, 1 + (l+1) \cdot\frac{1}{2})$. We make the following three claims:
\begin{enumerate}[label=(K1)]
\item \label{item:pairing} \emph{Each point $i \in\{1, 2, ..., \lfloor c \rfloor\}$ (left from $c$) can be paired to a point $j \in \{\lceil c \rceil, \lceil c \rceil + 1,..., K\}$ such that all the pairs are mutually disjoint and for each pair $(i ,j)$ we have that $ c - i < j - c$.} 
\end{enumerate}
\emph{Proof of (K1):} To construct the pairing, observe that, by the choice of $c$, $c$ is between two consecutive integer points and is strictly closer to one of them. If it is closer to the left point, then the pairing is $(\lfloor c\rfloor, \lceil c \rceil)$, $(\lfloor c\rfloor-1, \lceil c \rceil+1)$,..., $(1, 2 \lfloor c \rfloor)$. If $c$ is closer to the right point, then the pairing is $(\lfloor c\rfloor, \lceil c \rceil + 1)$, $(\lfloor c\rfloor-1, \lceil c \rceil+2)$,..., $(1, 2 \lfloor c \rfloor + 1)$.  Such pairings must exist as $c < \frac{K+1}{2}$. The pairings for $K=12$ and cases: $c \in (5, 5.5)$ and $c \in (5.5, 6)$ are illustrated in Fig.~\ref{fig:pairings}. Q.E.D.

\begin{enumerate}[label=(K2)]
\item \label{item:alphak-in-terms-of-delta}\emph{In the optimal power allocation that corresponds to a given $c$ and for any $i, j\in\{1,...,K\}$, if $|i-c| < |j - c|$, then $\alpha_i > \alpha_j$}.In other words, the smaller the distance between $k \in \{1,..., K\}$ and $c$, the larger the $\alpha_k$.
\end{enumerate}
\emph{Proof of (K2):} The proof has two parts. First, assume that $|i-c| = 0$ and observe $\alpha_j$ for $|j-c|>0$. From (\ref{eq:alpha-k}):
\begin{myalign}
\alpha_i = \alpha_K (1 + \alpha_K \gamma_{mm} (K-c)^2), \text{ and}\notag
\end{myalign}
\begin{myalign}
\alpha_j &= \frac{-1 + \sqrt{1 + 4 \alpha_K(1+ \alpha_K \gamma_{mm}(K-c)^2)\gamma_{mm}(j-c)^2}}{2\gamma_{mm} (j-c)^2}\notag\\
&= \frac{-1 + \sqrt{1 + 4 \alpha_i \gamma_{mm}(j-c)^2}}{2\gamma_{mm} (j-c)^2}.\notag
\end{myalign}
Using simple algebraic transformations:
\begin{myalign}
&\alpha_j < \alpha_i\notag\\
\Leftrightarrow & \frac{-1 + \sqrt{1 + 4 \alpha_i \gamma_{mm}(j-c)^2}}{2\gamma_{mm} (j-c)^2} < \alpha_i\notag\\
\Leftrightarrow & \sqrt{1 + 4 \alpha_i \gamma_{mm}(j-c)^2} < 1 + 2\alpha_i \gamma_{mm} (j-c)^2,\notag
\end{myalign}
we get that $\alpha_j < \alpha_i$ by squaring both sides of the last term, as $|j-c|>0$ implies $(2\alpha_i \gamma_{mm} (j-c)^2)^2>0$.

Second, assuming that $|k-c|>0$ and taking the first derivative of $\alpha_k$ with respect to $(k-c)^2$, we show that $\alpha_k$ decreases as $(k-c)^2$ (and consequently $|k-c|$) increases. Let $\Delta = (k-c)^2$. Then, as:
\begin{myalign}
&\frac{d \alpha_k}{d \Delta} =  \frac{d}{d \Delta} \left(\frac{-1}{2\gamma_{mm}\Delta} + \frac{\sqrt{1 + 4 \alpha_K(1+ \alpha_K \gamma_{mm}(K-c)^2)\gamma_{mm}\Delta}}{2\gamma_{mm} \Delta}\right)\notag\\
 &= \frac{1}{2\gamma_{mm}\Delta^2} - \frac{1 + 2 \alpha_K(1+ \alpha_K \gamma_{mm}(K-c)^2)\gamma_{mm}\Delta}{2\gamma_{mm}\Delta^2\sqrt{1 + 4 \alpha_K(1+ \alpha_K \gamma_{mm}(K-c)^2)\gamma_{mm}\Delta}}\notag,
\end{myalign}
it follows that 
$\frac{d \alpha_k}{d \Delta} < 0$, 
since \\
$
\sqrt{1 + 4 \alpha_K(1+ \alpha_K \gamma_{mm}(K-c)^2)\gamma_{mm}\Delta} < \\1 + 2 \alpha_K(1+ \alpha_K \gamma_{mm}(K-c)^2)\gamma_{mm}\Delta
$
. Q.E.D.

\begin{enumerate}[label=(K3)]
\item \label{item:dr-increases-as-alpha-decreases} \emph{As $|k-c|$ increases, $\left|\frac{\partial r_{k}}{\partial c}\right| = \frac{2\gamma_{mm}\alpha_k |k-c|}{1 + \gamma_{mm}\alpha_k (k-c)^2}$ decreases.}
\end{enumerate}
\emph{Proof of (K3):} Observe that:
\begin{myalign}
\frac{\partial}{\partial \alpha_k}\left|\frac{\partial r_{i, k}}{\partial c}\right| 
&= \frac{2\gamma_{mm} |k-c|}{(1 +\gamma_{mm}\alpha_k (k-c)^2)^2} > 0.\notag
\end{myalign}
We had from \ref{item:alphak-in-terms-of-delta} that $\frac{d \alpha_k}{d |k-c|}<0$, and therefore:
\begin{myalign}
\frac{\partial}{\partial |k-c|}\left|\frac{\partial r_{k}}{\partial c}\right| =  \frac{\partial}{\partial \alpha_k}\left|\frac{\partial r_{k}}{\partial c}\right|\cdot \frac{d \alpha_k}{d |k-c|} < 0, \text{ Q.E.D.}\notag
\end{myalign}

Using (\ref{eq:high-sinr-ci-deriv}), we can write $\frac{\partial r_i}{\partial c}$ as: 
\begin{myalign}
\frac{\partial r_i}{\partial c} &= \sum_{k=1}^K \frac{2 \gamma_{mm} \alpha_k (k-c)}{1 + \gamma_{mm} \alpha_k(k-c)^2}\notag\\
& = \sum_{i=1}^{\lfloor c \rfloor} \frac{2 \gamma_{mm} \alpha_i (i-c)}{1 + \gamma_{mm} \alpha_i(i-c)^2} + \sum_{j=\lfloor c \rfloor + 1}^K \frac{2 \gamma_{mm} \alpha_j (j-c)}{1 + \gamma_{mm} \alpha_j(j-c)^2}.\notag
\end{myalign}
If $c \in [1, \frac{K+1}{2})$, then, from \ref{item:pairing}, each term $i$ in the left summation 
can be paired to a term $j$ in the right summation
, such that all the pairs are disjoint and for each pair $(i, j)$: $|i-c|<|j-c|$. From \ref{item:dr-increases-as-alpha-decreases}, for each such pair $(i, j)$: $\frac{2 \gamma_{mm} \alpha_i |i-c|}{1 + \gamma_{mm} \alpha_i(i-c)^2} < \frac{2 \gamma_{mm} \alpha_j |j-c|}{1 + \gamma_{mm} \alpha_j(j-c)^2}$. As all the terms in the left summation are negative, and all the terms in the right summation are positive, it follows that:
\begin{myalign}
\frac{\partial r}{\partial c} &= \sum_{i=1}^{\lfloor c \rfloor} \frac{2 \gamma_{mm} \alpha_i (i-c)}{1 + \gamma_{mm} \alpha_i(i-c)^2} + \sum_{j=\lfloor c \rfloor + 1}^K \frac{2 \gamma_{mm} \alpha_j (j-c)}{1 + \gamma_{mm} \alpha_j(j-c)^2}\notag\\
& = -\sum_{i=1}^{\lfloor c \rfloor} \frac{2 \gamma_{mm} \alpha_i |i-c|}{1 + \gamma_{mm} \alpha_i(i-c)^2} + \sum_{j=\lfloor c \rfloor + 1}^K \frac{2 \gamma_{mm} \alpha_j |j-c|}{1 + \gamma_{mm} \alpha_j(j-c)^2} > 0.\notag
\end{myalign}

Proving that $\frac{\partial r}{\partial c} < 0$ for $c \in (\frac{K+1}{2}, K]$ is symmetrical to the proof that $\frac{\partial r}{\partial c} > 0$ for $c \in [1, \frac{K+1}{2})$. As $\frac{\partial r}{\partial c} = 0$ for $c = \frac{K+1}{2}$, at the globally maximum $r$ we have that $c = \frac{K+1}{2}$.
\end{IEEEproof}

\end{document}